\documentclass[12pt]{article}
\usepackage[utf8]{inputenc}
\usepackage[top=50pt,bottom=50pt,left=68pt,right=66pt]{geometry}
\usepackage{amsmath}
\usepackage{bm}
\usepackage{booktabs}
\usepackage{amssymb}
\usepackage[title]{appendix}
\usepackage{mathrsfs}
\usepackage{cite}
\usepackage{float}
\usepackage{xcolor}
\usepackage{multirow}
\usepackage{graphicx,caption,subcaption}
\usepackage[space]{grffile}
\usepackage{ upgreek }

\renewcommand{\arraystretch}{1.3}
\definecolor{darkblue}{rgb}{0.1,0.1,.7}
\usepackage[breaklinks=true,linktocpage=true,
colorlinks=true,urlcolor=darkblue,linkcolor=darkblue,
citecolor=darkblue,pdfpagelabels=true,hypertexnames=true,
plainpages=false,naturalnames=false,]{hyperref}

\usepackage{datetime}
\newdateformat{monthyeardate}{%
  \monthname[\THEMONTH], \THEYEAR}
\date{\monthyeardate\today}

\def\ca{{\cal A}}

\def\ce{{\cal E}}

\def\co{{\cal O}}
\def\cp{{\cal P}}

\def\car{{\cal R}}

\begin{document}

\renewcommand{\arraystretch}{1.3}
\newcommand{\R}[1]{\textcolor{red}{#1}} 
\newcommand{\Eval}[1]{\langle #1 \rangle} 
\newcommand{\RM}[1]{\mathrm{#1}}
\thispagestyle{empty}

{\hbox to\hsize{\vbox{\noindent\monthyeardate\today}}}

\noindent
\vskip2.0cm
\begin{center}

{\Large\bf Towards Stochastic Inflation in Higher-Curvature\\
\vspace{5pt}
Gravity}

\vglue.3in

Yermek Aldabergenov,${}^{a,}$\footnote{ayermek@fudan.edu.cn (corresponding author)} Ding Ding,${}^{a,}$\footnote{ding\_ding@fudan.edu.cn/iasdding@ust.hk} Wei Lin,${}^{a,}$\footnote{wlin24@m.fudan.edu.cn} Yidun Wan${}^{a,b,}$\footnote{ydwan@fudan.edu.cn (corresponding author)}
\vglue.1in

${}^a$~{\it State Key Laboratory of Surface Physics, Center for Astronomy and Astrophysics, Department of Physics, Center for Field Theory and Particle Physics, and Institute for Nanoelectronic devices and Quantum computing, Fudan University,
 2005 Songhu Road, Shanghai 200433, China}\\
${}^b$~{\it Shanghai Research Center for Quantum Sciences, 99 Xiupu Road, Shanghai 201315, China}\\
\vglue.1in

\end{center}

\vglue.3in

\begin{center}
{\Large\bf Abstract}
\vglue.2in
\end{center}

We study stochastic inflation in the presence of higher-curvature terms non-minimally coupled to the inflaton. Focusing on quadratic curvature invariants, we single out the Gauss--Bonnet term which is known to avoid ghosts, while having non-trivial effects on the background and scalar mode evolution when coupled to the scalar field. Stochastic Klein--Gordon and Langevin equations are derived in the presence of the Gauss--Bonnet coupling, and their slow-roll and ultra-slow-roll limits are studied. By using first-passage time method, scalar power spectrum and PBH mass fraction are estimated in these limits. Stochastic evolution of a Gauss--Bonnet-coupled spectator field in de Sitter vacuum is also discussed.

\newpage

\tableofcontents

\setcounter{footnote}{0}

\section{Introduction}

Stochastic approach to inflation incorporates quantum effects to background evolution by parametrizing them as a classical stochastic force (some early works can be found in \cite{starobinsky1986stochastic,Nambu:1987ef,Nambu:1988je,Kandrup:1988sc,Nakao:1988yi,Nambu:1989uf,Mollerach:1990zf,Linde:1993xx,Starobinsky:1994bd}). Stochastic approach is especially relevant for scalar fields whose potentials are so flat that quantum diffusion becomes important. Such scenarios include ultra-slow-roll (USR) inflation and evolution of massless/light scalars in de Sitter vacuum. We aim to extend stachastic inflationary formalism in the presence of higher-curvature corrections to general relativity (GR). As will be motivated below, we focus on the Gauss--Bonnet (GB) term, coupled to a scalar field (inflaton or a spectator field). Such coupling modifies the dynamics of the scalar field, and admits new USR solutions for example. We obtain the following key results:
\begin{itemize}
    \item Stochastic Langevin equations in the presence of GB-inflaton coupling, by using spatially flat gauge and consistently eliminating metric perturbations.
    \item Analytical slow-roll (SR) and USR expressions for the scalar power spectrum with leading-order noise corrections.
    \item Numerical results of Primordial Black Hole mass fractions in stochastic-noise-dominated USR inflation regimes.
    \item Stochastic evolution of a GB-coupled spectator field in de Sitter vacuum.
\end{itemize}

{\bf Background and motivation}. The paradigm of inflation has displayed tremendous success in explaining
the origin of cosmic structure using only a relative small number of assumptions.
Among them are the quantum initial conditions of primordial fluctuations,
where the uncertainty principle ensures a non-vanishing initial quantum
fluctuation,
from which the stars and galaxies are seeded. Yet, one has to find a certain definite observational signature that favors
a \textit{bona fide} quantum distribution of primordial fluctuations. An inexhaustive list of attempts 
to directly analyze the quantumness of structural origin of our universe
include: search for cosmic bell inequalities 
\cite{Campo:2005quantum,Campo:2005inflationary,Gallicchio:2013testing,Maldacena:2015amodel},
ruling out uniquely classical features in non-Gaussian correlators \cite{Green:2020signals},
and adopting tools from quantum optics and information
\cite{Nambu:2011classical,Lim:2014quantum,martin2016quantum}.
The typical theoretical foundation used in this context is 
cosmological perturbation theory (CPT) \cite{Mukhanov:1990theory},
where gauge invariant degrees of freedom in fluctuations are isolated
and treated as perturbations on an FLRW background.
These perturbations are then quantized and analyzed using tools from
curved spacetime QFT.

Despite CPT's success,
in its most commonly used version
the quantum backreaction of small-scale modes 
on the background geometry has been neglected---not without its price---it 
effectively rules out an avenue of exploration
when one's goal is to verify the quantumness of the early universe.
The assumption that 
the background evolves classically and independently
from the quantum perturbations
is valid in most cases of inflation.
Nevertheless, there are regimes such as the USR phase of inflation,
where the force provided by the classical background diminishes to the point
where quantum backreaction of small-scale modes cannot be ignored,
and thus, invalidating standard CPT assumptions.
It just so happens that rich and potentially observable signals
may arise in the USR regime \cite{Ivanov:1997nonlinear,Garcia-Bellido:2017primordial},
which would be unwise to ignore.

To amend this oversight,
the stochastic inflation (SI) formalism has been developed.
In its standard version, instead of perturbing around a homogeneous background,
a (time-dependent) window function is introduced to isolate the small-scale (UV)
modes from the large-scale (IR) ones.
If the amplitude of the UV modes are small compared with the
IR ones, it can be described as noise correction to the stochastic
evolution of the IR modes.
This gives rise to a Langevin equation when one performs this UV-IR split
in a Klein-Gordon (KG) equation on fixed FLRW background.
Leading order backreaction is then captured by the inflow of UV
modes into IR ones as the former get stretched by inflation.

The remedy provided by SI is not completely flawless.
In particular, the consistency between SI and CPT,
often centered around gauge-invariance, has been up for debate recently\cite{Pattison:2019stochastic,Artigas:2021Hamiltonian,Artigas:2023kyo}.
To address this, analysis have been carried out on the gauge-fixing procedure \cite{Artigas:2021Hamiltonian,Artigas:2023kyo},
on a complete GR derivation of stochastic formalism \cite{Pinol:2020cdp,Launay:2024qsm},
and on revisiting the separate universe approach (SUA)
\cite{Salopek:1990nonlinear,Wands:2000anew}, which is a crucial assumption of SI 
\cite{starobinsky1986stochastic,Vennin:2015correlation}.

In our paper, we wish to extend the application of SI and the analysis of its consistency to modified higher-curvature gravity. Higher-curvature corrections to GR can arise from quantum corrections and affect the dynamics of the early universe. In principle, various higher-order curvature invariants can be constructed from scalar curvature, Ricci tensor, and Riemann tensor; however, such theories in general suffer from Ostrogradski instability due to having higher time derivatives of the metric. One possible solution to this instability is to restrict the theories to at most two derivatives of the metric. This famously leads to Lovelock gravity \cite{Lovelock:1971yv} (see also Ref. \cite{Padmanabhan:2013xyr} for a review), whose leading correction to general relativity is the GB term,
\begin{equation}
    R^2_{\rm GB}\equiv R^2-4R_{\mu\nu}R^{\mu\nu}+R_{\mu\nu\rho\sigma}R^{\mu\nu\rho\sigma}~.
\end{equation}
In four spacetime dimensions, Lovelock gravity reduces to general relativity plus the GB term, which is an Euler density. Nontheless, once matter fields are included, they can couple to the GB term and hence modify the corresponding equations of motion (EoM). For example, in the case of the inflaton field, this coupling changes both the background equations and EoM for primordial perturbations.

The GB term can also be motivated from the point of view of effective field theory of inflation, where perturbative curvature corrections to an inflationary theory in GR can be reduced to the GB term and the gravitational Chern--Simons (CS) term \cite{Weinberg:2008hq}. Moreover, the GB term has been shown to appear in string effective actions as the leading $\alpha'$ (inverse string tension) or loop corrections, where it couples to the moduli fields and the dilaton \cite{Antoniadis:1993jc,Kawai:1998ab,Kawai:1999pw}.

These developments have lead to extensive studies of GB-scalar couplings in cosmology, e.g., in the context of inflaton \cite{Hwang:1999gf,Cartier:2001is,Guo:2006ct,Satoh:2008ck,Guo:2009uk,Guo:2010jr,Jiang:2013gza,Koh:2014bka,Yi:2018gse,Odintsov:2018zhw,Odintsov:2019clh,Gao:2020cvb,Kawai:2021bye,Kawai:2023nqs,Addazi:2024gew,Aldabergenov:2025oys} and quintessence dark energy \cite{Nojiri:2005vv,Calcagni:2005im,Nojiri:2006je,Tsujikawa:2006ph,Koivisto:2006ai,Neupane:2006dp,Granda:2017oku,Chatzarakis:2019fbn,Pozdeeva:2019agu,Vernov:2021hxo,MohseniSadjadi:2023amn,Pinto:2024dnm,TerenteDiaz:2023iqk,TerenteDiaz:2023kgc}. It has also been shown that inflationary (quasi-de Sitter) solutions in the presence of the GB term admit (GB-induced) USR regimes. This can lead, e.g., to the amplification of primordial perturbations and subsequent generation of primordial black holes (PBH) \cite{Kawai:2021edk,Kawaguchi:2022nku,Ashrafzadeh:2023ndt,Solbi:2024zhl}. Thus, it is important to adapt the tools from SI to incorporate higher curvature gravity, and to check the consistency between CPT and SI in this case. Our paper aims to provide an initial analysis in this regard. The consistency of higher-curvature corrections in a quantum backreaction description, such as SI, is of particular interest. On the theoretical side,
gauge related issues are even more non-trivial,
as higher-derivative terms alter the constraints that generate gauge transformations
and also challenge the validity of SUA. On the observational side, it would be interesting for proponents of higher-curvature gravity to see if their models might survive observational constraints when quantum backreaction of UV modes is included.

Our paper is organized as follows. We first discuss quadratic curvature corrections in Sec. \ref{Sec_quadratic_corrections}, where we write down our starting Lagrangian and equations of motion. In Sec. \ref{Sec_GB_inflation} we introduce and review GB-coupled inflation. Although in the literature, scalar modes of GB-coupled theories are usually discussed in uniform-field gauge, a more convenient choice for stochastic inflation is spatially flat gauge. Thus, in the end of Sec. \ref{Sec_GB_inflation} we re-derive the scalar mode equation by using spatially flat gauge, and show that it matches the result from uniform-field gauge. Our main results are presented in Sec. \ref{Sec_stochastic}, where we verify the validity of separate universe approach in GB-coupled inflation, and derive stochastic equations of motion in this case. In Sec. \ref{Sec_FPT} we use first-passage time method to find analytical expressions for scalar power spectrum in SR and USR limits as well as PBH mass fractions in stochastic-noise-dominated USR regimes. In Sec. \ref{Sec_spectator} we discuss the case of a GB-coupled spectator field in de Sitter background, and its stochastic evolution. This is followed by a discussion section.

\section{Inflation with quadratic curvature terms}\label{Sec_quadratic_corrections}

Consider a scalar field theory coupled to quadratic curvature invariants, where as ingredients we use the scalar curvature $R$, Ricci tensor $R_{\mu\nu}$, and Riemann tensor $R_{\mu\nu\rho\sigma}$. We write the Lagrangian as~\footnote{We use Planck units, $M_P=1$, unless otherwise stated, and $R_{\mu\nu}={R^{\rho}}_{\mu\rho\nu}$.}
\begin{equation}
\sqrt{-g}^{\,-1}\mathcal{L} = \tfrac{1}{2}R-\tfrac{1}{2}(\partial \phi)^2-\tfrac{1}{8}\sum_i\xi_i(\phi)I_i-V(\phi)~,\label{L_general}
\end{equation}
where $V(\phi)$ is a scalar potential for the real scalar $\phi$. The quadratic curvature invariants are
\begin{gather}\label{I_1_4}
\begin{gathered}
    I_1=R^2~,~~~I_2=R_{\mu\nu}R^{\mu\nu}~,~~~I_3=R_{\mu\nu\rho\sigma}R^{\mu\nu\rho\sigma}~,\\
    I_4=R\tilde R\equiv\tfrac{1}{2}\epsilon^{\mu\nu\rho\sigma}R_{\mu\nu\kappa\lambda}{R^{\kappa\lambda}}_{\rho\sigma}~,
\end{gathered}
\end{gather}
and the corresponding non-minimal coupling functions are $\xi_i(\phi)$. We use the convention where $\varepsilon_{\mu\nu\kappa\lambda}$ is the Levi-Civita symbol, while $\epsilon_{\mu\nu\kappa\lambda}=\sqrt{-g}\varepsilon_{\mu\nu\kappa\lambda}$ is the Levi-Civita tensor with $\epsilon^{\mu\nu\kappa\lambda}=\varepsilon^{\mu\nu\kappa\lambda}/\sqrt{-g}$.

As mentioned in the introduction, a general higher-curvature theory given by the Lagrangian \eqref{L_general} introduces additional degrees of freedom and suffers from various instabilities such as the Ostrodgradski ghost and gradient and/or tachyonic instabilities. In order to avoid these problems, we can put the quadratic invariants into the Gauss--Bonnet combination,
\begin{equation}
    R^2_{\rm GB}\equiv R^2-4R_{\mu\nu}R^{\mu\nu}+R_{\mu\nu\rho\sigma}R^{\mu\nu\rho\sigma}~.
\end{equation}
This term by itself is a topological density in the action, but it is not when coupled to for example a scalar field, and the EoM will be modified---without the introduction of higher derivatives of the metric or the scalar field.

Another topological term that can appear in the action is the gravitational Chern--Simons term $I_4=R\tilde R$ given in Eq. \eqref{I_1_4}. Unlike the GB term, the CS term (coupled to the scalar field) does not affect the background EoM or scalar perturbations, and only affects tensor mode propagation. (The CS term also introduces third derivative of the metric in the Einstein equations for tensor modes, which needs to be suppressed.) Therefore, we can ignore the CS term when discussing stochastic inflation, and focus on the GB term, where we fix the Lagrangian as
\begin{equation}\label{L_sGB}
\sqrt{-g}^{\,-1}\mathcal{L} = \tfrac{1}{2}R-\tfrac{1}{2}(\partial \phi)^2-\tfrac{1}{8}\xi(\phi) R_{GB}^{2}-V(\phi)
~.
\end{equation}
Here, $\xi(\phi)$ is a general GB coupling function. The corresponding Einstein equations read
\begin{align}
\begin{split}\label{EFE}
    R_{\mu\nu}&-\tfrac{1}{2}g_{\mu\nu}R+\xi_{,\phi}\big[(R_{\mu\nu}-\tfrac{1}{2}g_{\mu\nu}R)\Box\phi+\tfrac{1}{2}R\nabla_\mu\nabla_\nu\phi+g_{\mu\nu}R^{\alpha\beta}\nabla_\alpha\nabla_\beta\phi\\
    &\hspace{2.8cm}-R_{\mu\alpha}\nabla^\alpha\nabla_\nu\phi-R_{\nu\alpha}\nabla^\alpha\nabla_\mu\phi+R_{\mu\alpha\beta\nu}\nabla^\alpha\nabla^\beta\phi\big]\\
    &+\xi_{,\phi\phi}\big[(R_{\mu\nu}-\tfrac{1}{2}g_{\mu\nu}R)\partial\phi\partial\phi+\tfrac{1}{2}R\partial_\mu\phi\partial_\nu\phi+g_{\mu\nu}R^{\alpha\beta}\partial_\alpha\phi\partial_\beta\phi\\
    &\hspace{1.3cm}-R_{\mu\alpha}\partial^\alpha\phi\partial_\nu\phi-R_{\nu\alpha}\partial^\alpha\phi\partial_\mu\phi+R_{\mu\alpha\beta\nu}\partial^\alpha\phi\partial^\beta\phi\big]=T_{\mu\nu}~.
\end{split}
\end{align}
The Klein--Gordon equation derived from \eqref{L_sGB} is
\begin{equation}\label{KG_general}
    \nabla_\mu\nabla^\mu\phi-\tfrac{1}{8}\xi_{,\phi}R^2_{\rm GB}-V_{,\phi}=0~.
\end{equation}
Equations \eqref{L_sGB}, \eqref{EFE}, and \eqref{KG_general} will be our starting point of discussion.

\section{Gauss--Bonnet-coupled inflation}\label{Sec_GB_inflation}
In this section we will review GB-coupled inflation. We derive EoM for the scalar perturbation in uniform-field gauge, which is the most popular choice in GB literature, and in spatially flat gauge, which is more convenient in stochastic formalism.

\subsection{Background equations and slow-roll conditions}

Background equations are given by ($\bar\phi$ is the homogeneous part of $\phi$)
\begin{align}\label{backg_EOM_phi}
    \ddot{\bar\phi}+3H\dot{\bar\phi}+V_{,\phi}+3\xi_{,\phi} H^2(H^2+\dot{H}) &=0~,\\
    \label{backg_EOM_Hdot}
    2\dot H(1-\dot\xi H)-\ddot\xi H^2+\dot\xi H^3+\dot{\bar\phi}^2 &=0~,\\
    3H^2(1-\dot\xi H)-\tfrac{1}{2}\dot{\bar\phi}^{2}-V &=0~,\label{backg_EOM_H}
\end{align}
where $V$, $\xi$, and their derivatives should be understood as background quantities.

It is convenient to introduce the following slow-roll parameters,
\begin{equation}\label{SR_pars}
    \epsilon=-\frac{\dot H}{H^2}~,~~~\eta=\frac{\dot\epsilon}{H\epsilon}~,~~~\omega=\dot\xi H~,~~~\sigma=\frac{\dot\omega}{H\omega}~,
\end{equation}
which render the background equations as
\begin{align}\label{backg_EOM_phi_N}
    \bar\phi_{,NN}+(3-\epsilon)\bar\phi_{,N}+V_{,\phi}H^{-2}+3\xi_{,\phi} H^2(1-\epsilon) &=0~,\\
    \label{backg_EOM_Hdot_N}
    \bar\phi_{,N}^2-2\epsilon+\omega+\epsilon\omega-\omega\sigma &=0~,\\
    3H^2(1-\omega)-\tfrac{1}{2}H^2\bar\phi_{,N}^2-V &=0~,\label{backg_EOM_H_N}
\end{align}
where for convenience we switched to the e-fold time $N$ by using $\dot N=H$. The conditions for slow-roll inflation can be written as
\begin{equation}   |\epsilon|,|\eta|,|\omega|,|\sigma|\ll 1~,\label{GB_SR}
\end{equation}
It should also be mentioned that $\epsilon$ can in principle turn negative during possible slow-roll violations (due to the GB contribution), as can be seen from Eq. \eqref{backg_EOM_Hdot_N}.

\subsection{Scalar perturbation in uniform-field gauge}

In GB-coupled inflation, scalar perturbations are often studied in the uniform-field gauge, $\delta\phi=0$. We first derive the usual (linear) scalar perturbation and then move on to the stochastic approach, where the stochastic noise will be determined from the solution to the scalar mode equation.

In the uniform-field gauge, the genuine scalar perturbation is associated with the spatial metric perturbation $\zeta$. It is introduced by the ADM metric (spatial indices are raised and lowered with $\delta_{ij}$)
\begin{equation}\label{ADM_metric}
    ds^2=-(1+\alpha)^2dt^2+a^2e^{2\zeta}\delta_{ij}(a^{-2}\partial^i\beta dt+dx^i)(a^{-2}\partial^j\beta dt+dx^j)~,
\end{equation}
where $\alpha$ is the perturbation of the lapse function, and $\partial_i\beta$ is the shift vector. After linearizing the Einstein equations \eqref{EFE}, we get the constraint equations for $\alpha$ and $\beta$:
\begin{align}
    &(2V-6H^2\omega)\alpha+(2-3\omega)H\frac{\partial_i^2}{a^2}\beta-3(2-3\omega)H\dot\zeta+2(1-\omega)\frac{\partial_i^2}{a^2}\zeta=0~,\\
    &(2-3\omega)H\alpha-2(1-\omega)\dot\zeta=0~,
\end{align}
from $00$ and $0i$ components of the Einstein equations, respectively. The perturbations $\alpha$ and $\beta$, found from the equations above, can be inserted into the Einstein equations for $i\neq j$ and leads to
\begin{equation}
    \alpha+\dot\beta+\Big(1-\frac{\omega\sigma}{1-\omega}\Big)H\beta+\frac{1-\epsilon\omega-\omega\sigma}{1-\omega}\zeta=0~,
\end{equation}
which results in the equation of motion for the (scalar) $\zeta$-perturbation. In Fourier space, it reads
\begin{equation}\label{EOM_zeta_t}
    \ddot\zeta_k+\Big(H+\frac{(Z^2)_{,t}}{Z^2}\Big)\dot\zeta_k+C_s^2\frac{k^2}{a^2}\zeta_k=0~.
\end{equation}
where we define
\begin{align}
    Z^2 &\equiv \frac{a^2(1-\omega)^2}{(1-\tfrac{3}{2}\omega)^2}\Big[\frac{\dot{\bar\phi}^2}{H^2}+\frac{3\omega^2}{2(1-\omega)}\Big]~,\label{Z_def}\\
    C_s^2 &\equiv 1-2a^2\omega^2\,\frac{\epsilon+\frac{1}{4}\omega(1-5\epsilon-\sigma)}{Z^2(1-\tfrac{3}{2}\omega)^2}~.\label{C_s_def}
\end{align}
Here $C_s$ is the effective sound speed of the $\zeta$-perturbation. Alternatively, Eq. \eqref{EOM_zeta_t} can be obtained directly from the quadratic action for the $\zeta$-perturbation, see e.g. \cite{Kawai:2021bye}.

It is handy to rewrite \eqref{EOM_zeta_t} as
\begin{equation}\label{EOM_u_tau}
    u_k''-\frac{Z''}{Z}u_k+C_s^2k^2u_k=0~,
\end{equation}
where $u_k\equiv Z\zeta_k$, and $'\equiv\partial_\tau$ with conformal time $\tau$. 

\textbf{Constant-roll and slow-roll limits}. Analytical solution to \eqref{EOM_u_tau} can be found in the constant-roll limit~\footnote{We refer to the case with $\epsilon={\rm const.}$ as ``constant-roll", although some authors use this term for the constant $\eta$ case.}, when $\epsilon,\omega\simeq {\rm const}$, which leads to
\begin{equation}
    \frac{Z''}{Z}\simeq\frac{2-\epsilon}{(1-\epsilon)^2\tau^2}~,~~~\tau\simeq-\frac{1}{(1-\epsilon)aH}~,~~~C_s\simeq {\rm const}~.
\end{equation}

The (physically relevant) solution then reads \cite{Cartier:2001is,Guo:2006ct,Guo:2009uk}
\begin{equation}
    u_k\simeq \tfrac{1}{2}\sqrt{\pi|\tau|}e^{i\pi(1+2\nu_s)/4}H^{(1)}_{\nu_s}(C_sk|\tau|)~,
\end{equation}
where $\nu_s^2\equiv \frac{1}{4}+\frac{2-\epsilon}{(1-\epsilon)^2}$, and $H^{(1)}_{\nu_s}(C_sk|\tau|)$ is the Hankel function of the first kind. In the superhorizon limit, $C_sk|\tau|\ll 1$, we can write
\begin{equation}
    H_{\nu_s}^{(1)}(C_sk|\tau|)\simeq\frac{J_{-\nu_s}(C_sk|\tau|)-e^{-i\pi\nu_s}J_{\nu_s}(C_sk|\tau|)}{i\sin(\pi\nu_s)}~,~~~J_{\nu_s}(C_sk|\tau|)\simeq\frac{(C_sk|\tau|/2)^{\nu_s}}{\Gamma(1+\nu_s)}~,
\end{equation}
where $J_{\nu_s}(C_sk|\tau|)$ is the Bessel function of the first kind (its superhorizon approximation is shown), and $\Gamma$ is the gamma function.

The power spectrum of $\zeta$ takes the form
\begin{equation}
    P_\zeta=\frac{k^3|u_k|^2}{2\pi^2Z^2}\Big|_{C_sk=aH}\simeq\frac{2^{2\nu_s-3}a^2H^2C_s^{-3}Z^{-2}}{\pi\sin^2(\pi\nu_s)\Gamma^2(1-\nu_s)(1-\epsilon)}~,
\end{equation}
where we used $|\tau|\simeq\frac{1}{(1-\epsilon)aH}$. Expanding this expression in $\epsilon$ and $\omega$, together with using $C_s\simeq 1$, $Z^2/a^2\simeq \dot{\bar\phi}^2/H^2$, and $\nu_s\simeq 3/2$, results in the slow-roll limit,
\begin{equation}\label{P_zeta_SR}
    P_\zeta\simeq\frac{H^4}{4\pi^2\dot{\bar\phi}^2}~.
\end{equation}
This coincides with the standard result of slow-roll inflation, but when expressing $\dot{\bar\phi}^2$ in terms of the slow-roll parameter $\epsilon$, one has to take into account GB corrections as shown in Eq. \eqref{backg_EOM_Hdot} or \eqref{backg_EOM_Hdot_N}. In the constant-roll case, we have $\sigma\simeq 0$, so only $\omega$-correction is present: $\dot{\bar\phi}^2/H^2\simeq 2\epsilon-\omega$.

\textbf{Ultra-slow-roll limit}. As in general relativity, USR inflation in GB models is characterized by the equation $\ddot{\bar\phi}+3H\dot{\bar\phi}\simeq 0$, which is obtained from \eqref{backg_EOM_phi} when the slope of the effective potential vanishes,
\begin{equation}
    V_{,\phi}^{\rm eff}=V_{,\phi}+3\xi_{,\phi}H^4\simeq 0~.
\end{equation}
In terms of the slow-roll parameters, it is convenient to introduce the velocity-based slow-roll parameter,
\begin{equation}
    \ce\equiv\bar\phi_{,N}^2/2~.
\end{equation}
We can then write (in terms of the e-fold time),
\begin{equation}
    2\bar\phi_{,NN}/\bar\phi_{,N}=\ce_{,N}/\ce\simeq -6~,
\end{equation}
during USR, whose solution can be written as
\begin{equation}\label{phi_cl_USR}
    \phi(N)\simeq \phi(N_{\rm in})+\tfrac{1}{3}\phi_{,N}(N_{\rm in})(1-e^{-3(N-N_{\rm in})})~,
\end{equation}
where $N_{\rm in}$ is the time at the start of USR.

Let us also show that in general, $\eta\equiv\epsilon_{,N}/\epsilon\neq -6$ during USR in GB-coupled inflation. We will use Eq. \eqref{backg_EOM_Hdot_N} in the form
\begin{equation}\label{CE_eq}
    \ce=\epsilon-\omega(1+\epsilon)+\omega_{,N}~.
\end{equation}
By taking its derivative, and using $\ce_{,N}/\ce\simeq -6$ and $|\epsilon|,|\omega|,|\omega_{,N}|,|\omega_{,NN}|\ll 1$~\footnote{We require all these slow-roll parameters to be small in order to keep the velocity parameter $\ce$ small, as required by USR. This can be shown by taking derivatives of \eqref{CE_eq}.}, we get
\begin{equation}
    \eta\simeq -6+3\frac{\omega}{\epsilon}-\frac{5\omega_{,N}}{2\epsilon}-\frac{\omega_{,NN}}{2\epsilon}~.
\end{equation}
As can be seen, GB contributions are in the form of the ratios $\omega/\epsilon$, $\omega_{,N}/\epsilon$, and $\omega_{,NN}/\epsilon$. This means that even if $\omega$, $\omega_{,N}$, and $\omega_{,NN}$ are small, their ratios to $\epsilon$ can be large, so that $\eta\neq -6$.

Considering the evolution of the scalar mode, given by \eqref{EOM_u_tau}, we can find that under USR, $Z''/Z\simeq 2/\tau^2\simeq 2a^2H^2$, as in GR. Unlike in GR, however, the sound speed $C_s$ can in principle deviate from unity in specific circumstances. For example, if $Z^2/a^2$ becomes much smaller than $\omega^2\epsilon$ or $\omega^3$, such that the second term of \eqref{C_s_def} becomes large. In such cases, Eq. \eqref{EOM_u_tau} is to be solved numerically. Otherwise, if $C_s\simeq 1$, the solution to \eqref{EOM_u_tau} coincides with the GR result, namely
\begin{equation}
    u_k\simeq \frac{e^{-ik\tau}}{\sqrt{2k}}\Big(1+\frac{i}{k|\tau|}\Big)~,
\end{equation}
and the power spectrum is given by \eqref{P_zeta_SR}.

\subsection{Scalar perturbation in spatially flat gauge}

A more natural gauge choice for stochastic inflation is spatially flat gauge, where we set $\zeta=0$ in the metric tensor \eqref{ADM_metric}, while $\delta\phi$ acts as the genuine scalar perturbation. 

In order to derive the quadratic action for the scalar perturbation $\delta\phi$, we use the metric and its inverse, expanded as
\begin{gather}\label{SF_metric}
\begin{gathered}
    g_{00}=-(1+2\alpha+\alpha^2)+a^{-2}\partial_i\beta\partial^i\beta~,~~~g_{0i}=\partial_i\beta~,~~~g_{ij}=a^2\delta_{ij}~,\\
    g^{00}=-(1-2\alpha+3\alpha^2)~,~~~g^{0i}=a^{-2}(1-2\alpha)\partial^i\beta~,~~~g^{ij}=a^{-2}\delta^{ij}-a^{-4}\partial^i\beta\partial^j\beta~.
\end{gathered}
\end{gather}
The scalar curvature and the GB term expanded up to the second order in $\alpha$ and $\beta$ can be found in Appendix \ref{App_A}.

Using the perturbed scalar $\phi(t,x)=\bar\phi(t)+\delta\phi(t,x)$, and the metric \eqref{SF_metric}, the quadratic action derived from \eqref{L_sGB} takes the form
\begin{align}\label{L_quad_SF}
\begin{aligned}
    a^{-3}\mathcal{L}=-(3H^2-\tfrac{1}{2}\dot{\bar\phi}^2-6\dot\xi H^3)\alpha^2-H(2-3\dot\xi H)\alpha\frac{\partial_i^2}{a^2}\beta+\tfrac{1}{2}\delta\dot\phi^2-(\dot{\bar\phi}+3\xi_{,\phi}H^3)\alpha\delta\dot\phi\\
    -(3\xi_{,\phi\phi}H^3\dot{\bar\phi}+V_{,\phi})\alpha\delta\phi+\xi_{,\phi}H^2\alpha\frac{\partial_i^2}{a^2}\delta\phi+(\dot{\bar\phi}+\xi_{,\phi}H^3-\xi_{,\phi\phi}H^2\dot{\bar\phi})\delta\phi\frac{\partial_i^2}{a^2}\beta\\
    -\xi_{,\phi}H^2\delta\dot\phi\frac{\partial_i^2}{a^2}\beta+\tfrac{1}{2}\delta\phi\frac{\partial_i^2}{a^2}\delta\phi-\tfrac{1}{2}\big[V_{,\phi\phi}+3\xi_{,\phi\phi}H^4(1-\epsilon)\big]\delta\phi^2~.
\end{aligned}
\end{align}

After eliminating $\alpha$ and $\beta$, integrating by parts, and some algebra, as outlined in Appendix \ref{App_A}, the final Lagrangian for $\delta\phi$ reads
\begin{equation}\label{L_dphi_final}
    \mathcal{L}=\frac{a^3}{2}\Big[X_1\delta\dot\phi^2+X_2\delta\phi\frac{\partial^2_i}{a^2}\delta\phi+\tfrac{1}{2}(a^3\sqrt{\ce})^{-1}\partial_t\Big(a^3\frac{\dot\ce}{\sqrt{\ce}}X_1\Big)\delta\phi^2\Big]~,
\end{equation}
where the following definitions are used,
\begin{align}\label{X12_def}
\begin{aligned}
    X_1 &\equiv\Big[1+\frac{3\omega^2}{4\ce(1-\omega)}\Big]\frac{(1-\omega)^2}{(1-\frac{3}{2}\omega)^2}~,\qquad\qquad\ce\equiv\frac{\dot{\bar\phi}^2}{2H^2}~,\\
    X_2 &\equiv\Big[1-\tfrac{1}{2}\omega+\frac{3\omega^2}{4\ce}(1-\epsilon)+\frac{3\omega^3\sigma}{8\ce(1-\frac{3}{2}\omega)}\Big]\frac{1}{1-\frac{3}{2}\omega}~.
\end{aligned}
\end{align}

From \eqref{L_dphi_final} follows the linearized equation of motion for the scalar mode $\delta\phi_k$:
\begin{equation}\label{dphi_EOM}
    \delta\ddot\phi_k+\Big(3H+\frac{\dot X_1}{X_1}\Big)\delta\dot\phi_k+\Big[\frac{X_2}{X_1}\frac{k^2}{a^2}-\tfrac{1}{2}(a^3\sqrt{\ce}X_1)^{-1}\partial_t\Big(a^3\frac{\dot\ce}{\sqrt{\ce}}X_1\Big)\Big]\delta\phi_k=0~.
\end{equation}

Let us show the relation between the EoM \eqref{dphi_EOM} for $\delta\phi$ and the EoM \eqref{EOM_zeta_t} for the spatial curvature perturbation $\zeta$ in the uniform-field gauge. The perturbation $\delta\phi$ can be related (in spatially flat gauge) to the gauge-invariant comoving curvature perturbation $\car$ as $\delta\phi=\sqrt{2\ce}\car$, while in the uniform-field gauge we have $\zeta=\car$. Therefore, one can show that Eq. \eqref{EOM_zeta_t} coincides with \eqref{dphi_EOM} after using $\delta\phi=\sqrt{2\ce}\zeta$, and notice that (up to the background equations)
\begin{equation}
    X_1=\frac{Z^2}{2a^2\ce}~,~~~\frac{X_2}{X_1}=C_s^2~,
\end{equation}
where $Z$ and $C_s$ are given by \eqref{Z_def} and \eqref{C_s_def}.

\section{Stochastic Gauss--Bonnet-coupled inflation}\label{Sec_stochastic}

In this section we apply stochastic formalism to GB-coupled inflation. We will first check the validity of SUA in the presence of the GB term, and then derive stochastic Langevin equations for the inflaton. An example of a USR regime induced by the GB-term will be discussed as well.

\subsection{Separate universe approach}

Stochastic formalism of inflation relies on the SUA, which states that each super-Hubble patch evolves as a separate locally homogeneous and isotropic universe, such that super-Hubble perturbations act like local shifts in the FLRW background field values (see e.g. \cite{Vennin:2020kng} for more detailed discussion). 

To demonstrate the consistency between the SUA and CPT equations of motion (up to gradient terms)
without assuming a specific gauge,
\begin{equation}
    ds^2=-(1+2\alpha)dt^2
    +\partial_i\beta dt dx^i
    +a^2[(1+2\zeta)\delta_{ij}+\partial_i\partial_jE]dx^idx^j,
\end{equation}
we recall that the SUA, at an operational level, assumes local
coordinates and field values being
related to global ones via the perturbations
\begin{equation}
	dt\rightarrow (1+\alpha)dt\,,\ 
	\phi\rightarrow \bar{\phi}+\delta\phi\,,\ 
	\delta H=\frac{V_{,\phi} \delta\phi+\dot{\bar{\phi}}
	\delta\dot{\phi}
	-\dot{\bar{\phi}}^2\alpha}{6H}\,
	\label{}
\end{equation}
to a local KG equation on fixed FLRW.
(One can include the gradient-mass term for the field, but this term
becomes negligible in the leading-order
gradient expansion assumed by the standard
SI.)

The background KG equation (in GR) when perturbed via the SUA becomes~\footnote{
Here we have set $M_P=1$. To restore the mass dimension, the inverse $2H$
in the equation below becomes $2HM_P^2$.}
\begin{equation}
	\delta\ddot{\phi}+\Big(3H+\frac{\dot{\phi}^2}{2H}\Big)\dot{\delta\phi}
	+\Big(\frac{\dot{\phi}}{2H}V_{,\phi}+V_{,\phi\phi}\Big)\delta\phi
	=\dot{\phi}\dot{\alpha}
	+\Big(2\ddot{\phi}+3H\dot{\phi}+\frac{\dot{\phi}^3}{2H}\Big)\alpha\,.
	\label{KG-SUA-GR}
\end{equation}
We want to match this to the CPT field equation (up to gradient terms).

The KG equation (keeping the gradient-mass term)
for perturbations in CPT reads
\begin{equation}
	\delta\ddot{\phi}+3H\delta\dot{\phi}
	+\Big(-\frac{\partial_i^2}{a^2}+V_{,\phi\phi}\Big)\delta\phi
	=-2V_{,\phi}\alpha+\dot{\phi}\dot{\alpha}-3\dot{\phi}\dot{\zeta}
	-\dot{\phi}\partial_i^2(\dot{E}-B/a)\,.
	\label{KG-CPT-GR}
\end{equation}
One can recognize that the last two terms of the right-hand side (RHS) also appear in combination
in the CPT Hamiltonian constraint but missing a $\dot{\phi}$ factor.
Multiplying the CPT Hamiltonian constraint with $\dot{\phi}$
and
after rearranging, one has
\begin{equation}
	-3\dot{\phi}\dot{\zeta}-\dot{\phi}\partial_i^2(\dot{E}-\beta/a^2)
	=-\frac{1}{2H}[\dot{\phi}^2\delta\dot{\phi}-\dot{\phi}^3\alpha
	+\dot{\phi}V_{\phi}\delta\phi]
	-\frac{\dot{\phi}}{a^2H}\partial_i^2\zeta-3H\alpha\dot{\phi}\,.
	\label{}
\end{equation}
Plugging this equation into \eqref{KG-CPT-GR}, one obtains
\begin{equation}
	\delta\ddot{\phi}+(3H+\frac{\dot{\phi}^2}{2H})\delta\dot{\phi}
	+(V_{,\phi\phi}+\frac{V_{,\phi}\dot{\phi}}{2H})\delta\phi
	-\dot{\phi}\dot{\alpha}+(2V_{,\phi}+3H\dot{\phi}-\dot{\phi}^3/(2H))
	\alpha
	=\partial_i^2(\frac{\delta\phi}{a^2}-\frac{\dot{\phi}}{Ha^2}\zeta)\,.
	\label{}
\end{equation}
Using the background KG equation
$2V_{,\phi}=-2\ddot{\phi}-6H\dot{\phi}$, one finds that \eqref{KG-CPT-GR} above matches the SUA one \eqref{KG-SUA-GR} up to $O(\partial_i^2(\dots))$ gradient terms.

Now, we should check that the SUA indeed holds in our Gauss--Bonnet-coupled inflation. For simplicity, we work in the spatially
flat gauge where the inflaton field serves as the gauge-invariant
variable to be quantized.

By working in the spatially flat gauge, we will show that super-Hubble perturbations act as local shifts in background field values, i.e.,
\begin{equation}\label{def_loc_shift}
	dt \to \big[ {1 + \alpha(t)} \big]dt,~~~\bar \phi  \to \bar \phi  + \delta \phi(t)~.
\end{equation}
The shifts \eqref{def_loc_shift} to the FLRW background equation  \eqref{backg_EOM_phi} of $\bar\phi$ lead to
\begin{equation}\label{shi_backg_EOM_phi}
	\begin{gathered}
		- \dot \alpha \dot {\bar \phi}  - 2\alpha \ddot {\bar \phi}  + \delta \ddot \phi  + 3\delta H\dot {\bar \phi}  + 3H({ - \alpha \dot {\bar \phi}  + \delta \dot \phi } )  \hfill \\
		+12{H^3}\delta H{\xi _{,\phi }}\left( {1 - \epsilon } \right) + 3{H^4}{\xi _{,\phi \phi }}\delta \phi \left( {1 - \epsilon } \right) - 3{H^4}{\xi _{,\phi }}\delta \epsilon  + {V_{,\phi \phi }}\delta \phi  = 0 \hfill~,
	\end{gathered}
\end{equation}
at the leading order in the shifts, where the Hubble shift $\delta H$ can be calculated through the shift of $H$ through the background equation \eqref{backg_EOM_H} as
\begin{equation}\label{shi_backg_EOM_H}
	\delta H = \frac{{( {3{\xi _{,\phi \phi }}\dot {\bar \phi} {H^3} + {V_{,\phi }}} )\delta \phi  - ( {3{\xi _{,\phi }}\dot {\bar \phi} {H^3} + {\dot {\bar \phi}^2}} )\alpha  + ( {\dot {\bar \phi}  + 3{\xi _{,\phi }}{H^3}} )\delta \dot \phi }}{{6H - 9{\xi _{,\phi }}\dot {\bar \phi} {H^2}}}~,
\end{equation}
and we have $\delta \epsilon  =  -H^{-2}\delta \dot H + 2\dot H H^{-3}\delta H$. The shift $\delta {\dot{H}}$~\footnote{Not to be confused with $d(\delta H)/dt$, although for the scalar perturbation/shift $\delta\phi$, we use $d\delta\phi/dt\equiv\delta\dot\phi$ for notational convenience.}, however, needs to be calculated from $\delta H$ directly, as
\begin{equation}\label{shi_dot_H}
	\delta \dot H =  - \alpha \dot H + \frac{d}{{dt}}\left( {\delta H} \right)~,
\end{equation} 
rather than from the background equation \eqref{backg_EOM_Hdot}. Therefore, one can show that $\delta \phi$ satisfies 
\begin{equation}\label{shi_backg_dphi_EOM}
	\delta\ddot\phi+\Big(3H+\frac{\dot X_1}{X_1}\Big)\delta\dot\phi-\frac{1}{2}(a^3\sqrt{\ce}X_1)^{-1}\partial_t\Big(a^3\frac{\dot\ce}{\sqrt{\ce}}X_1\Big)\delta\phi=0~,
\end{equation}
by eliminating $\alpha$ by the constraint \eqref{C_beta_CPT}, and using background equations \eqref{backg_EOM_phi}--\eqref{backg_EOM_H} after some algebra. We see that Eq. \eqref{shi_backg_dphi_EOM} for the background field shift is indeed consistent with Eq. \eqref{dphi_EOM} for the field perturbation in the spatially flat gauge, if we take long-wavelength limit, $k\ll aH/C_s$. This confirms that the long-wavelength perturbations indeed act like local shifts of the background field values, such that the SUA holds for our Gauss--Bonnet-coupled inflation. Thus, when applying stochastic formalism to GB-coupled inflation, we can use the coarse-graining condition $k\ll aH/C_s$ as the criterion for the split of IR and UV modes.

\subsection{Stochastic formalism: general strategy}

In stochastic inflation, we split the fields into IR and UV parts, such as
\begin{equation}\label{phi_c_q}
    \phi(t,x)=\phi_c(t)+\phi_q(t,x)~,
\end{equation}
with the IR/classical part $\phi_c$, and the UV/quantum part $\phi_q$. We treat $\phi_c$ as an effective background comprised of long wavelength modes (in particular, we assume that
$\partial_i \phi_c$ is negligible
at the leading order of gradient expansion), while $\phi_q$ acts as a perturbation around that background. The difference with CPT is that the Fourier transform of $\phi_q$ includes a time-dependent window function,
\begin{equation}\label{phi_q_quantized}
    \phi_q=\int\frac{d^3\bf{k}}{(2\pi)^{3/2}}W\Big(\frac{k}{k_\Sigma}\Big)(e^{-i{\bf kx}}\delta\phi_k\hat a_k+e^{i{\bf kx}}\delta\phi_k^*\hat a_k^\dagger)~,
\end{equation}
where $k_\Sigma\equiv\Sigma aH/C_s$ is the coarse-graining wavenumber with $\Sigma\ll 1$. The window function $W(k/k_\Sigma)$ selects short wavelength modes with $k\geq k_\Sigma$. We stick to the simplest window function---the step function, $W(x)=\theta(x-1)$, which will produce white noise in the stochastic equations. The mode function $\delta\phi_k$ is assumed to satisfy its CPT equation of motion, which in the GB-coupled inflation is given by Eq. \eqref{dphi_EOM}.

Metric tensor can be split into IR and UV parts similarly:
\begin{equation}
	g_{\mu\nu}=(g_c)_{\mu\nu}+(g_q)_{\mu\nu}\,.
	\label{}
\end{equation}
The perturbations $\alpha$, $\beta$, and $\zeta$, introduced in the ADM metric \eqref{ADM_metric}, act as the $q$-variables ($\alpha=\alpha_q$, $\beta=\beta_q$, $\zeta=\zeta_q$), while the homogeneous part of the metric can be identified with the IR part, because we ignore the spatial gradients of $(g_c)_{\mu\nu}$. The Fourier transform of $\alpha_q$, $\beta_q$, and $\zeta_q$ also includes the window function $W(k/k_\Sigma)$. For instance,
\begin{equation}
    \alpha_q=\int\frac{d^3\bf{k}}{(2\pi)^{3/2}}W\Big(\frac{k}{k_\Sigma}\Big)(e^{-i{\bf kx}}\alpha_k+{\rm c.c.})~.
\end{equation}

In order to write down the stochastic equation of motion for $\phi_c$, will first set $\zeta_q=0$ by choosing the spatially flat gauge, and then eliminate the non-dynamical perturbations $\alpha_q$ and $\beta_q$ (their Fourier modes $\alpha_k$ and $\beta_k$ satisfy the usual CPT constraint equations). To do this consistently, we will need to expand the Lagrangian \eqref{L_sGB} up to the second order in the $q$-variables, while keeping non-linearities in the IR sector. The presence of the window function in the Fourier transform of the $q$-variables leads to the mixing between the background and the perturbations ($c$- and $q$-variables). This means that we should keep the zeroth, first, and second order terms (in the $q$-variables) in the Lagrangian when deriving the equations of motion, without assuming that they decouple from each other. This is in contrast to CPT, where the zeroth order (background) Lagrangian can be separated from the second order Lagrangian, while its first order part vanishes on-shell (up to total derivatives).

Once the stochastic Klein-Gordon equation (for $\phi_c$) is obtained, one can derive the Langevin equations by introducing the canonical momentum of $\phi_c$.

\subsection{Stochastic equations}

By using the spatially flat gauge \eqref{SF_metric}, and IR-UV split $\phi(t,x)=\phi_c(t)+\phi_q(t,x)$, we expand the Lagrangian \eqref{L_sGB} up to the second order in the $q$-fields (since $\alpha=\alpha_q$ and $\beta=\beta_q$ as argued above, we drop subscript ``$q$" for these variables),
\begin{align}\label{L_stochastic}
\begin{aligned}
    a^{-3}{\cal L} &= 3H^2(2-\epsilon)-3\xi H^4(1-\epsilon)+\tfrac{1}{2}\dot\phi_c^2-V\\
    &+\dot\phi_c\dot\phi_q-3\xi_{,\phi}H^4(1-\epsilon)\phi_q-V_{,\phi}\phi_q+C_H\alpha\\
    &-(3H^2-\tfrac{1}{2}\dot\phi_c^2-6\dot\xi H^3)\alpha^2-(2-3\dot\xi H)H\alpha\frac{\partial_i^2}{a^2}\beta+\tfrac{1}{2}\dot\phi_q^2-(\dot\phi_c+3\xi_{,\phi}H^3)\alpha\dot\phi_q\\
    &-(3\xi_{,\phi\phi}H^3\dot\phi_c+V_{,\phi})\alpha\phi_q+\xi_{,\phi}H^2\alpha\frac{\partial_i^2}{a^2}\phi_q+(\dot\phi_c+\xi_{,\phi}H^3-\xi_{,\phi\phi}H^2\dot\phi_c)\phi_q\frac{\partial_i^2}{a^2}\beta\\
    &-\xi_{,\phi}H^2\dot\phi_q\frac{\partial_i^2}{a^2}\beta+\tfrac{1}{2}\phi_q\frac{\partial_i^2}{a^2}\phi_q-\tfrac{1}{2}\big[V_{,\phi\phi}+3\xi_{,\phi\phi}H^4(1-\epsilon)\big]\phi_q^2~,
\end{aligned}
\end{align}
where $\xi$, $V$, and their derivatives should be understood as functions of $\phi_c$, and $C_H$ is the first background Friedmann equation,
\begin{equation}
    C_H\equiv 3H^2(1-\dot\xi H)-\tfrac{1}{2}\dot\phi_c^2-V~.
\end{equation}
One consequence of the IR-UV mixing is that $C_H$ does not independently vanish, but is equal to a noise term, as will be shown below.

The Lagrangian \eqref{L_stochastic} is so structured that the first line is zeroth order in the $q$-variables, the second line is first order, and the last three lines consist of the quadratic terms.

Our first goal is to consistently eliminate
the metric perturbations $\alpha$ and $\beta$. Although time derivatives of $\alpha$ and $\beta$ are absent from the Lagrangian, they couple to the time derivative of $\phi_q$, whose Fourier transform \eqref{phi_q_quantized} includes the window function. Therefore, care must be taken when deriving the constraint equations in the stochastic formalism.

Variation of \eqref{L_stochastic} w.r.t. $\alpha$ and $\beta$ yields
\begin{align}
\begin{split}\label{C_alpha}
    C_H-2(3H^2-\tfrac{1}{2}\dot\phi_c^2-6\dot\xi H^3)\alpha-(2-3\dot\xi H)H\frac{\partial_i^2}{a^2}\beta-(\dot\phi_c+3\xi_{,\phi}H^3)\dot\phi_q\\
    -(3\xi_{,\phi\phi}H^3\dot\phi_c+V_{,\phi})\phi_q+\xi_{,\phi}H^2\frac{\partial_i^2}{a^2}\phi_q=0~,
\end{split}\\
\begin{split}\label{C_beta}
    \partial^2_i\big\{(2-3\dot\xi H)H\alpha-(\dot\phi_c+\xi_{,\phi}H^3-\xi_{,\phi\phi}H^2\dot\phi_c)\phi_q+\xi_{,\phi}H^2\dot\phi_q\big\}=0~.
\end{split}
\end{align}
Expanding \eqref{C_alpha} in Fourier modes, and using CPT constraint equations \eqref{C_alpha_CPT} and \eqref{C_beta_CPT} for $\alpha_k$ and $\beta_k$, we get
\begin{equation}\label{C_H_eq}
    C_H=-(\dot\phi_c+3\xi_{,\phi}H^3)H\Xi_\phi~,
\end{equation}
where $\Xi_\phi$ stands for the noise term defined as
\begin{equation}
    \Xi_\phi\equiv -H^{-1}\int\frac{d^3\bf{k}}{(2\pi)^{3/2}}\dot W\Big(\frac{k}{k_\Sigma}\Big)(e^{-i{\bf kx}}\delta\phi_k\hat a_k+e^{i{\bf kx}}\delta\phi_k^*\hat a_k^\dagger)~.
\end{equation}
This noise term comes from the $\dot\phi_q$ in \eqref{C_alpha}.

Consider now the second constraint given by \eqref{C_beta}. In standard CPT, the term within the curly brackets would vanish because any spatially homogeneous part would decouple from the first order perturbations. This is not the case in stochastic inflation, and the general solution to \eqref{C_beta} is 
\begin{equation}\label{C_beta_h}
    (2-3\dot\xi H)H\alpha-(\dot\phi_c+\xi_{,\phi}H^3-\xi_{,\phi\phi}H^2\dot\phi_c)\phi_q+\xi_{,\phi}H^2\dot\phi_q=h(t)~,
\end{equation}
where $h(t)$ is some homogeneous function. After Fourier expansion of the RHS of \eqref{C_beta_h}, we obtain
\begin{equation}\label{h_eq}
    h(t)=-\xi_{,\phi}H^3\Xi_\phi~.
\end{equation}
Note that this term does not appear in GR, as it is proportional to the derivative of the GB coupling $\xi(\phi_c)$ (it can also be expected to appear in more general modified gravities).

After using \eqref{C_H_eq} and \eqref{h_eq} in Eqs. \eqref{C_alpha} and \eqref{C_beta_h}, we obtain the solutions for $\alpha$ and $\beta$ as
\begin{align}\label{alpha_expr}
    \alpha &=\frac{\dot\phi_c+\xi_{,\phi}H^3-\xi_{,\phi\phi}H^2\dot\phi_c}{(2-3\dot\xi H)H}\phi_q-\frac{\xi_{,\phi}H}{2-3\dot\xi H}(\dot\phi_q+H\Xi_\phi)~,\\
    \frac{\partial_i^2}{a^2}\beta &=\frac{2-2\dot\xi H+3\xi_{,\phi}^2H^4}{(2-3\dot\xi H)^2H}\Big[(\ddot\phi_c+H\epsilon\dot\phi_c)\phi_q-\dot\phi_c\dot\phi_q-H\dot\phi_c\Xi_\phi\Big]+\frac{\xi_{,\phi}H}{2-3\dot\xi H}\frac{\partial_i^2}{a^2}\phi_q~.\label{beta_expr}
\end{align}

Before proceeding with the KG equation, it is convenient to switch to the local e-fold time $N$, and introduce the canonical momentum $\Pi=\phi_{,N}$ of the inflaton, which can be split as $\Pi=\Pi_c+\Pi_q$, with
\begin{equation}
    \Pi_q=\int\frac{d^3\bf{k}}{(2\pi)^{3/2}} W\Big(\frac{k}{k_\Sigma}\Big)(e^{-i{\bf kx}}\delta\Pi_k\hat a_k+{\rm h.c.})~.
\end{equation}
The Fourier modes of the canonical pair are related to each other as $\delta\Pi_k=\delta\phi_{k,N}$. Putting these relations together in the definition of the momentum, $\Pi=\phi_{,N}$, yields the first Langevin equation,
\begin{equation}\label{Langevin_momentum}
    \phi_{c,N}=\Pi_c+\Pi_q-\phi_{q,N}=\Pi_c+\Xi_\phi~.
\end{equation}

Now we can write the (stochastic) Friedmann equations in terms of the momentum $\Pi_c$ by using \eqref{Langevin_momentum}. In the first Friedmann equation \eqref{C_H_eq}, when using the momentum instead of velocity, the noise on the RHS cancels out and we get
\begin{equation}\label{C_H_Pi}
    3H^2(1-\xi_{,\phi} H^2\Pi_c-\tfrac{1}{6}\Pi_c^2)-V=0~.
\end{equation}
One of the advantages of using the e-fold time is that Eq. \eqref{C_H_Pi} is now a quadratic equation in $H^2$ (rather than a cubic equation in $H$, as was the case when using physical time $t$), whose solution reads
\begin{equation}\label{H2_sol}
    H^2=\frac{2V}{3-\tfrac{1}{2}\Pi_c^2}\Bigg\{1+\sqrt{1-\frac{4\xi_{,\phi}\Pi_c V}{3(1-\tfrac{1}{6}\Pi_c^2)^2}}\Bigg\}^{-1}~.
\end{equation}

The second Friedmann equation can be obtained by varying the Lagrangian \eqref{L_stochastic} w.r.t. the scale factor and eliminating $\alpha$ and $\beta$,
\begin{align}
\begin{aligned}
    \epsilon(2-\xi_{,\phi}H^2\phi_{c,N})-\phi^2_{c,N}-\xi_{,\phi}H^2\phi_{c,N}+(\xi_{,\phi}H^2\phi_{c,N})_{,N}\\
    =-\big[\phi_{c,N}-H^{-1}(\xi_{,\phi}H^3)_{,N}\big]\Xi_\phi+\xi_{,\phi}H^2\Xi_{\phi,N}~,
\end{aligned}
\end{align}
where $\epsilon=-H_{,N}/H$. For future convenience, we introduce the momentum-dependent noise-less slow-roll parameter
\begin{equation}\label{epsilon_Pi_def}
    \epsilon_\Pi\equiv\frac{1}{2-3\xi_{,\phi}H^2\Pi_c}(\Pi_c^2+\xi_{,\phi}H^2\Pi_c-\xi_{,\phi}H^2\Pi_{c,N}-\xi_{,\phi\phi}H^2\Pi_c^2)~, 
\end{equation}
which is related to the Hubble slow-roll parameter as~\footnote{The goal is to write the slow-roll parameter $\epsilon$ in terms of the momentum $\Pi_c$ rather than the velocity $\phi_{c,N}$, such that the second Langevin equation does not contain the velocity (through $\epsilon$) and is only coupled to the field.}
\begin{equation}\label{epsilon_Pi_eq}
    \epsilon=\epsilon_\Pi+(\phi_{c,N}+\xi_{,\phi}H^2-\xi_{,\phi\phi}H^2\phi_{c,N})\frac{\Xi_\phi}{2-3\omega}~.
\end{equation}
In the coefficients of noise terms, we do not distinguish between $\phi_{c,N}$ and $\Pi_c$ because the difference will contribute to the quadratic order in the noise.

The stochastic KG equation can be obtained by varying \eqref{L_stochastic} w.r.t. $\phi_q$, and using the constraints \eqref{alpha_expr} and \eqref{beta_expr}, as well as Eq. \eqref{epsilon_Pi_eq}. The result is (after Fourier expansion of $\phi_q$)
\begin{align}\label{KG_stochastic_phi_N}
\begin{aligned}
    \phi_{c,NN}+(3-\epsilon_\Pi)\phi_{c,N}+H^{-2}V_{,\phi}+3\xi_{,\phi}H^2(1-\epsilon_\Pi)\\
    =\Xi_{\phi,N}+X_1\Xi_\Pi+\Big[3-\epsilon-\frac{X_1\ce_{,N}\omega}{4\ce(1-\omega)}\Big]\Xi_\phi~,
\end{aligned}
\end{align}
where $X_1$ is defined in \eqref{X12_def},
and the noise terms are defined as
\begin{align}
    \Xi_\phi &\equiv -\int\frac{d^3\bf{k}}{(2\pi)^{3/2}}W_{,N}\Big(\frac{k}{k_\Sigma}\Big)(e^{-i{\bf kx}}\delta\phi_k\hat a_k+{\rm h.c.})~,\\
    \Xi_\Pi &\equiv -\int\frac{d^3\bf{k}}{(2\pi)^{3/2}}W_{,N}\Big(\frac{k}{k_\Sigma}\Big)(e^{-i{\bf kx}}\delta\phi_{k,N}\hat a_k+{\rm h.c.})~.
\end{align}

Klein--Gordon equation \eqref{KG_stochastic_phi_N} can be readily decomposed into the Langevin equations for the canonical pair $(\phi_c,\Pi_c)$:
\begin{align}
    \phi_{c,N} &=\Pi_c+\Xi_{\phi}~,\label{Langevin_phi_q}\\
    \Pi_{c,N} &= -(3-\epsilon_\Pi)\Pi_c-H^{-2}V_{,\phi}-3\xi_{,\phi}H^2(1-\epsilon_\Pi)+X_1\Xi_{\Pi}-\frac{X_1\ce_{,N}\omega}{4\ce(1-\omega)}\Xi_\phi~,\label{Langevin_Pi_q}
\end{align}
where both $H$ and $\epsilon_\Pi$ can be eliminated by their noise-less expressions \eqref{H2_sol} and \eqref{epsilon_Pi_def}. We express noise correlators as
\begin{equation}
    \langle\Xi_I(N)\Xi_J(N')\rangle\equiv {\cal A}_{IJ}\delta(N'-N)~,
\end{equation}
where $I,J$ denote $\phi$ or $\Pi$. The amplitudes $\ca_{IJ}$ are computed as
\begin{align}
    {\cal A}_{\phi\phi} &= \frac{k_{\Sigma}^2}{2\pi^2}|k_{\Sigma,N}||\delta\phi_{k_{\Sigma}}|^2~,\label{A_phiphi}\\
    {\cal A}_{\phi\Pi} &= \frac{k_{\Sigma}^2}{2\pi^2}|k_{\Sigma,N}|(\delta\phi^*_{k,N}\delta\phi_{k})_{k=k_\Sigma}~,\label{A_phiPi}\\
    {\cal A}_{\Pi\Pi} &= \frac{k_{\Sigma}^2}{2\pi^2}|k_{\Sigma,N}||\delta\phi_{k,N}|^2_{k=k_{\Sigma}}~.\label{A_PiPi}
\end{align}
Since $k_{\Sigma,N}=k_\Sigma(1-\epsilon-C_{s,N}/C_s)$, time dependence of the sound speed $C_s$ affects the noise terms directly, through $k_{\Sigma,N}$, and also indirectly, through the evolution of the linearized perturbation $\delta\phi_k$.

During SR ($\epsilon,|\eta|,|\omega|,|\sigma|\ll 1$), Eqs. \eqref{Langevin_phi_q} and \eqref{Langevin_Pi_q} are simplified as
\begin{equation}\label{Langevin_SR}
    \phi_{c,N}\simeq \Pi_c+\Xi_{\phi}~,~~~\Pi_{c,N}\simeq -3\Pi_c-H^{-2}V^{\rm eff}_{,\phi}+\Xi_{\Pi}~,
\end{equation}
where the slope of the effective potential is $V^{\rm eff}_{,\phi}= V_{,\phi}+3\xi_{,\phi}H^4$.

In the constant-roll case ($\epsilon,\omega={\rm const.}$), $\delta\phi_k$ can be solved analytically, and the only non-vanishing noise amplitude is
\begin{equation}
    {\cal A}_{\phi\phi}\simeq \frac{2^{2\nu_s-2}(1-\epsilon)\ce C_s^{-2\nu_s}}{\pi\sin^2(\pi\nu_s)\Gamma^2(1-\nu_s)a^2}k_{\Sigma}^{3-2\nu_s}|\tau|^{1-2\nu_s}=\frac{H^2}{4\pi^2}+\co(\epsilon,\omega)~,
\end{equation}
where the last equality shows the SR limit. Thus, we conclude that during SR, our stochastic equations resemble the usual GR ones, except for the contribution of the GB function to the effective potential (this contribution can be large even during SR). The Hubble function, however, does not receive such GB corrections in the SR limit: $3H^2\simeq V$.

For the USR case with $V^{\rm eff}_{,\phi}\simeq 0$, and assuming $C_s\simeq 1$ (implying $X_1\simeq 1$) and $|C_{s,N}/C_s|\ll 1$, we get the same result as during SR,
\begin{equation}\label{GR_amplitude}
\ca_{\phi\phi}\simeq H^2/(4\pi^2)~,
\end{equation}
with vanishing $\ca_{\phi\Pi}$ and $\ca_{\Pi\Pi}$. Therefore, the Langevin equations reduce to
\begin{equation}\label{Langevin_USR}
    \phi_{c,N}\simeq \Pi_c+\Xi_{\phi}~,~~~\Pi_{c,N}\simeq -3\Pi_c~.
\end{equation}
Although the effective potential must be flat ($V^{\rm eff}_{,\phi}\simeq 0$) for USR regime, the potential $V$ may not. This is an important difference with USR in GR.

The Langevin equations \eqref{Langevin_USR} can be written as
\begin{equation}
    \phi_{c,N}\simeq \phi_{c,N}(N_{\rm in})e^{-3(N-N_{\rm in})}+\Xi_\phi~,
\end{equation}
where $N_{\rm in}$ is the time at the beginning of USR. In the absence of noise, $\phi_c$ reduces to its classical solution \eqref{phi_cl_USR}. The initial velocity $\phi_{c,N}(N_{\rm in})$ should be found from the evolution of the field prior to USR.

It should be noted that our assumptions $C_s\simeq 1$ and $C_{s,N}/C_s\ll 1$ could be violated during USR. In general, this would require us to solve the EoM for $\delta\phi_k$ numerically. The numerical results would modify the noise amplitudes.

\subsection{An example: Gauss--Bonnet-induced ultra-slow-roll}\label{GB_example}

As an example, let us estimate stochastic effects on background dynamics in the GB-induced USR model in Ref. \cite{Kawai:2021edk}. In this model, the authors used a canonical kinetic term, inflationary scalar potential of the form $V=\Lambda^4[1+\cos(\phi/f)]$, and GB coupling $\xi=\xi_0\tanh{[\xi_1(\phi-\phi_{\rm cr})]}$. Here, $\phi_{\rm cr}$ is the critical point of the effective potential, where USR takes place for appropriately chosen parameters $\xi_0$ and $\xi_1$.

\begin{figure}
\centering
  \centering
  \includegraphics[width=.7\linewidth]{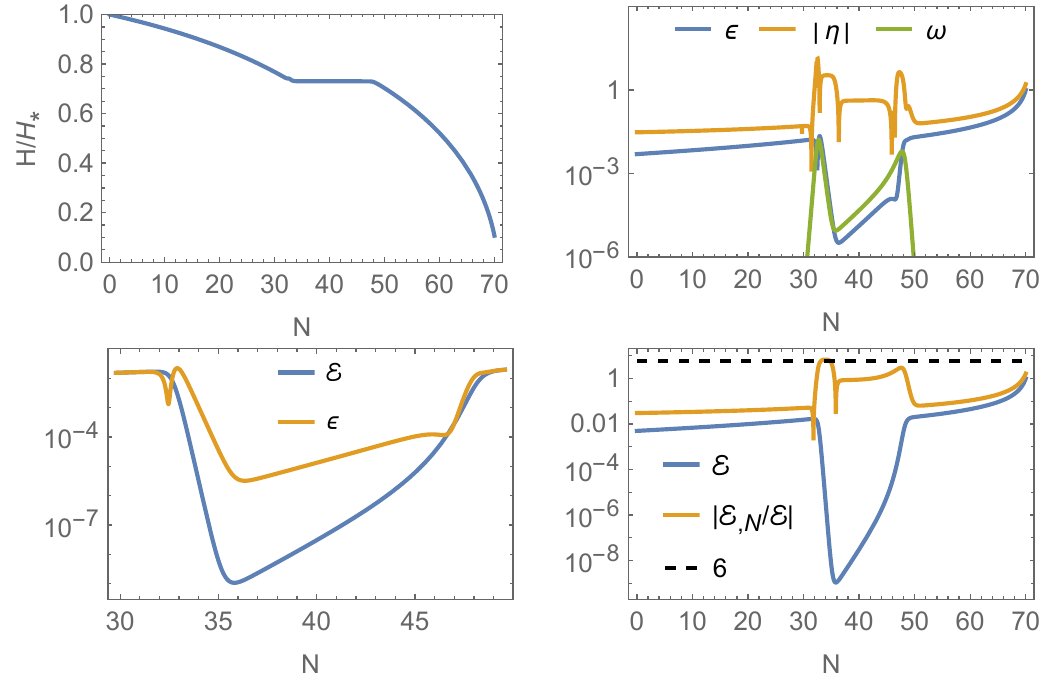}
\captionsetup{width=1\linewidth}
\caption{Top-left: Hubble function normalized by its value at the CMB scale $H_*$. Top-right: Hubble slow-roll parameters $\epsilon,\eta$ and the GB slow-roll parameter $\omega$. Bottom-left: the comparison between $\epsilon$ and the field-velocity-based slow-roll parameter $\ce\equiv\bar\phi_{,N}^2/2$. Bottom-right: $\ce$ and its rate of change $\ce_{,N}/\ce$. Last $70$ e-folds are shown.}\label{Fig_GB_USR}
\end{figure}

Ultra-slow-roll phase of the aforementioned model can generate PBH as dark matter while evading the current observational constraints. For example, the parameter set (in Planck units)~\footnote{In \cite{Kawai:2021edk}, the authors choose $\xi_0=3.022\times 10^7$ for the PBH dark matter scenario, which in our numerical analysis leads to the solution being trapped in USR well indefinitely. This disagreement may arise from the difference in numerical methods used (we use \textit{Mathematica} solver). When reproducing the results of \cite{Kawai:2021edk} for the PBH fraction, by using Press--Schechter \cite{Press:1973iz} formalism (without stochastic effects), we also used a higher value of the threshold amplitude of perturbations, $\delta_c=0.45$ (as opposed to $\delta_c=1/3$), motivated by numerical estimates of Ref. \cite{Musco:2018rwt}. For the details of the Press--Schechter method applied to this model, see \cite{Kawai:2021edk} and Refs. therein.}
\begin{gather}
\begin{gathered}
    \Lambda=0.0065~,~~~f=7~,~~~\phi_{\rm cr}=13~,\\
    \xi_0=3.003925\times 10^7~,~~~\xi_1=15~,
\end{gathered}
\end{gather}
leads to all dark matter in the form of PBHs. Numerically solving the background equations, we obtain the evolution of the Hubble function and the slow-roll parameters as shown in Fig. \ref{Fig_GB_USR}.

Top-right plot of Fig. \ref{Fig_GB_USR} shows the evolution of Hubble slow-roll parameters $\epsilon$ and $\eta$, as well as the GB slow-roll parameter $\omega$, which becomes important around the USR phase. The evolution of the slow-roll parameter $\epsilon$ is somewhat different from the usual expectation from PBH-producing single-field models. Small oscillatory features can be seen around $N=30$ and $N=50$. Nontheless, the expected evolution can be seen if we instead consider the field-velocity-based slow-roll parameter $\ce$ introduced earlier. The evolution of $\ce$ and its rate of change are shown in the bottom-right plot. In the absence of the GB coupling, we have $\ce=\epsilon$, but this no longer holds in the GB-coupled inflation. In particular, from the top-right and bottom-right plots, one can notice that while $\epsilon$ at its minimum is larger than $10^{-6}$, $\ce$ reaches a much lower value of $\sim 10^{-9}$. This can be explained by the precise cancellation of $\epsilon$ with the GB corrections in Eq. \eqref{backg_EOM_Hdot_N} during USR phase.

In the bottom-right plot of Fig. \ref{Fig_GB_USR}, one can identify a brief USR phase when $\ce_{,N}/\ce\simeq -6$. This is followed by a second slow-roll phase when $\ce$ increases again. To be more precise, we define the beginning and end of USR by $\ce_{,N}=0$ ($\ce_{,N}/\ce$ goes from positive to $-6$ and back to positive). The change in the inflaton during USR is $\Delta\phi_{\rm usr}\approx 0.2$, and the duration of USR in the classical limit is $\Delta N_{\rm usr}^{\rm cl}\approx 4.02$ e-folds.

The USR phase is expected to be affected by stochastic noise the most because the effective potent has a vanishing slope. During the USR phase we consider the Langevin equations \eqref{Langevin_USR} with $V_{,\phi}^{\rm eff}=0$ and the noise amplitude $\ca_{\phi\phi}\simeq H^2/(4\pi^2)$. The noise amplitude coincides with its GR expression because $C_s\simeq X_1\simeq 1$ and $|C_{s,N}/C_s|\ll 1$ in this model, as we confirmed numerically. Consequently, the stochastic equations reduce to their standard USR form from GR, and the existing analytical methods can be readily applied~\footnote{There is one subtlety related to the non-vanishing of $V_{,\phi}$ in GB-induced USR inflation. We compute the mean number of e-folds and the power spectrum in GB-induced USR inflation in the next section, by including the correction due to non-vanishing $V_{,\phi}$. For the model discussed here, this correction is suppressed.}. For example, we can use small noise approximation of Ref. \cite{Firouzjahi:2018vet} in calculating the mean number of e-folds during USR,
\begin{equation}\label{mean_N_USR}
    \langle \Delta N_{\rm usr}\rangle
    \simeq \Delta N_{\rm usr}^{\rm cl}\Big[1+\frac{\kappa^2}{6}+\frac{\kappa^4}{36}(5+9\Delta N_{\rm usr}^{\rm cl})+\frac{\kappa^6}{72}\big(17+77\Delta N_{\rm usr}^{\rm cl}+60(\Delta N_{\rm usr}^{\rm cl})^2\big)\Big]~,
\end{equation}
where
\begin{equation}
    \kappa^2\equiv 9e^{6\Delta N_{\rm usr}^{\rm cl}}\frac{H^4}{4\pi^2\dot{\bar\phi}_{\rm in}^2}\simeq 9e^{6\Delta N_{\rm usr}^{\rm cl}}P_{R,s}^{\rm cl}\simeq 9P_{R,e}^{\rm cl}~,
\end{equation}
is the ratio of quantum kicks to classical drift, which is taken as an expansion parameter ($\dot{\bar\phi}_{\rm in}$ and $\dot{\bar\phi}_e$ are field velocity values at the beginning and end of USR). The last equality shows $\kappa$ in terms of the classical curvature power spectrum at the end of USR, i.e., at its peak. We can estimate $P_{R,e}^{\rm cl}$ from the background solution of Fig. \ref{Fig_GB_USR} (this yields $P_{R,e}^{\rm cl}\approx 0.005$~\footnote{A more precise value of $P^{\rm cl}_{R,e}$ can be obtained from numerical solution to Mukhanov--Sasaki equation, however, the difference is inconsequential for our purposes.}), and use $\Delta N_{\rm usr}^{\rm cl}\approx 4.02$. The resultant mean number of e-folds from \eqref{mean_N_USR} is $\langle \Delta N_{\rm usr}\rangle\approx 4.07$. The difference, albeit small, can lead to overproduction of PBHs. This is because the power spectrum grows during USR, and the PBH abundance is sensitive to even small changes in the power spectrum's peak value. 

The authors of \cite{Firouzjahi:2018vet} also estimated the power spectrum with stochastic corrections,
\begin{equation}\label{P_stoch_USR}
    P_{R,e}=P_{R,e}^{\rm cl}+9(1+5\Delta N_{\rm usr}^{\rm cl})(P_{R,e}^{\rm cl})^2+\co((P_{R,e}^{\rm cl})^3)~.
\end{equation}
Although the power spectrum is small, $P_{R,e}^{\rm cl}\approx 0.005$, the leading-order correction in \eqref{P_stoch_USR} becomes large, $\sim \co(P_{R,e}^{\rm cl})$. Thus, \eqref{P_stoch_USR} gives only qualitative result indicating underestimation of the power spectrum in purely classical analysis. More precise calculations involve numerical simulation of GB-coupled Langevin equations, which is beyond the scope of this work; however, in the next section, we consider a different approach to analytically calculate the power spectrum during SR and USR, as well as the effects of stochastic inflation on the distribution of large perturbations which collapse into PBH.

\section{First-passage time for Gauss--Bonnet-coupled inflation}\label{Sec_FPT}

\par In this section, we start from reviewing the concept of first-passage time analysis and how this analysis is related to power spectrum. After that, we apply first-passage time method to GB-coupled inflation, obtaining the analytical expressions of scalar power spectrums with leading-order noise corrections in SR and USR limits. Finally, we numerically calculate PBH mass fraction in stochastic-noise-dominated USR regimes.

In  stochastic $\delta N$ formalism, the first-passage time $\mathcal{N}$ refers to the number of e-folds it takes for a given stochastic trajectory of the inflaton field configuration, starting from a given initial field configuration $\mathbf{\Phi}=\mathbf{\Phi}_{\text{initial}}$, to first meet a specified field configuration boundary $\partial\Omega_{-}$\cite{Vennin:2020kng}. Here, the field configuration can contain both the inflaton field $\phi_i$ and the corresponding momentum $\pi_i$, i.e., $\mathbf{\Phi}=\left(\phi_1,\pi_1,\phi_2,\pi_2,...,\phi_i,\pi_i,...\right)$ (in this work we limit our discussion to one scalar field, or one canonical pair). The stochasticity of a trajectory in configuration space comes from quantum origin of the scalar fluctuations. Once the boundary $\partial\Omega_{-}$ is fixed, the statistics related to $\mathcal{N}$ will define the initial field configuration $\mathbf{\Phi}_{\text{initial}}$, and $\left\langle \mathcal{N} \right\rangle=\left\langle \mathcal{N} \right\rangle\left( \mathbf{\Phi}_{\text{initial}} \right)$. For simplicity, the subscript ``initial" will be dropped, i.e. $\left\langle \mathcal{N} \right\rangle=\left\langle \mathcal{N} \right\rangle\left( \mathbf{\Phi}_{\text{initial}} \right)\equiv \left\langle \mathcal{N} \right\rangle\left( \mathbf{\Phi} \right)$. In practice, the boundary $\partial\Omega_{-}$ is usually set as the end of inflation $\partial \Omega_{-}=\left\{\boldsymbol{\Phi} \mid \epsilon(\mathbf{\Phi})=1\right\}$. In this case, observables like power spectrum can be related to the statistics of the first-passage time \cite{Vennin:2020kng,bullock1997non,Vennin:2015correlation,Vennin_2017,Assadullahi_2016}.

Using the vector notation introduced above, the Langevin equations formally become
\begin{equation}\label{Gen_Langevin_eq}
	\frac{{d{\mathbf{\Phi }}}}{{dN}} = {\mathbf{F}}\left( {\mathbf{\Phi }} \right) + {\mathbf{G}}\left( {\mathbf{\Phi }} \right) \cdot {\boldsymbol{\chi }}~,
\end{equation}
where ${\mathbf{F}}\left( {\mathbf{\Phi }} \right)$ can be interpreted as the classical dynamics of a field configuration, and ${\mathbf{G}}\left( {\mathbf{\Phi }} \right) \cdot {\boldsymbol{\chi }}$ can be viewed as stochastic noises. Here, stochastic noises $\chi_i$ (the components of $\boldsymbol{\chi}$) are defined to satisfy $\left\langle {{\chi _i}\left( {{{\mathbf{x}}_i},N} \right)} \right\rangle  = 0$ and $\left\langle {{\chi _i}\left( {{{\mathbf{x}}_i},N} \right){\chi _j}\left( {{{\mathbf{x}}_j},N'} \right)} \right\rangle  = {\delta _{ij}}\delta \left( {N - N'} \right)$. From the vector Langevin equation \eqref{Gen_Langevin_eq}, one can construct Fokker--Planck and adjoint Fokker--Planck operators as \cite{Vennin:2020kng,Vennin_2025}

\begin{align}\label{Gen_FP_O}
		{\mathcal{L}_{{\text{FP}}}}({\mathbf{\Phi }}) & =  - \frac{\partial }{{\partial {\Phi _i}}}\left[ {{F_i}({\mathbf{\Phi }}) + \gamma {G_{\ell j}}({\mathbf{\Phi }})\frac{{\partial {G_{ij}}({\mathbf{\Phi }})}}{{\partial {\Phi _\ell }}}} \right] + \frac{1}{2}\frac{{{\partial ^2}}}{{\partial {\Phi _i}\partial {\Phi _j}}}{G_{i\ell }}({\mathbf{\Phi }}){G_{j\ell }}({\mathbf{\Phi }}) ~, \\ \label{Gen_FP_O_A}
		{\mathcal{L}_{{\text{FP}}}}^\dag ({\mathbf{\Phi }}) & = {F_i}({\mathbf{\Phi }})\frac{\partial }{{\partial {\Phi _i}}} + \gamma{G_{ij}}({\mathbf{\Phi }})\frac{{\partial {G_{\ell j}}({\mathbf{\Phi }})}}{{\partial {\Phi _\ell }}}\frac{\partial }{{\partial {\Phi _i}}} + \frac{1}{2}{G_{i\ell }}({\mathbf{\Phi }}){G_{j\ell }}({\mathbf{\Phi }})\frac{{{\partial ^2}}}{{\partial {\Phi _i}\partial {\Phi _j}}} ~,  
\end{align}
where $\gamma$ is a constant with a value between 0 and 1. The parameter $\gamma$ represents how the mean values of the fields in the interval between $N$ and $N+\delta N$ is related to $\mathbf{\Phi}\left(N\right)$ and $\mathbf{\Phi}\left(N+\delta N\right)$. In single field case, the dependence on $\gamma$ is highly suppressed \cite{Vennin:2020kng}, so we simply set it to zero. With the help of Fokker--Planck and adjoint Fokker--Planck operators, one can describe the evolution of the probability density of the system by
\begin{equation}
	\frac{\partial}{\partial N} P\left(\mathbf{\Phi}, N \mid \mathbf{\Phi}^{\mathrm{in}}, N_{\mathrm{in}}\right)=\mathcal{L}_{\mathrm{FP}}(\mathbf{\Phi}) P\left(\mathbf{\Phi}, N \mid \mathbf{\Phi}^{\mathrm{in}}, N_{\mathrm{in}}\right)~,
\end{equation}
given the initial condition $\mathbf{\Phi}^{\text{in}}$ at some initial time $N_{\text{in}}$.
Then, the $n$-th moment of $\mathcal{N}$ can be related to its $(n-1)$-th moment via
\begin{equation}\label{Gen_nN}
	\mathcal{L}_{{\text{FP}}}^\dag \left( \mathbf{\Phi } \right) \cdot \left\langle {{\mathcal{N}^n}} \right\rangle \left( \mathbf{\Phi }\right) =  - n\left\langle {{\mathcal{N}^{n - 1}}} \right\rangle \left( \mathbf{\Phi } \right)~.
\end{equation}
In particular, we notice that $\left\langle {{\mathcal{N}^{0}}} \right\rangle=1$. Therefore, all the $n$-th moments of $\mathcal{N}$ can in principle be calculated via \eqref{Gen_nN}. Nevertheless, solving $\left\langle {{\mathcal{N}^{n}}} \right\rangle $ through this process is complicated because we need to solve $n$ differential equations. To reduce the burden, one can introduce a characteristic function \cite{Vennin:2020kng,Vennin_2025}
\begin{equation}\label{def_chi_N}
	\chi_{\mathcal{N}}(t, \mathbf{\Phi}) \equiv\left\langle e^{i t \mathcal{N}(\mathbf{\Phi})}\right\rangle~,
\end{equation}
which satisfies
\begin{equation}\label{Gen_EoM_chi}
	\mathcal{L}_{\mathrm{FP}}^{\dagger} \cdot \chi_{\mathcal{N}}(t, \boldsymbol{\Phi})=-i t \chi_{{\mathcal{N}}}(t, \boldsymbol{\Phi})~.
\end{equation}
Therefore, one can solve $\chi_{\mathcal{N}}$ through \eqref{Gen_EoM_chi}, and $\left\langle {{\mathcal{N}^{n}}} \right\rangle $ can be obtained by taking $n$ partial derivatives of $\chi_{\mathcal{N}}$ as
\begin{equation}\label{Gen_N_n_chi}
\langle {\mathcal{N}^n}\rangle ({\mathbf{\Phi }}) = {i^{ - n}}{\left. {\frac{{{\partial ^n}{\chi _\mathcal{N}}(t,{\mathbf{\Phi }})}}{{\partial {t^n}}}} \right|_{t = 0}}~.
\end{equation} 

The power spectrum can be related to the $n$-th moment of $\mathcal{N}$ as
\begin{equation}\label{Gen_pow_spec}
{\mathcal{P}_\zeta } = \frac{{{\text{d}}\left\langle {\delta {\mathcal{N}^2}} \right\rangle }}{{{\text{d}}\langle \mathcal{N}\rangle }}~,
\end{equation}
where 
\begin{equation}
	\left\langle {\delta \mathcal{N}^2} \right\rangle=	\left\langle {\mathcal{N}^2} \right\rangle - 	{\left\langle \mathcal{N} \right\rangle}^2~.
\end{equation}

\subsection{First-passage time for standard slow-roll case}

Four conditions, ${{\dot {\bar \phi} }^2} \ll V$, $|{\ddot {\bar \phi} }| \ll 3H\dot{\bar\phi}$, $|{\dot \xi}|H \ll 1$, and $| {\ddot \xi } |\ll| {\dot \xi }|H$ (equivalent to the conditions \eqref{GB_SR}) are imposed for standard SR inflation with GB correction\cite{Guo:2010jr}. With these conditions, the background equation \eqref{backg_EOM_phi} can be simplified as
\begin{align}\label{SSR_backg_EOM_H}
	{H^2} &\simeq \tfrac{1}{3}V~,\\
	\label{SSR_backg_EOM_phidot}
	H\dot {\bar \phi}  &\simeq  - \tfrac{1}{3}({{V_{,\phi }} + 3{\xi _{,\phi }}{H^4}})~,\\ \label{SSR_backg_EOM_Hdot}
	\dot H &\simeq  - \tfrac{1}{2}\dot \xi {H^3} - \tfrac{1}{2}{{\dot {\bar \phi} }^2}.~ 
\end{align}

Using \eqref{SSR_backg_EOM_phidot} and the fact that ${\cal A}_{\phi\phi}\simeq H^2/4\pi$, ${\cal A}_{\phi\Pi} \ll {\cal A}_{\phi\phi}$, and ${\cal A}_{\Pi\Pi} \ll {\cal A}_{\phi\phi}$ in the spatially flat gauge with standard slow-roll conditions, one can write the Langevin equation as
\begin{equation}\label{SSR_Lang_eq}
	\frac{{d\phi_{\rm c} }}{{dN}}= - \frac{{{v_{,\phi }}}}{v} - \frac{{{{\tilde \xi }_{,\phi }}v}}{3} + \frac{H}{{2\pi }}\chi,
\end{equation}
where $\displaystyle v = \frac{V}{{24{\pi ^2}}}$, $\tilde \xi  = 24{\pi ^2}\xi $, and $\chi$ is the (previously defined) stochastic variable associated with a normalized Gaussian distribution. Therefore, the Fokker-Planck and adjoint Fokker-Planck operators \eqref{Gen_FP_O} and \eqref{Gen_FP_O_A} can expressed as
\begin{align}\label{SSR_FP_O}
		{\mathcal{L}_{\text{FP}}} & = \frac{\partial }{{\partial \phi }}\left( {\frac{{{v_{,\phi }}}}{v} + \frac{{{{\tilde \xi }_{,\phi }}v}}{3}} \right) + \frac{{{\partial ^2}}}{{\partial {\phi ^2}}}v ~, \\ \label{SSR_FP_O_A}
		\mathcal{L}_{\text{FP}}^\dag & =  - \left( {\frac{{{v_{,\phi }}}}{v} + \frac{{{{\tilde \xi }_{,\phi }}v}}{3}} \right)\frac{\partial }{{\partial \phi }} + v\frac{{{\partial ^2}}}{{\partial {\phi ^2}}}~.  
\end{align}
Hence, the stationary solution to the Fokker-Planck equation ${\partial _N}P = {\mathcal{L}_{FP}}P=0$ will be
\begin{equation}
	{P_{{\text{stat}}}} \propto\frac{1}{v}~{{e^{\frac{1}{v} + \frac{1}{3}\int {{{\tilde \xi }_{,\phi }}vd} \phi }}}~.
\end{equation}
There is a correction related to $\xi$, compared to the result of traditional inflation without GB correction. By defining the effective potential in GB-coupled inflation as $1/v_{\rm eff}(\phi)=1/v(\phi)+\tilde{\xi}(\phi)/3$, the mean first-passage time can be also solved from  \eqref{Gen_nN} as
\begin{equation}\label{Gen_mN}
	\langle \mathcal{N}\rangle (\phi ) = \int_{{\phi _1}}^\phi  {dz_1} \int_{z_1}^{\bar \phi \left( {{\phi _1},{\phi _2}} \right)} {dz_2} \frac{1}{{v(z_2)}}\exp \left[ {\frac{1}{{{v_{{\text{eff}}}}(z_2)}} - \frac{1}{{{v_{{\text{eff}}}}(z_1)}}} \right]~.
\end{equation}
Here, $\phi_1$ and $\phi_2$ are two different ending points of inflation. The integration constant $\bar \phi \left( {\phi _1},{\phi _2}\right)$ is set to satisfy the boundary condition $\langle \mathcal{N}\rangle (\phi_1)=\langle \mathcal{N}\rangle (\phi_2)=0$, which holds by the definition of $\mathcal{N}$. It can be verified \eqref{Gen_mN} will boil down to the standard purely classical result in some ``classical limit". This can be done by using so-called ``saddle-point" expansion to calculate the integration in \eqref{Gen_mN} (See Appendix \ref{Saddle-point} for detail), which results in 
\begin{equation}\label{mN_LO}
	 \langle \mathcal{N}\rangle (\phi ) \simeq \int_{{\phi _{{\text{end}}}}}^\phi  {dz} \frac{{3v\left( z \right)}}{{3v'\left( z \right) + \tilde \xi '\left( z \right){v^2}\left( z \right)}}~.\
\end{equation}

 Notice that the mean first-passage time $\langle \mathcal{N}\rangle$ \eqref{mN_LO} is exactly the classical deterministic number of e-folds to reach the end of inflation obtained from classical EoM \eqref{SSR_backg_EOM_phidot} directly. Therefore, classical trajectory appears as a saddle-point limit of the mean stochastic trajectory. Nevertheless, due to the validity of  ``saddle-point" expansion, the classical trajectory can be taken as a good approximation to the mean stochastic one iff both $v$ and $\xi$ satisfy ``classicality" criteria defined as
 \begin{align}
	\eta _{{\text{cl}}}^v &= \left| {2v - \frac{{v''{v^2}}}{{v{'^2}}}} \right|~,\\
	\eta _{{\text{cl}}}^\xi  &= \left| {\frac{{\tilde \xi ''}}{{\tilde \xi {'^2}}}} \right|~,
\end{align}
satisfy $\eta _{{\text{cl}}}^v \ll 1$ and $\eta _{{\text{cl}}}^\xi \ll 1$ (see Appendix \ref{Saddle-point} for detail).

The first-order quantum correction to the classical trajectory can be obtained by using the second-order Taylor expansion of $1/v$ and $\tilde{\xi}$ in the integral \eqref{Kernal_int} as

\begin{equation}
	\langle \mathcal{N}\rangle_{{\eta _{{\text{cl}}}^v \ll 1,\eta _{{\text{cl}}}^\xi  \ll 1}}   = \int_{{\phi _{{\text{end}}}}}^\phi  {dz} \frac{{3v\left( z \right)}}{{3v'\left( z \right) + \tilde \xi '\left( z \right){v^2}\left( z \right)}} \times \left( {1 +\delta {{\langle \mathcal{N}\rangle }^{{\text{SSR 1st cf}}}}\left( z \right)} \right)~,
\end{equation}
where the first-order correction factor $\delta{\langle \mathcal{N}\rangle ^{{\text{SSR 1st cf}}}}$ is
\begin{equation}
    \delta{\langle \mathcal{N}\rangle ^{{\text{SSR 1st cf}}}} =  - \frac{{3v\left( { - 3{{v'}^2} + {v^2}v'\tilde \xi ' + 3v{v^{\prime \prime }} + {v^3}{{\tilde \xi }^{\prime \prime }}} \right)}}{{{{\left( {3v' + {v^2}\tilde \xi '} \right)}^2}}}~.
\end{equation}

Mean number of e-folds squared $\langle \mathcal{N}^2\rangle$ can be solved from $\langle \mathcal{N}\rangle$ via \eqref{Gen_nN} by setting $n=2$ and using the adjoint Fokker-Planck operators given in \eqref{SSR_FP_O}. The result is
\begin{equation}
    \langle {\mathcal{N}^2}\rangle (\phi ) = 2\int_{{\phi _1}}^\phi  {dz_1} \int_{z_1}^{\bar \phi \left( {{\phi _1},{\phi _2}} \right)} {dz_2} \frac{{\langle \mathcal{N}\rangle (\phi )}}{{v(z_2)}}\exp \left[ {\frac{1}{{{v_{{\text{eff}}}}(z_2)}} - \frac{1}{{{v_{{\text{eff}}}}(z_1)}}} \right]~.
\end{equation}
The power spectrum can thus be obtained from \eqref{Gen_pow_spec} as
\begin{equation}
	{\mathcal{P}_\zeta }({\phi _*}) = 2I{\left( {{\phi _*}} \right)^{ - 1}}\int_{{\phi _*}}^{{{\bar \phi }_2}} {dz} \left[ {I{{\left( z \right)}^2}\exp \left[ {\frac{1}{{{v_{{\text{eff}}}}\left( z \right)}} - \frac{1}{{{v_{{\text{eff}}}}({\phi _*})}}} \right]} \right]~,
\end{equation}
where the integration $I(x)$ is
\begin{equation}
    I\left( z \right) = \int_z^{\bar \phi } {dz'} \frac{1}{{v(z')}}\exp \left[ {\frac{1}{{{v_{{\text{eff}}}}\left( z' \right)}} - \frac{1}{{{v_{{\text{eff}}}}({\phi _*})}}} \right]~.
\end{equation}
By using the saddle-point expansion with the first-order Taylor expansion, the power spectrum becomes
\begin{equation}\label{SSR_pow_spec_0th}
	{\mathcal{P}_\zeta }({\phi _*}) =  {\frac{{18v{{({\phi _*})}^3}}}{{{{\left( {3v'({\phi _*}) + \tilde \xi ({\phi _*})v{{({\phi _*})}^2}} \right)}^2}}}}~.
\end{equation}
It can be seen that this power spectrum \eqref{SSR_pow_spec_0th} is indeed the ``classical" one \eqref{P_zeta_SR} by using the slow-roll background equation \eqref{SSR_backg_EOM_H} and \eqref{SSR_backg_EOM_phidot}.

The power spectrum with the first-order quantum correction can also be obtained by using the second-order Taylor expansion as
\begin{equation}
	{\mathcal{P}_\zeta }({\phi _*}){|_{\eta _{{\text{cl}}}^v \ll 1,\eta _{{\text{cl}}}^\xi  \ll 1}}={\frac{{18v{{({\phi _*})}^3}}}{{{{\left( {3v'({\phi _*}) + \tilde \xi ({\phi _*})v{{({\phi _*})}^2}} \right)}^2}}}}  \times \left( {1 +\delta {\mathcal{P}_\zeta }^{{\text{SSR 1st cf}}}\left( {{\phi _*}} \right)} \right)~,
\end{equation}
which is the ``classical" result multiplied by the stochastic correction factor defined as
\begin{equation}\label{1st_P_SSR_cf}
    \delta{{\mathcal{P}_\zeta }^{{\text{SSR 1st cf}}}\left( {{\phi _*}} \right)}= - \frac{{3v\left( { - 15{{v'}^2} + 3{v^2}v'\tilde \xi ' + 12v{v^{\prime \prime }} + 4{v^3}{{\tilde \xi }^{\prime \prime }}} \right)}}{{{{\left( {3v' + {v^2}\tilde \xi '} \right)}^2}}}~.
\end{equation}
In the case discussed in  \ref{GB_example}, the noise correction is small, \eqref{1st_P_SSR_cf}
 $\delta{{\mathcal{P}_\zeta }^{{\text{SSR 1st cf}}}}\simeq O(10^{-10})$.

\subsection{First-passage time for ultra-slow-roll case}

\begin{figure}
	\centering
	\centering
	\includegraphics[width=.5\linewidth]{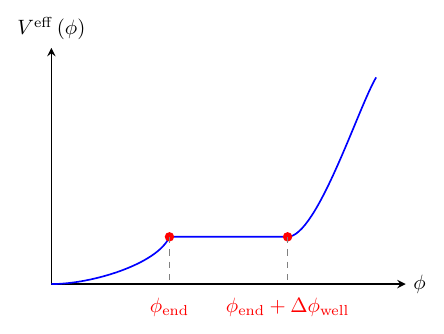}
	\captionsetup{width=1\linewidth}
	\caption{Schematic representation of the USR inflation scenario. The inflation experiences a $V_{,\phi}^{\rm eff} =0$ USR region with width $\Delta \phi_{\rm well}$ before the end of inflation. For the evolution of slow-roll parameters, see for example Fig. \ref{Fig_GB_USR}.}\label{Fig_Veff_USR}
\end{figure}

In the Standard slow-roll limit, the $\ddot{\bar\phi}$ in \eqref{backg_EOM_phi} is ignored, such that we can have 1D Langevin equation \eqref{SSR_Lang_eq}. In the USR case where $V_{,\phi}^{\rm eff}\simeq 0$, however, full 2D Langevin equations need to be used since the second derivative of $\phi$ cannot be neglected. Therefore, the analysis in SSR could not be directly used. Nevertheless, we can analyze the classical limit of USR with the help of characteristic function \eqref{def_chi_N}. Let us consider the following scenario (see Fig. \ref{Fig_Veff_USR}): As the inflaton rolls down, standard slow-roll occurs first. Then, before the inflaton reaches the end of inflation $\phi_{\text{end}}$, it falls into the $V^{\rm eff}_{,\phi} =0$ USR region of the width $\Delta \phi_{\text{well}}$. 

During the USR, the Langevin equation \eqref{Langevin_SR} can be simplified as 
\begin{equation}\label{Langevin_USR_O}
	\begin{aligned}
		 \phi_{c,N}&=\Pi_c+\frac{H}{2\pi}\chi~,\\ \Pi_{c,N}&=-3\Pi_c ~,
	\end{aligned}
\end{equation}
where we use ${\cal A}_{\phi\phi}\simeq H^2/4\pi$ , ${\cal A}_{\phi\Pi} \ll {\cal A}_{\phi\phi}$, and $\  {\cal A}_{\Pi\Pi} \ll {\cal A}_{\phi\phi}$ . During the USR stage, we still have $3H^2\simeq V$. Therefore, equation \eqref{Langevin_USR_O} can be written as 
\begin{equation}\label{Langevin_USR_nv}
	\begin{gathered}
		\frac{{{\text{d}}x}}{{\;{\text{d}}N}} =  - 3y + \sqrt {2 \nu \left( x \right)}\chi (N), \hfill \\
		\frac{{{\text{d}}y}}{{\;{\text{d}}N}} =  - 3y, \hfill \\ 
	\end{gathered} 
\end{equation}
where we define normalized field $x=(\phi-\phi_{\rm end})/\Delta \phi_{\rm well}$, normalized momentum field $y=-\Pi_c/\left(3\Delta \phi_{\rm well}\right)$, and the normalized field $\nu(x)=v(x)/\Delta\phi_{\rm well}^2=V(\phi)/(24\pi^2 \Delta\phi_{\rm well}^2)$. Hence, the USR starts when $x=1$ and ends when $x=0$. During the USR phase, $x$ ranges from 0 to 1. Formally, Langevin equations \eqref{Langevin_USR_nv} are the same as the ones for USR inflation in GR. Nevertheless, when USR limit can be taken in GR, the potential $\nu$ should be exactly flat, i.e., $\nu^{\prime}(x)=0$. In gravitational theory with a GB correction, the USR stage has a vanishing slope of the effective potential $V_{,\phi}^{\rm eff}=V_{,\phi}+3\xi_{,\phi}H^4$ rather than the original potential $V_{,\phi}$. The slope $V_{,\phi}$ itself could have a non-zero value. Since in this case $V_{,\phi}=-3\xi_{,\phi}H^4$, the non-zero $\nu^{\prime}(x)$ here can be viewed as the GB correction effect in USR inflation.

The dynamics of characteristic function \eqref{Gen_EoM_chi} corresponding to Langevin equation \eqref{Langevin_USR_nv} should be 
\begin{equation}\label{EoM_USR_chi}
	\left[ {\nu \left( x \right)\frac{{{\partial ^2}}}{{\partial {x^2}}} - 3y\left( {\frac{\partial }{{\partial x}} + \frac{\partial }{{\partial y}}} \right) + it} \right]{\chi _\mathcal{N}}(x,y,t) = 0.
\end{equation}

As a second-order partial differential equation, \eqref{EoM_USR_chi} is generally difficult to solve analytically. Nonetheless, through the procedure introduced in \cite{Vennin:2020kng}, one can have an analytical solution around the classical limit where the stochastic effect is small. As discussed in \cite{Vennin:2020kng}, the classical limit can be taken if 
\begin{equation}\label{USR_cl_c}
	y_{\text {in }} \gg \max \left(1, \sqrt{\nu}\right),y_{\rm{in}}\gg x_{\rm{in}}.
\end{equation}

To be specific, by \eqref{Gen_Langevin_eq} as well as \eqref{Gen_FP_O_A} and \eqref{EoM_USR_chi}, one may notice the contribution of $\nu{\partial_x}^2$ in \eqref{EoM_USR_chi} fully comes from the stochastic noise effect. Therefore, at the \textbf{leading order} in the classical limit we have
\begin{equation}\label{EoM_chi_USR_cl_LO}
	\left[ { - 3y\left( {\frac{\partial }{{\partial x}} + \frac{\partial }{{\partial y}}} \right) + it} \right]{{\chi _\mathcal{N}}}^{\rm (LO)}(x,y,t) = 0~.
\end{equation}
Equation \eqref{EoM_chi_USR_cl_LO} can be solved with the boundary condition ${{\chi _\mathcal{N}}}^{\rm (LO)}(0,y,t)=1$ as
\begin{equation}\label{chi_USR_cl_LO}
	{{\chi _\mathcal{N}}}^{\rm (LO)}(x,y,t)=\left(\frac{y}{y-x}\right)^{i t / 3}.
\end{equation}
The mean first-passage time $\left\langle \mathcal{N} \right\rangle$ obtained from \eqref{chi_USR_cl_LO} via \eqref{Gen_N_n_chi} should be
\begin{equation}\label{N_USR_cl_LO}
	\langle\mathcal{N}\rangle_{\rm{(LO)}}=-\frac{1}{3} \ln \left(1-\frac{x_{\mathrm{in}}}{y_{\mathrm{in}}}\right).
\end{equation}
Here, the subscript ``in" means initial. Note that \eqref{N_USR_cl_LO} indeed matches the number of e-folds from the classical equation $\ddot{\bar \phi} +3H\dot{\bar\phi}=0$.  The pure classical result can be obtained from $\left\langle \mathcal{N}^{n} \right\rangle=\left\langle \mathcal{N} \right\rangle^{n}$ by using \eqref{chi_USR_cl_LO}. This is not surprising because we got rid of the stochastic effect in \eqref{EoM_chi_USR_cl_LO}. Moreover, the GB correction $\nu^{\prime}$ is excluded in \eqref{N_USR_cl_LO}, so no GB correction at the pure classical limit. This is also expected because the GB effect is contained in $\nu(x)$, which only contributes to the stochastic noise in the Langevin equation \eqref{Langevin_USR_nv}.

In the expansion around the classical limit, $\chi_{\mathcal{N}}$ should differ only slightly from the pure classical result \eqref{chi_USR_cl_LO}. Therefore, the next-to-leading order result can be obtained by evaluating the second order derivative contribution in \eqref{EoM_USR_chi} with the leading order results. That is,
\begin{equation}\label{EoM_chi_USR_NLO}
	\left[-3 y\left(\frac{\partial}{\partial x}+\frac{\partial}{\partial y}\right)+i t\right] {\chi_{\mathcal{N}}}^{\rm (NLO)}(x, y, t)=-\nu(x) \frac{\partial^2}{\partial x^2}\left(\frac{y}{y-x}\right)^{i t / 3} .
\end{equation}
With the boundary condition $\chi_{\mathcal{N}}(0, y, t)=1$, Eq. \eqref{EoM_chi_USR_NLO} has the solution
\begin{equation}\label{chi_USR_cl_NLO}	
{\chi_{\mathcal{N}}}^{\rm (NLO)}(x, y, t) = {\left( {\frac{y}{{y - x}}} \right)^{\frac{{it}}{3}}} \times \left( {1 + \delta{\chi_\mathcal{N}}^{{\text{1st cf}}}\left( {x,y,t} \right)} \right)~,	
\end{equation}
where the first-order noise-correction factor is 
\begin{equation}
   \delta {\chi _\mathcal{N}}^{{\text{1st cf}}}\left( {x,y,t} \right) = \frac{{ - {t^2} + 3it}}{{27{{\left( {y - x} \right)}^2}}}\left( {\nu \left( x \right)\ln \left( {\frac{y}{{y - x}}} \right) - \int_0^x {{{\nu ^\prime }\left( {x'} \right)\ln \left( {\frac{y}{{y - x'}}} \right)}d} x'} \right)~.
\end{equation}
Since $\nu^{\prime}$ appears in \eqref{chi_USR_cl_NLO}, GB correction takes effect at next-to-leading order. From \eqref{chi_USR_cl_NLO}, the mean first-passage time is the sum of the classical result and some first order stochastic noise correction with GB term effect:
\begin{equation}\label{USR_N_cl_NLO}
    \langle \mathcal{N}\rangle_{\rm{(NLO)}}= - \frac{1}{3}\ln \left( {1 - \frac{{{x_{{\text{in}}}}}}{{{y_{{\text{in}}}}}}} \right){\text{ + }}{\delta\left\langle \mathcal{N} \right\rangle ^{{\text{USR 1st cf}}}}\left( {{x_{{\text{in}}}},{y_{{\text{in}}}}} \right)~,
\end{equation}
where the first-order noise correction is
\begin{equation}
    \delta \left\langle \mathcal{N} \right\rangle^{\text{USR 1st cf}} = \frac{{ - \nu {{\left( {{x_{{\text{in}}}}} \right)}}\ln \left( {1 - \frac{{{x_{{\text{in}}}}}}{{{y_{{\text{in}}}}}}} \right) + \int_0^{{x_{{\text{in}}}}} { {{\nu ^\prime }\left( {x'} \right)\ln \left( {1 - \frac{{x'}}{{{y_{{\text{in}}}}}}} \right)} d} x'}}{{9{{\left( {{y_{{\text{in}}}} - {x_{{\text{in}}}}} \right)}^2}}}~.
\end{equation}
  With conditions \eqref{USR_cl_c}, one can expect that the stochastic correction effect in \eqref{USR_N_cl_NLO} is indeed small. In  the limit \eqref{USR_cl_c}, one may find that the leading-order correction of \eqref{mean_N_USR}, $\kappa^2\Delta N_{\rm usr}^{\rm cl}/6$, matches the following part of stochastic correction in \eqref{USR_N_cl_NLO}, 
\begin{equation}
    -\frac{\nu \left( {{x_{{\text{in}}}}} \right)\ln \left( {1 - \frac{{{x_{{\text{in}}}}}}{{{y_{{\text{in}}}}}}} \right)}{{{9{{\left( {{y_{{\text{in}}}} - {x_{{\text{in}}}}} \right)}^2}}}}~,
\end{equation}
up to $\co(\nu/{y_{\rm in}}^2)$, with $3H^2\simeq V$ and $\Pi_{\rm c}\simeq H^{-1}\dot{\bar\phi}$. The GB term induces noise corrections in \eqref{USR_N_cl_NLO}, which is absent in the GR result \eqref{mean_N_USR}.

From \eqref{USR_N_cl_NLO}, one may find that when $\nu^{\prime}=0$, stochastic noise prolongs the USR phase.  When $\nu^{\prime}>0$, GB correction reduces the resistance from stochastic noise. This result is reasonable because positive $\nu^{\prime}$ offers smaller quantum noise in the Langevin equation \eqref{Langevin_USR_nv} as $\phi$ rolls down to $\phi_{\rm end}$.

Around the classical limit \eqref{USR_cl_c}, the power spectrum with the leading-order noise corrections can be obtained from $\langle \mathcal{N}^2 \rangle$ calculated via \eqref{chi_USR_cl_NLO} (see Appendix \ref{Pow_spec_USR_FPT} for detail): 
\begin{equation}\label{Pow_spec_USR_cl_NLO_LO}
 		\mathcal{P}_{\zeta{\rm (NLO)}}\approx\frac{2\nu( x_{\rm in})}{{9{y_{\rm in}}^2}}\left( 1 + \frac{{{x_{{\text{in}}}}}}{{{3y_{{\text{in}}}}}} + \int_0^{{x_{{\text{in}}}}} {\frac{{{\nu ^\prime }\left( x \right)}}{{\nu \left( {{x_{{\text{in}}}}} \right)}}\frac{x}{{{y_{{\text{in}}}}}}}dx - \frac{{\nu \left( {{x_{{\text{in}}}}} \right)}}{{3{y_{{\text{in}}}}^2}}+\co\left(\left(\frac{x_{\rm in}}{y_{\rm in}}\right)^2\right) \right)~,
 \end{equation}
where we assume that $x_{\rm in}/y_{\rm in}$ and $\nu/{y_{\rm in}}^2$ are infinitesimal of the same order. The leading order term in power spectrum 
\eqref{Pow_spec_USR_cl_NLO_LO}, $2\nu(x_{\rm in})/(9{y_{\rm in}}^2)$, is indeed the power spectrum \eqref{P_zeta_SR} obtained by using $3H^2\approx V$ and $\Pi_{\rm c}\approx H^{-1}\dot{\bar\phi}$. The remaining terms can be interpreted as the leading-order noise corrections affected by the GB term.

\subsection{Primordial black holes from stochastic
GB effects}
The ``opposite" limit of the previous
section is the stochastic limit.
This is the limit where the velocity,
represented by $y$, is made small
by the Hubble friction when the classical force is nearly vanishing. The quantized inhomogeneities then backreact stochastically on the classical
background. At the level of Fokker-Planck solutions, non-negligible corrections
to the Gaussian curvature distribution
appear. In particular, the tail portion
is corrected from a Gaussian tail to an exponential tail. Such a correction will affect PBH production because the tail of the distribution determines how likely (or unlikely)
rare events like large curvatures required for initiating collapse occur. In this section
we will show how a GB coupling affects the tail part of the distribution by using the techniques developed in \cite{Ezquiaga:2019ftu}.

\subsubsection{Small-velocity solution to the characteristic function}
Our starting point is using
the small velocity expansion
of the characteristic function
$\chi(x,y,t)$:
\begin{equation}\label{y-exp}
    \chi(x,y,t)\approx
    \chi_0(x,t)+yf(x,t)\,,
    \ \chi_0(x,t)\equiv \chi(x,0,t)\,.
\end{equation}
Plugging this into
\eqref{EoM_USR_chi},
we obtain the two equations
at the leading order
governing $f(x,t)$ and $\chi_0(x,t)$:
\begin{align}
    &\nu(x)\chi_{0,xx}(x,t)
    =-it\chi_0(x,t)\,,\label{chi0-eq}\\
    &\nu(x)f_{,xx}(x,t)-3\chi_{0,x}(x,t)
    +(it-3)f(x,t)=0
    \label{f-eq}\,.
\end{align}
Here, the comma subscripts $(\dots)_{,x}$ denote derivatives with repect to $x$.

A further simplification
occurs in the USR limit.
The slope of the classical potential
becomes small. In the case
of standard GR, at leading order the slope can be taken as zero (provided the USR stage is short and an initial $y$ exists). Things simplify greatly and an exact solution could be found. 
When a GB coupling is present instead, a leading order USR limit translates to a vanishing slope of effective potential $V^{\rm eff}_{,\phi}=0$ rather than $V_{,\phi}=0$.
Nevertheless, we expect a small-slope expansion of $V$ to remain justifiable. The reason is that a constant effective potential implies $V_{,\phi}\sim \xi_{\phi}$,
which we assume to be small for the case of small GB coupling and slow-roll regime.

As a concrete example,
we return to our model introduced
in Section \ref{GB_example}.
The smallness of the slope
of $\nu(x)$ can be seen from the expansion around the end of USR $x=x_{\rm end}$:
\begin{equation}
    	\nu(x)\approx
     \nu(x_{\rm end})-\varepsilon\tilde{\Lambda}^4\sin(\phi_{\rm end}/f)x
	+O(\varepsilon^2)\,.
\end{equation}
Here, the two newly introduced
(positive) parameters
$\tilde{\Lambda}^4=\Lambda^4/(24\pi^2\Delta\phi_{\rm well})$
and $\varepsilon=\Delta\phi_{\rm well}/f\approx 1/35$.
(Notice that $\nu(x_{\rm end})$
also constains a factor of
$\tilde{\Lambda}^4$;
and the $\sin$ function is positive
for the USR regime in our model.)
Utilizing the smallness of $\varepsilon$, we have at $O(\varepsilon)$-order:
\begin{equation}
	\chi_{0,xx}(x,t)+it[\frac{1}{\nu(x_{\rm end})}+
	\frac{\varepsilon\tilde{\Lambda}^4}{\nu(x_{\rm end})^2}
	\sin(\frac{\phi_{\rm end}}{f})x]\chi_0(x,t)=0\,.
	\label{approx-chi0-eq}
\end{equation}
This is solved by Airy functions.
For the boundary conditions representing a deterministic PDF
(at end of USR $\varphi=\varphi_{\rm end}$)
and a reflective boundary at the onset of USR:
we demand $\chi(0,0,t)=1, \chi'(1,0,t)=0$
\cite{Ezquiaga:2019ftu}.
We thus have the solution
\begin{equation}
\chi_0\equiv\chi (x,0,t)=
	\frac{\text{Ai}'\left(\frac{a (-i) t-i b \varepsilon  t}{(-i b t \varepsilon )^{2/3}}\right) \text{Bi}\left(\frac{a (-i) t-i b x \varepsilon  t}{(-i b t \varepsilon )^{2/3}}\right)-\text{Bi}'\left(\frac{a (-i) t-i b \varepsilon  t}{(-i b t \varepsilon )^{2/3}}\right) \text{Ai}\left(\frac{a (-i) t-i b x \varepsilon  t}{(-i b t \varepsilon )^{2/3}}\right)}{\text{Ai}'\left(\frac{a (-i) t-i b \varepsilon  t}{(-i b t \varepsilon )^{2/3}}\right) \text{Bi}\left(-\frac{i a t}{(-i b t \varepsilon )^{2/3}}\right)-\text{Ai}\left(-\frac{i a t}{(-i b t \varepsilon )^{2/3}}\right) \text{Bi}'\left(\frac{a (-i) t-i b \varepsilon  t}{(-i b t \varepsilon )^{2/3}}\right)}\,.
	\label{chi0-full}
\end{equation}
In the above, the newly introduced positive constants 
(evaluated at $\phi_{\rm end}\sim 13$) are
\begin{equation}
	a:=\frac{1}{\nu(x_{\rm end})}\,,
	\ b:=\frac{\tilde{\Lambda}^4}{\nu(x_{\rm end})^2}
	\sin(\frac{\phi_{\rm end}}{f})=-\frac{\nu'}{\varepsilon\nu^2}\,.
	\label{}
\end{equation}
These $a$ and $b$ parameters
are closely related to how \eqref{approx-chi0-eq} is written but clog
up the notation for the solution.
To avoid clutter,
we also introduce the new quantities (functions of $t$)
$\alpha=-iat/(-ibt\varepsilon)^{2/3}\sim O(\varepsilon^{-2/3})$
and $\beta=-ib\varepsilon t/(-ibt\varepsilon)^{2/3}\sim O(\varepsilon^{1/3})$
for future convenience\footnote{We leave all cubic roots alone for now to avoid dealing with multivalued functions at this stage.
}.
This notation allows $\chi_0$ to be re-written as
\begin{equation}\label{chi0-sol-alpha}
    \chi_0=
    \frac{{\rm Ai}'(\alpha+\beta){\rm Bi}(\alpha+\beta x)-{\rm Bi}'(\alpha+\beta){\rm Ai}(\alpha+\beta x)}{{\rm Ai}'(\alpha+\beta){\rm Bi}(\alpha)-{\rm Bi}'(\alpha+\beta){\rm Ai}(\alpha)}
\end{equation}

The solution \eqref{chi0-full} cannot be directly used without contradicting
the correct bookkeeping in $O(\varepsilon)$ (flatness of $\nu$).
To see this, recall that we have only expanded $\nu(x)$
up to $O(\varepsilon)$. So, the $\chi_0(x,t)/\nu(x_{\rm end})^2$ term in \eqref{approx-chi0-eq}
can at most be consistently expanded up to $O(\varepsilon)$.
Namely, we can only at most keep the $O(\varepsilon)$ part of $\chi_0(x,t)$.
With this in mind,
we turn to the equation for $f(x,t)$ (written using $a$ and $b$ parameters):
\begin{equation}
	f''(x,t)+(it-3)(a+\varepsilon b x)f(x,t)=3(a+\varepsilon b x)\chi_{0,x}(x,t)\,,
	\label{approx-f-eq}
\end{equation}
The homogeneous part of this
ordinary differential equation
is the same as the equation for
$\chi_0$ but with a shift
$it\to it-3$.
So we still expect the solution
to be in terms of Airy functions
with a slight change in the argument inside the functions.
Indeed, we have the solution
(satisfying previously given boundary conditions):
\begin{equation}\label{f-sol}
    f(x,t)
    =-y_1(x)\int^x_0 \frac{y_2(x') S(x')}{W(x')}dx'
    -y_1(x)C_1
    +y_2(x)\int^x_0 \frac{y_1(x')S(x')}{W(x')}dx'
    +y_2(x)C_2\,.
\end{equation}
The newly introduced quantities are the independent solutions $y_1,y_2$ 
to the homogeneous equation of $f(x,t)$, the Wronskian
$W(x)$, the source term $S(x)$ and integration
constants
\begin{align*}
C_1=& -\frac{y_2(0)[(y_1'(1)\int^1_0 \frac{y_2(x') S(x')}{W(x')}dx'
-y_2'(1)\int^1_0 \frac{y_1(x') S(x')}{W(x')}dx']}{y_1'(1)y_2(0)-y_1(0)y_2'(1)}\\
C_2=&-\frac{y_1(0)[(y_1'(1)\int^1_0 \frac{y_2(x') S(x')}{W(x')}dx'
-y_2'(1)\int^1_0\frac{y_1(x') S(x')}{W(x')}dx']}{y_1'(1)y_2(0)-y_1(0)y_2'(1)}\\
    y_1(x)=&{\rm Ai}(\bar{\alpha}+\bar{\beta}x)\\
    y_2(x)=&{\rm Bi}(\bar{\alpha}+\bar{\beta}x)
    \\
    W(x)=&y_1(x)y_{2,x}(x)-y_2(x) y_{1,x}(x)=\bar{\beta}/\pi\\
    S(x)=& 3(a+\varepsilon b x) \chi_{0,x}\\
\end{align*}
Here $\bar{\alpha}=\alpha |_{it\rightarrow it-3}$
and similarly for $\bar{\beta}$.

\subsubsection{Tail distribution from the characteristic function}
Following \cite{Ezquiaga:2019ftu},
we look for in $\chi$ its simple poles (simple because \eqref{EoM_USR_chi} is linear in $t$) on the lower imaginary $t$-axis---at $it=\Lambda_n\in \mathbb{R}_{+}$ ($\Lambda_n$ here is not related to $\tilde{\Lambda}$)---and the associated
residues $a_n$ to approximate the tail distribution of e-folds realized by configuration $(x,y)$:
\begin{equation}\label{Eexp_PDF}
    P_{x,y}[\mathcal{N}]
    =\sum_n a_n(x,y)e^{-\Lambda_n\mathcal{N}}\,.
\end{equation}
Here, residues $a_n$ can be related to $\chi_{\mathcal{N}}$ as 
\begin{equation}\label{Gen_Res}
    a_n(x,y)=-i\left[\frac{\partial}{\partial t} \chi_{\mathcal{N}}^{-1}\left(x,y,t=-i \Lambda_n \right)\right]^{-1}\, .
\end{equation}
This form of tail distribution and the declared properties of $\Lambda_n$ can be justified from the definition of the characteristic equation as a Fourier transform of $P_{x,y}$ in $\mathcal{N}$, the positivity and reality conditions of the distribution, and the linearity of $it$ in \eqref{EoM_USR_chi}.
 
Fortunately, in our small-$y$ aproximation we only need to find the poles of $\chi_0$.
We get the $f(x,t)$ poles for free.
To see this, note that the poles of $f(x,t)$ comprise one set of poles directly from $\chi_0$ (inherited from the source term) and another set of poles from the homogeneous part of its equation. The similarity between the homogeneous part of $f(x,t)$'s equation \eqref{approx-f-eq} and $\chi_0$'s equation \eqref{approx-chi0-eq} tells us that the other set of poles are simply $\chi_0$ poles but shifted by $it\to it-3$.
The same conclusion can be obtained by inspecting the solution\footnote{
The solution cannot be used
for the case of $it=0$ or $it-3=0$.
It makes no sense to speak
of $t$-poles if $t=0$
in \eqref{Gen_EoM_chi}.
}
\eqref{f-sol}.

To find the poles of $\chi_0$,
we expand the denominator of \eqref{chi0-sol-alpha} in small $\varepsilon$:
\begin{equation} \label{chi0-denom}
\text{Denom}[\chi_0]
	\approx
	-\frac{1}{\pi}\cosh(\sqrt{\alpha}\beta)
	+(-\frac{\beta^2}{4\sqrt{\alpha}\pi}+\frac{1}{4\pi\alpha^{3/2}})
	\sinh(\sqrt{\alpha}\beta)
	-\frac{\beta}{4\pi\alpha}\cosh(\sqrt{\alpha}\beta)+O(\varepsilon^2)\,.
\end{equation}
We assume that a small slope of $\nu(x)$ induces a correction of the $it$-pole from the constant $\nu(x)$ case
given in \cite{Ezquiaga:2019ftu}:
\begin{equation}
(-it)^{1/2}=
\sqrt{\alpha}\beta\nu(x_{\rm end})^{1/2}=
\nu(x_{\rm end})^{1/2}\Big[i(1/2+n)\pi+\varepsilon\delta_n\Big]\,.
\end{equation}
That is, the roots of $\chi_0$ is $O(\varepsilon)$
away from the unperturbed roots (i.e. when $\nu'=0$ exactly).
In our approximation, the effect of $\varepsilon \delta_n$ is only manifest in the
first term of \eqref{chi0-denom}
because that is the only term not multiplied with a $O(\varepsilon)$ factor (from combinations of $\alpha,\beta$).
We solve $\varepsilon \delta_n$
by setting \eqref{chi0-denom}
to zero:
\begin{align*}
	\varepsilon\delta_n=& \sqrt{\frac{-it}{\nu}}(\frac{\nu'}{4\nu}-\frac{\nu'}{4(-it)})
	+O(\varepsilon^2)\\
	\approx& i\theta_0(\frac{\nu'}{4\nu}+\frac{\nu'}{4\theta_0^2\nu})+O(\varepsilon^2)\,,
\end{align*}
where $\theta_0=(\frac{1}{2}+n)\pi$ and we have used $\nu'\sim O(\varepsilon)$.
All $\nu,\nu'$ are understood to be evaluated at $x_{\rm end}$.
Recalling that poles are at $\sqrt{-\Lambda_n/\nu(x_{\rm end})}=\sqrt{-it/\nu(x_{\rm end})}=i\theta_0+\varepsilon\delta_n$,
We have
\begin{equation}\label{chi0-pole}
	\Lambda_n=\nu(x_{\rm end}) \theta_0^2(1+\frac{\nu'}{2\nu}(1+\frac{1}{\theta_0^2}))\,.
\end{equation}

To find the second set of poles of $f(x,t)$ besides the one directly from $\chi_0$, we simply need to shift the poles of \eqref{chi0-pole} by $\Lambda_n\to \Lambda_n+3$. (This shifting is consistent with the shifting in the standard GR case \cite{Vennin:2020kng}.)

With the poles found for $f(x,t)$ and $\chi_0(x,t)$, we will turn to their residues which determine the tail behavior of $P_{x,y}(\mathcal{N})$
and PBH mass fraction.

\subsubsection{PBH mass fraction from characteristic function}

PBHs are formed from coarse-grained inflationary curvature perturbations whose amplitudes surpass a critical threshold $\zeta_c$ and subsequently collapse upon horizon re-entry in the post-inflationary universe in uniform energy gauge. Therefore, given a coarse-grained curvature perturbation distribution $P(\zeta_{\text{cg}})$, the corresponding PBH mass fraction shall be calculated as\cite{Vennin:2020kng}
\begin{equation}\label{Gen_beta_f}
	\beta_{\mathrm{f}}(M)=\int_{\zeta_{\mathrm{c}}}^{\infty} P\left(\zeta_{\mathrm{cg}}\right) \mathrm{d} \zeta_{\mathrm{cg}}\, ,
\end{equation}
where the PBH mass $M$ is  the mass contained in a Hubble patch at the time of formation. In stochastic formalism, the amount of expansion realized along a given trajectory is $\mathcal{N}$. Therefore, coarse-grained curvature perturbations in spatially flat gauge are indeed $\zeta_{\text{cg}}=\mathcal{N}-\left\langle \mathcal{N}\right\rangle$\cite{Vennin:2020kng}. Hence, PBH mass fraction \eqref{Gen_beta_f} corresponding to tail distribution form PDF \eqref{Eexp_PDF} will be 
\begin{equation}\label{tail-expPDF-betaf}
	{\beta _{\rm{f}}} = \sum\limits_n {\frac{{{a_n}}}{{{\Lambda _n}}}} {e^{ - {\Lambda _n}\left( {{\zeta _{\rm{c}}} + \langle {\cal N}\rangle } \right)}}\, .
\end{equation}
If the poles $\Lambda_n$ are ordered as $0<\Lambda_0<\Lambda_1<\cdots<\Lambda_n$, then PBH mass fraction $\beta_{\rm{f}}$ \eqref{tail-expPDF-betaf} can be approximately as 
\begin{equation}\label{beta_f_Approx}
	\beta_{\rm{f}}\sim \frac{a_0}{\Lambda_0}e^{ - {\Lambda _0}\left( {{\zeta _{\rm{c}}} + \langle {\cal N}\rangle } \right)}
\end{equation}
since the remaining terms are generally exponentially suppressed relative to the leading one. 

In the zero initial momentum case whose corresponding characteristic function is $\chi_0$ given in \eqref{chi0-sol-alpha}, the mean first passage time  can be calculated up to $O\left( \varepsilon \right)$ via \eqref{Gen_N_n_chi} as
\begin{equation}\label{Zeroy_mN}
	{\left\langle {\cal N} \right\rangle _{y = 0}} = \frac{x}{{{\nu _0}}}\left( {1 - \frac{x}{2} - \frac{{\nu _0^\prime }}{{2{\nu _0}}} + \frac{{{x^3}\nu _0^\prime }}{{6{\nu _0}}}} \right)\, ,
\end{equation}
where we denote $\nu_0\equiv \nu(x_{\rm{end}})$ and $\nu _0^\prime \equiv -\varepsilon\tilde{\Lambda}^4\sin(\phi_{\rm end}/f)$. The PBH mass fraction $\beta_{\rm{f}}$ \eqref{beta_f_Approx} up to $O\left( \varepsilon \right)$ will be
\begin{equation}\label{beta_f_0y}
	\beta_{\rm{f}}\sim \left(\frac{{4\cos \left[ {\frac{1}{2}\pi \left( {1 - x} \right)} \right]}}{\pi } + \delta_{\rm{fac}}\right)\mathrm{exp}\left[{{\nu _0}\frac{{{\pi ^2}}}{4}{\zeta _{\rm{c}}} - \frac{{{\pi ^2}x}}{4}\left( {1 - \frac{x}{2}} \right)+\delta_{\rm{exp}}}\right]\, ,
\end{equation}
where the GB-coupling correction is
\begin{align}
	\delta_{\rm{fac}}&=\frac{{\nu _0^\prime }}{{{\nu _0}}}\left( {\frac{{\left( { - 4 + {\pi ^2}} \right)\mathrm{c}\left( x \right)}}{{{\pi ^3}}} + \frac{1}{\pi }\left( {\left( { - 1 + x} \right)\mathrm{c}\left( x \right) + \left( {\frac{2}{\pi } + \frac{1}{2}\pi \left( {1 - x} \right)} \right)x\mathrm{s}\left( x \right)} \right)} \right)\, , \\
	\delta_{\rm{exp}} & = - \left( {\frac{{{\pi ^2}}}{4} + 1} \right)\frac{{\nu _0^\prime {\zeta _{\rm{c}}}}}{2} - \frac{{\nu _0^\prime x}}{{2{\nu _0}}}\left( {1 - \frac{x}{2} - \frac{{{\pi ^2}x}}{8} + \frac{{{\pi ^2}{x^3}}}{{12}}} \right)\, .		
\end{align}
  Here for notational convenience, we define $\mathrm{c}\left( x \right)\equiv \cos \left[ {\frac{1}{2}\pi \left( {1 - x} \right)} \right]$ and $\mathrm{s}\left( x \right)\equiv \sin \left[ {\frac{1}{2}\pi \left( {1 - x} \right)} \right]$. In the absence of GB correction ($\propto \nu _0^\prime$), \eqref{beta_f_0y} is consistent with the result in GR case proposed in \cite{Vennin:2020kng}.
  
  \subsubsection{Results of First-passage time and PBH mass fraction}
  
So far we've got the analytical solution for $\chi_{\mathcal{N}}$, at least formally. Hence, PBH mass fraction $\beta_{\rm{f}}$ can be expressed in principle. Nevertheless, when substituting parameters introduced in Section \ref{GB_example}, both numerators and denominators are $O(\mathrm{exp}[{O(10^4\times (it))}])$ in the expressions \eqref{chi0-sol-alpha} and \eqref{f-sol}. This leads to numerical difficulty to get the first passage time as well as the PBH mass fraction corresponding to the example given in section \ref{GB_example} via the method proposed here. Moreover, in example \ref{GB_example}, the classical trajectory results in e-fold number  $\Delta N_{\rm usr}^{\rm cl}\approx 4.02$, which implies $y_{\rm{beg}}\sim O(1)$ at the beginning of USR region. This means that small velocity limit does not hold for the entire USR inflation. In spite of this, expressions \eqref{chi0-sol-alpha} and \eqref{f-sol} are valid for any case with small constant potential slope during USR. Thus, we may analyze qualitative properties of the stochastic effects by using the results of this subsection with arbitrary $\nu_0$ and small $\nu_0^{\prime}$.~\footnote{Given a specific model, $\nu_0^{\prime}$ only depends on the shape of the potential. In principle, the potential can be mirrored with respect to the $\phi=0$ axis, such that the inflaton evolves in the $\phi<0$ domain. This yields an inflationary dynamics equivalent to that obtained for $\phi>0$. Such transformation would, in principle, leads to a potential with the opposite sign of $V_{,\phi}$. Nevertheless, it does not mean $\nu_0^{\prime}$ with the opposite sign. In fact, in this scenario, $x$ should be defined as $x=(\phi_{\rm{end}}-\phi)/\Delta \phi_{\rm{well}}$ so that $x$ still ranges from 0 to 1. Therefore, $\nu^{\prime}\equiv \frac{d\nu}{dx}\propto\frac{dV}{d\phi}\frac{d\phi}{dx}\propto -V_{,\phi}$, which cancels out the  opposite sign generated by the mirror transformation.}
  
  \begin{figure}[htbp]
  	\centering
  	\begin{subfigure}{0.48\linewidth}
  		\includegraphics[width=\linewidth]{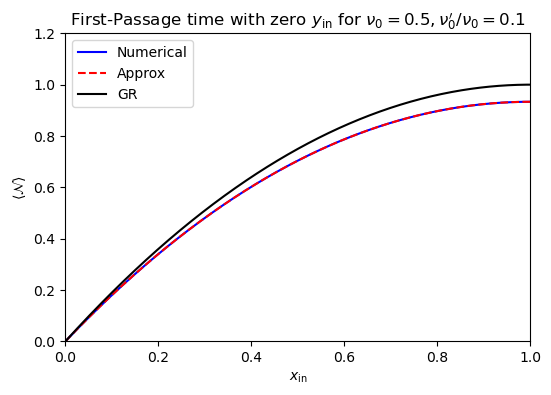}
  		\caption{The first-passage time with small $\nu_0^{\prime}>0$ for different initial field $x_{\rm{in}}$ in the absence of initial velocity. }
  	\end{subfigure}\hspace{10pt}
  	\begin{subfigure}{0.45\linewidth}
  		\includegraphics[width=\linewidth]{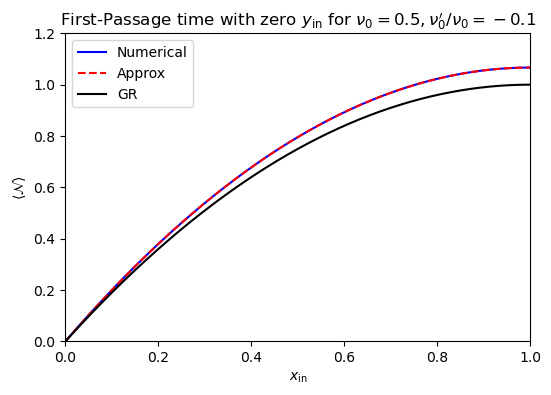}
  		\caption{The first passage time with small $\nu_0^{\prime}<0$ for different initial field $x_{\rm{in}}$ in the absence of initial velocity.}
  	\end{subfigure}
  	\captionsetup{width=1\linewidth}
  	\caption{The mean first-passage time with $\nu_0=0.5$ without initial velocity $y_{\rm{in}}$. The black line represents the results in GR case where $\nu_0^{\prime}=0$. The blue solid line means the results calculated directly from \eqref{chi0-sol-alpha} via \eqref{Gen_N_n_chi} numerically, while the dotted red one is the results from small $\varepsilon$ approximation expression \eqref{Zeroy_mN}, which shows the validity of \eqref{Zeroy_mN}. The positive $\nu_0^{\prime}$ results in a smaller $\left\langle \mathcal{N} \right\rangle$ and vice-versa. }\label{Zeroy_mN_Plot}
  \end{figure}
  
  While the classical trajectory leads to an infinite number of e-folds in the absence of an initial velocity within the USR regime, stochastic noise introduces random kicks that can eject the field from USR, giving rise to a finite first-passage time \eqref{Zeroy_mN}. Comparing with the results in GR, \eqref{Zeroy_mN} shows positive potential slope $\nu_0^{\prime}>0$ reduce the first-passage time and vice versa (see Fig. \ref{Zeroy_mN_Plot} for example). Physically, this makes sense because positive potential slope leads to larger stochastic noise during USR region.  
  
  \begin{figure}[htbp]
  	\centering
  	\includegraphics[width=0.5\textwidth]{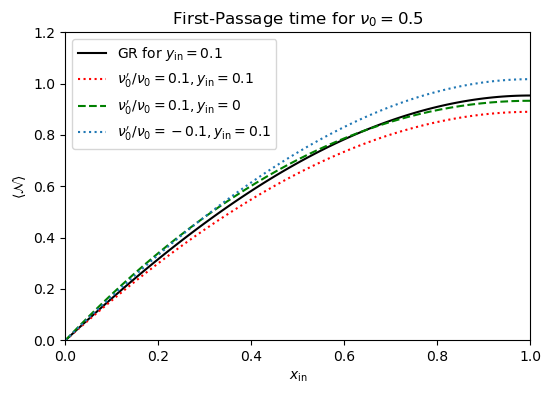}
  	\caption{The mean first-passage time for $\nu_0=0.5$ with non-zero velocity $y_{\rm{in}}$. The black line represents GR results with initial velocity $y_{\rm{in}}=0.1$. The green dashed line shows the results of $\nu_0^{\prime}/\nu_0=0.1$ with zero initial velocity. The red dotted line illustrates the results for $\nu_0^{\prime}/\nu_0=0.1$ with $y_{\rm{in}}=0.1$, while the blue line displays that for $\nu_0^{\prime}/\nu_0=-0.1$ case with the same velocity. Non-zero velocity results in smaller first-passage time. Positive $\nu_0^{\prime}$ leads to reduced first-passage time and vice-versa. }\label{01_mN_Plot}
  \end{figure}
  
  The mean first-passage time results with non-zero velocity are shown in Fig. \ref{01_mN_Plot}. The inclusion of initial velocity leads to a reduction in the first-passage time, which is consistent with physical intuition. Consistent with the velocity-free case, when a small finite velocity is included, a positive potential slope $\nu_0^{\prime}>0$ leads to a shorter first-passage time, and the opposite trend applies for a negative slope.
  
  \begin{figure}[htbp]
  	\centering
  	\begin{subfigure}{0.48\linewidth}
  		\includegraphics[width=\linewidth]{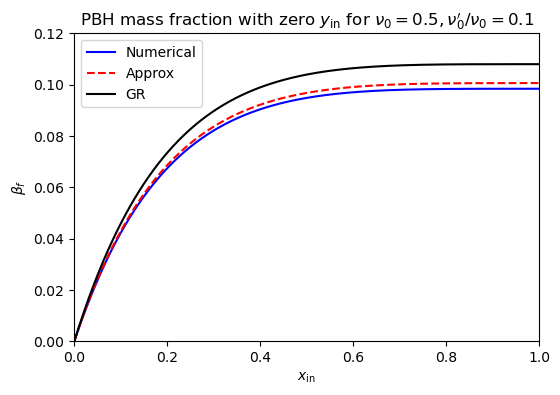}
  		\caption{PBH mass fraction for $\nu_0^{\prime}/\nu_0=0.1$ in the absence of initial velocity with different initial fields $x_{\rm{in}}$.}
  	\end{subfigure}\hspace{10pt}
  	\begin{subfigure}{0.45\linewidth}
  		\includegraphics[width=\linewidth]{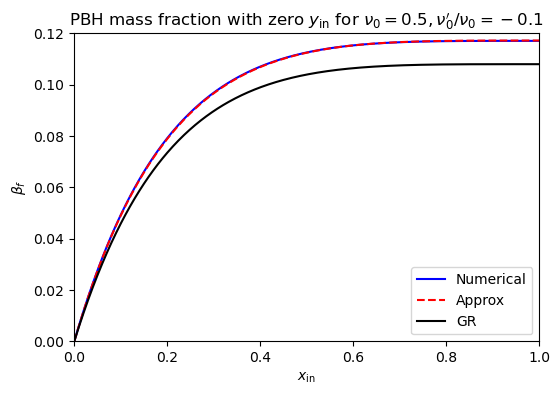}
  		\caption{PBH mass fraction for $\nu_0^{\prime}/\nu_0=-0.1$ in the absence of initial velocity with different initial fields $x_{\rm{in}}$.}
  	\end{subfigure}
  	\captionsetup{width=1\linewidth}
  	\caption{PBH mass fraction for $\nu_0=0.5$ without initial velocity (PBH fractions here and in subsequent plots use the results from Fokker--Planck equation). Here the critical curvature is taken as $\zeta_c=1$. The black solid line represents the results in GR case. The blue solid line means the results calculated numerically from \eqref{beta_f_Approx} by using \eqref{chi0-sol-alpha}, while the dotted red line is the results from small $\varepsilon$ approximation expression \eqref{beta_f_0y}, which shows the validity of \eqref{beta_f_0y}. Positive potential slope leads to smaller PBH mass fraction and vice-versa. }\label{Zeroy_betaf_Plot}
  \end{figure}

  \begin{figure}[htbp]
  	\centering
  	\includegraphics[width=0.55\textwidth]{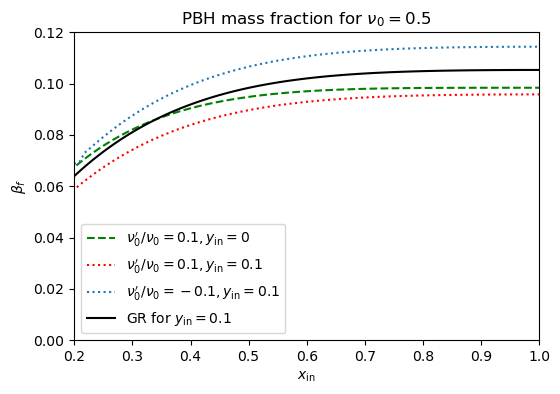}
  	\caption{The PBH mass fraction for $\nu_0=0.5$ with non-zero velocity $y_{\rm{in}}$. Here the critical curvature is taken as $\zeta_c=1$. The black line represents GR results with initial velocity $y_{\rm{in}}=0.1$. The green dashed line shows the results of $\nu_0^{\prime}/\nu_0=0.1$ with zero initial velocity. The red dotted line illustrates the results for $\nu_0^{\prime}/\nu_0=0.1$ with $y_{\rm{in}}=0.1$, while the blue line -- for $\nu_0^{\prime}/\nu_0=-0.1$ case with the same velocity. Non-zero velocity results in smaller mass fraction. Positive $\nu_0^{\prime}$ leads to reduced PBH mass fraction and vice-versa. }\label{betaf_01y}
  \end{figure}

In the absence of initial velocity, the classical power spectrum \eqref{P_zeta_SR} diverges, which results in PBH mass fraction $\beta_{\rm{f}}   \to \frac{1}{2}$\cite{Ashrafzadeh:2023ndt,Ashrafzadeh:2024oll}. The stochastic noise, however, will produce different $\beta_{\rm{f}}$. For example, Fig. \ref{Zeroy_betaf_Plot} shows PBH mass fraction $\beta_{\rm{f}}<0.5$ for $\nu_0=0.5$ when stochastic noise is taken into account. Moreover,  Fig. \ref{Zeroy_betaf_Plot} indicates positive potential slope $\nu_0^{\prime}>0$ results in smaller PBH mass fraction compared with this in GR case, while negative potential slope has the opposite effect. Indeed, these conclusions also hold for non-zero velocity case. (See Fig. \ref{betaf_01y} for example.)

\begin{figure}[htbp]
	\centering
	\includegraphics[width=0.55\textwidth]{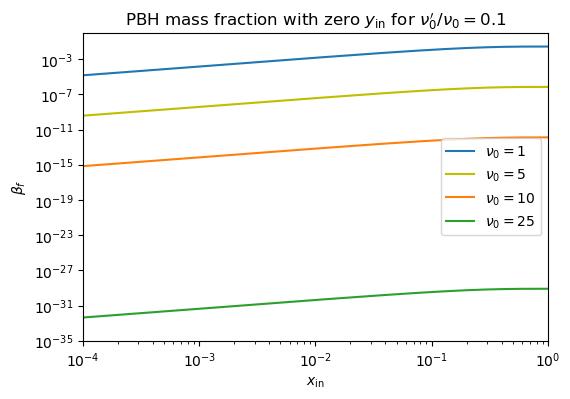}
	\caption{Velocity-free PBH mass fraction for $\nu_0^{\prime}/\nu_0=0.1$ with different $\nu_0$. Here the critical curvature is taken as $\zeta_c=1$. The order of magnitude of PBH mass fraction is not sensitive to the initial field value. Larger $\nu_0$ leads to smaller PBH mass fraction.}\label{vnuZeroyBetaf_01}
\end{figure}

\begin{figure}[htbp]
	\centering
	\begin{subfigure}[t]{0.48\linewidth}
		\includegraphics[width=\linewidth]{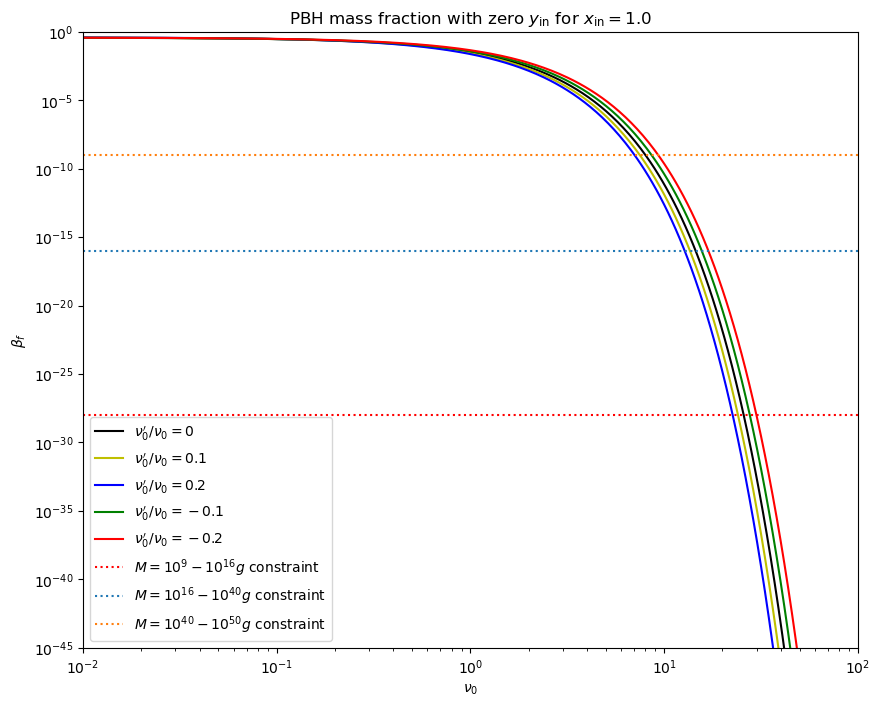}
		\caption{PBH mass fraction in the absence of initial velocity with fixed initial field $x_{\rm{in}}=1$ with $\nu_0$ ranging from $10^{-2}$ to $10^2$. Observational constraints of $\beta_{\rm{f}}(M)$ for different PBH mass are included. }
	\end{subfigure}\hspace{10pt}
	\begin{subfigure}[t]{0.48\linewidth}
		\includegraphics[width=\linewidth]{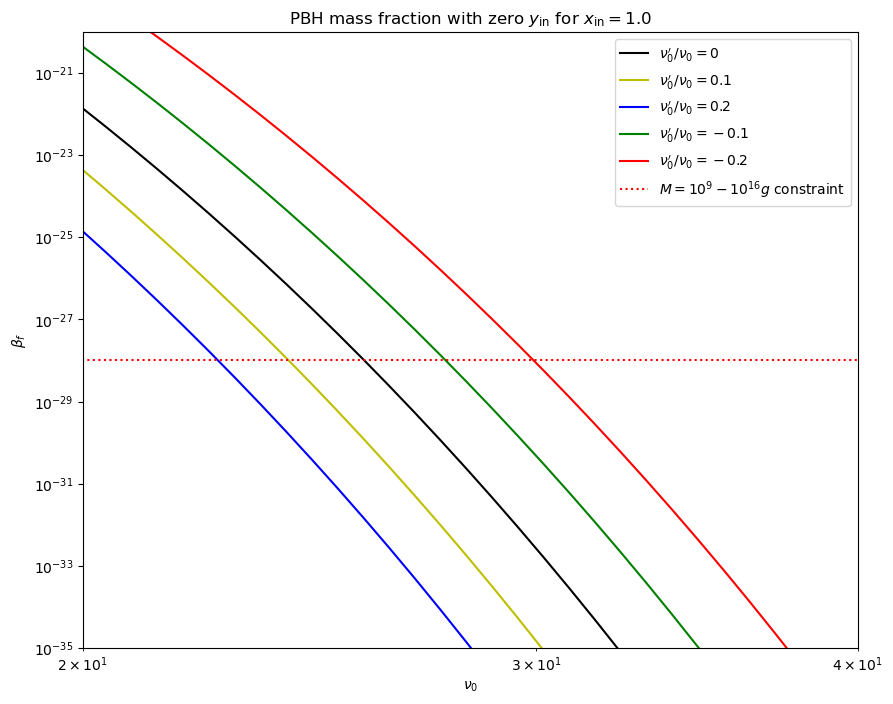}
		\caption{PBH mass fraction around the $\sim 10^{-28}$ exclusion bound. Compared to the GR constraints, positive $\nu_0^{\prime}$ allows smaller values of $\nu_0$ and vice-versa. The larger value of $\nu_0^{\prime}/\nu_0$ enables smaller values of $\nu_0$.}
	\end{subfigure}
	\captionsetup{width=1\linewidth}
	\caption{PBH mass fraction with zero initial velocity and fixed $x_{\rm{in}}=1$ for different $\nu_0$. Here the critical curvature is taken as $\zeta_c=1$. The black solid line represents the results in GR case. The dotted lines means observational constraints of $\beta_{\rm{f}}(M)$ for different PBH masses \cite{Carr:2020gox, Auffinger:2022khh}. For PBH within certain mass range, values of $\beta_{\rm{f}}$ above the corresponding line are excluded by observations. Larger value of $\nu_0$ leads to smaller $\beta_{\rm{f}}$. Positive $\nu_0^{\prime}$ results in reduced $\beta_{\rm{f}}$ compared to the GR results and vice-versa. To satisfy the observational constraint, positive $\nu_0^{\prime}$ allows smaller values of $\nu_0$ compared to the GR constraints, while for negative $\nu_0^{\prime}$, larger value of $\nu_0$ is needed.}\label{vnucxZeroyBetaf_01}
\end{figure}

Fig. \ref{vnuZeroyBetaf_01} indicates that the PBH mass fraction exhibits only weak dependence on the field, while being predominantly controlled by the parameter $\nu_0$. Increasing $\nu_0$ leads to a smaller PBH mass fraction. Therefore, observational constraints on the PBH mass fraction can be used to constrain the parameter $\nu_0$ as well as $\nu_0^{\prime}$. To be specific, PBH mass fraction results with zero initial velocity and fixed $x_{\rm{in}}=1$ with different $\nu_0$ are shown in Fig. \ref{vnucxZeroyBetaf_01}. Through this figure, one can draw the conclusion that in the stochastic-noise–dominated USR regime, compliance of our present model (not to be confused with Section \ref{GB_example}) with observational constraints necessitates $\nu_0 \gtrsim O(10)$, which would require a really small $\Delta \phi_{\rm well}$. Moreover, the presence of the GB coupling modifies the observational constraints on $\nu_0$. The positive value of $\nu_0^{\prime}$ allows smaller values of $\nu_0$ compared with constraints in GR, while larger values of $\nu_0$ are needed for negative $\nu_0^{\prime}$. Observational constraints on PBH mass fraction depend on corresponding masses of PBHs. Here, however, due to the computational difficulties involved, we do not perform calculations for specific inflationary models that satisfy the CMB observational constraints. Instead, we just treat the normalized potential $\nu_0$ and its slope $\nu_0^{\prime}$ as input parameters in the regime dominated by stochastic noise during USR inflation. Consequently, we are not able to determine the relation between the horizon-crossing values $x_{\rm{in}},y_{\rm{in}}$ and the resulting PBH mass $M$. Establishing such a correspondence requires further investigation with more advanced numerical methods.

\section{Spectator field in de Sitter vacuum}\label{Sec_spectator}

So far we discussed the case where $\phi$ is the inflaton. In this section we consider $\phi$ as a GB-coupled spectator field evolving during inflation, where we approximate the background by a de Sitter vacuum~\footnote{See Refs. \cite{Honda:2024evc, Nasuda:2025gyd} for recent discussions on some exact solutions for a stochastic spectator field in general relativity.}. The Lagrangian is given by \eqref{L_sGB}, as before.

We parametrize the scalar potential as
\begin{equation}\label{spectator_V}
    V=V_0+U(\phi)~,
\end{equation}
where $V_0$ represents the effective cosmological constant during inflation, and $U(\phi)$ is the possible potential for the spectator field, satisfying $|U|\ll V_0$.

After splitting $\phi$ into IR modes $\phi_c$ and UV modes $\phi_q$, we find the Hubble constraint equations (up to the noise corrections)
\begin{equation}\label{spectator_Friedmann}
    3H^2\simeq V_0~,~~~\phi_{c,N}^2\simeq -\omega+\omega_{,N}~,
\end{equation}
 under the conditions 
\begin{equation}\label{spectator_cond}
    |U|/V_0,~\phi_{c,N}^2,~|\omega|,~|\omega_{,N}|\ll 1~,
\end{equation}
where $\omega\equiv \xi_{,N} H^2$ is the GB slow-roll parameter. These conditions ensure no backreaction of $\phi$ on inflation, justifying its role as a spectator field.

Let us recall the equation for the linearized perturbation $\delta\phi_k=u_k/a$,
\begin{equation}\label{spectator_u_eq}
    u_k''-\frac{Z''}{Z}u_k+C_s^2u_k=0~,
\end{equation}
where $Z$ and $C_s$ are given by \eqref{Z_def} and \eqref{C_s_def}. Under the conditions \eqref{spectator_cond}, we get
\begin{equation}
    \frac{Z''}{Z}\simeq \Big(2+\frac{\phi_{c,NNN}}{\phi_{c,N}}+3\frac{\phi_{c,NN}}{\phi_{c,N}}\Big)a^2H^2~,~~~C_s\simeq 1~.
\end{equation}

The solution to \eqref{spectator_u_eq} (where $\phi_c$ evolution is derived in the classical limit) determines the noise terms of the Langevin equations \eqref{Langevin_phi_q} and \eqref{Langevin_Pi_q}. In general, the solution for $u_k$ and the corresponding noise terms should be found numerically, but in certain cases, we can obtain analytical results.

\subsection{The case with $V^{\rm eff}_{,\phi}=0$}

By using the potential \eqref{spectator_V}, and imposing
\begin{equation}\label{spectator_V_phi=0}
    V^{\rm eff}_{,\phi}=U_{,\phi}+\tfrac{1}{3}\xi_{,\phi}V_0^2=0~,
\end{equation}
we arrive at the scenario of effectively massless spectator field. We assume that $U_{,\phi}$ and $\xi_{,\phi}$ do not vanish separately, in order to exclude the well-known GR case.

The function $Z''/Z$ can be simplified by using the Klein--Gordon equation for $\phi_c$ in the classical limit, $\phi_{c,NN}\simeq -3\phi_{c,N}$, under the condition \eqref{spectator_V_phi=0}. This is the same equation as during USR inflation, and it leads to $Z''/Z\simeq 2a^2H^2$, such that Eq. \eqref{spectator_u_eq} is solved by
\begin{equation}\label{spectator_u_sol}
    \delta\phi_k=\frac{u_k}{a}\simeq\frac{e^{-ik\tau}}{a\sqrt{2k}}\Big(1+\frac{iaH}{k}\Big)~.
\end{equation}

In this case, the Langevin equations read
\begin{equation}\label{spectator_Langevin_1}
    \phi_{c,N}=\Pi_c+\Xi_\phi~,~~~\Pi_{c,N}=-3\Pi_c~,
\end{equation}
where $\Xi_\Pi$ vanishes. Here, we set $X_1\simeq 1$, which follows from conditions \eqref{spectator_cond}. Since $C_s\simeq 1$, we have $k_\Sigma\simeq\Sigma aH$ with $\Sigma\ll 1$. By using the solution \eqref{spectator_u_sol}, Eqs. \eqref{spectator_Langevin_1} reduce to
\begin{equation}\label{spectator_phi_stoch_1}
    \phi_{c,N}\simeq\phi_{c,N}(N_{\rm in})e^{-3(N-N_{\rm in})}+\frac{H}{2\pi}\chi~,
\end{equation}
where $\chi$ satisfies $\langle\chi(N)\chi(N')\rangle=\delta(N-N')$, and $N_{\rm in}$ is the initial time of the USR-like regime where $V^{\rm eff}_{,\phi}=0$.

As is clear from \eqref{spectator_phi_stoch_1}, the stochastic evolution of an effectively massless ($V^{\rm eff}_{,\phi}=0$) spectator field matches the result from GR, even if $U_{,\phi}$ and $\xi_{,\phi}$ are individually non-vanishing.

\subsection{The case with $U=0$}

A more interesting scenario arises if the spectator $\phi$ is massless (for simplicity we assume $U=0$), but its GB coupling is non-vanishing, $\xi_{,\phi}\neq 0$. In this case, the Klein--Gordon equation in the classical limit is
\begin{equation}\label{U=0_KG}
    \frac{\phi_{c,NN}}{\phi_{c,N}}\simeq -3-\frac{\xi_{,\phi}V_0}{\phi_{c,N}}\simeq\frac{3\omega_{,N}}{\omega-\omega_{,N}}~,
\end{equation}
where in the last step, we used $\omega=\xi_{,\phi}H^2\phi_{c,N}$ and Eq. \eqref{spectator_Friedmann}. In general, $\omega$ and its derivative will depend on the choice of $\xi(\phi)$, but there are two limits where the solution is insensitive to this choice: $|\omega|\ll |\omega_{,N}|$ and $|\omega|\gg |\omega_{,N}|$. In the former limit we have $\phi_{c,NN}\simeq -3\phi_{c,N}$, which is similar to the USR evolution. In the latter limit we have $|\phi_{c,NN}|\ll |\phi_{c,N}|$, which is similar to the usual slow-roll solution. In both of limits, the solution to the mode equation \eqref{spectator_u_eq} is the same and is given by \eqref{spectator_u_sol} (provided that $|\phi_{c,NNN}/\phi_{c,N}|\ll 1$ in the slow-roll-like case).

The Langevin equations for $U=0$ can be written as
\begin{align}
    \phi_{c,N} &\simeq \Pi_c+\frac{H}{2\pi}\chi~,\label{U=0_phi_eq}\\
    \Pi_{c,N} &\simeq -3\Pi_c-\xi_{,\phi}V_0\simeq -3\Pi_c+\frac{3\omega}{\omega-\omega_{,N}}\phi_{c,N}~,\label{U=0_Pi_eq}
\end{align}
where Eq. \eqref{spectator_Friedmann} is used. In the USR-like limit, $|\omega|\ll|\omega_{,N}|$, we have $\Pi_{c,N}\simeq -3\Pi_c$, which leads to the stochastic equation \eqref{spectator_phi_stoch_1}. In the SR-like limit one can ignore the acceleration term $\Pi_{c,N}$ in \eqref{U=0_Pi_eq} (which is justified by the slow-roll-like evolution \eqref{U=0_KG}), and take $\Pi_c\simeq -\xi_{,\phi}V_0/3$. Inserting it into \eqref{U=0_phi_eq} yields
\begin{equation}
    \phi_{c,N}\simeq -\tfrac{1}{3}\xi_{,\phi}V_0+\frac{H}{2\pi}\chi~,
\end{equation}
which describes SR-like stochastic evolution with the effective potential $V^{\rm eff}(\phi)=\xi(\phi)V_0^2/3$. In the absence of a scalar potential $U$, the GB-induced mass can in principle be used to stabilize the spectator field, if $\xi_{,\phi\phi}\gg H^{-1}$ (in Planck units).

\section{Discussion}

In this work, we studied stochastic inflationary dynamics in the presence of leading higher curvature corrections to general relativity, in the form of the Gauss--Bonnet invariant coupled to a scalar field. The latter can be associated with the inflaton or other massless scalar in de Sitter background. The motivation to pick the Gauss--Bonnet term is two-fold. On the one hand, it is known to produce a ghost-free, stable theory (under mild assumptions), even if the GB-scalar coupling becomes large during inflation. On the other hand, the GB term is predicted by effective field theory as well as string theory, making it a well-motivated test ground for testing possible effects beyond general relativity.

We first reviewed the dynamics of the background fields and scalar perturbations in uniform-field gauge, where the physical scalar perturbation is associated with the spatial metric perturbation which we call $\zeta$. Although in the literature, scalar perturbations in GB-coupled inflation are most often analyzed in uniform-field gauge, a more convenient choice for stochastic formalism is spatially flat gauge, where the gauge-invariant scalar perturbation can be directly related to the scalar field perturbation $\delta\phi$. We thus derived the equation of motion for $\delta\phi$ with GB coupling in spatially flat gauge, and confirmed that the result exactly matches the well-known equation for $\zeta$ (in uniform-field gauge) from the literature (see, for example, \cite{Kawai:2021bye} for the derivation).

By using spatially flat gauge, we derived stochastic Klein--Gordon and Langevin equations. In constant-roll ($\epsilon,\omega\simeq {\rm const.}$) limit, one can analytically solve the linear scalar mode equation, and find the analytical form of the stochastic noise amplitudes. In SR and USR limits, under additional conditions on the sound speed of scalar modes ($C_s\simeq 1$, $|C_{s,N}/C_s|\ll 1$), the Langevin equations almost coincide with the standard general relativity result, where the only difference is the contribution of the GB term to the slope of the effective potential.

By using first-passage time technique, we computed scalar power spectra in SR and USR regimes, with the leading-order noise correction first. Then we estimated PBH mass fraction in USR regimes in stochastic noise dominated case, showing the impact of the GB-coupling term on the parameter space allowed by observational constraints. Due to numerical difficulties, we treated normalized potential $\nu_0$ and its slope $\nu_0^{\prime}$ as constant input parameters in our estimations. A more thorough calculation of the PBH mass fraction deserves a separate study where more advanced numerical methods could be employed. Nevertheless, our estimations show that one can expect the results for PBH formation to differ noticeably from those obtained under the commonly adopted Gaussian assumption -- an approach in which the PBH mass fraction with a GB coupling has already been examined in Refs. \cite{Ashrafzadeh:2023ndt,Ashrafzadeh:2024oll,Solbi:2024zhl,Teimoori:2021thk}. This is because the stochastic formalism generically predicts probability distributions with exponentially suppressed tails \cite{Pattison_2017,Ezquiaga:2019ftu,Vennin:2020kng,Vennin_2025}, and the PBH mass fraction is highly sensitive to the behavior of the distribution in its far tail \cite{ozsoy2023inflation}. In addition to higher-curvature gravity, PBH can be formed in models with non-minimal derivative interactions \cite{Teimoori:2021thk,Heydari:2023rmq,Heydari:2024bxj}. Stochastic effects in such scenarios and their impact on the corresponding PBH mass fractions merit further investigation.

Finally, we considered stochastic evolution of a light GB-coupled spectator field in de Sitter background. We outlined some limits resembling SR and USR, where the amplitude of the stochastic noise matches the expected result $H/(2\pi)$. If a spectator field is massless, its GB coupling can induce an effective mass term to stabilize it, and suppress the stochastic noise. This could help to avoid large isocurvature perturbations during inflation.

\section*{Acknowledgements}

We are grateful to the referee for his/her constructive comments. DD is supported by China Postdoctoral Science Foundation (Certificate No.2023M730704). YW is supported by NSFC Grant No. 12475001, the Shanghai Municipal Science and Technology Major Project (Grant No. 2019SHZDZX01), Science and Technology Commission of Shanghai Municipality (Grant No. 24LZ1400100), and the Innovation Program for Quantum Science and Technology (No. 2024ZD0300101).  
YW is grateful for the hospitality of the Perimeter Institute during his visit, where the main part of this work is done. This research was supported in part by the Perimeter Institute for Theoretical Physics. Research at Perimeter Institute is supported by the Government of Canada through the Department of Innovation, Science and Economic Development and by the Province of Ontario through the Ministry of Research, Innovation and Science. 

\appendix

\section{Scalar perturbation in spatially flat gauge}\label{App_A}

The scalar curvature and the GB term expanded up to quadratic perturbations in the metric \eqref{SF_metric} reads (in the spatially flat gauge):
\begin{align}
\begin{aligned}\label{R_SF}
    R &=6H^2(2-\epsilon)(1-2\alpha+3\alpha^2)-6H(1-3\alpha)\dot\alpha-2(1-\alpha)\frac{\partial^2_i}{a^2}\alpha-2(1-2\alpha)\frac{\partial^2_i}{a^2}\dot\beta\\
    &-4H(1-2\alpha)\frac{\partial^2_i}{a^2}\beta+2\dot\alpha\frac{\partial^2_i}{a^2}\beta+6\frac{H}{a^2}\partial_i\alpha\partial^i\beta+\frac{1}{a^4}\Big[(\partial_i\partial_j\beta)^2 + \partial_i^2\beta\partial_j^2\beta  + 2\partial_i\beta\partial_j^2\partial_i\beta\Big]~,
\end{aligned}\\
\begin{aligned}\label{R_GB_SF}
    R_{\rm GB}^2 &=8H^2\bigg\{3H^2(1-\epsilon)(1-4\alpha+10\alpha^2)-3H(1-5\alpha)\dot\alpha-(1-3\alpha)\frac{\partial^2_i}{a^2}\alpha-(1-4\alpha)\frac{\partial^2_i}{a^2}\dot\beta\\
    &-2H(1-\epsilon)(1-4\alpha)\frac{\partial^2_i}{a^2}\beta+3\dot\alpha\frac{\partial^2_i}{a^2}\beta+\frac{2}{a^2}\partial_i\alpha\partial^i\alpha+3\frac{H}{a^2}\partial_i\alpha\partial^i\beta \\
    &+\frac{1}{a^4}\Big[(\partial_i\partial_j\beta)^2 + \partial_i\beta\partial^i\partial_j^2\beta-\tfrac{1}{2}(1-\epsilon)\Big((\partial_i\partial_j\beta)^2-\partial_i^2\beta\partial_j^2\beta\Big)\Big]\\
    &-\frac{1}{Ha^4}\big(\partial_i\partial_j\alpha\partial^i\partial^j\beta-\partial_i^2\alpha\partial_j^2\beta-\partial_i^2\dot\beta\partial_j^2\beta+\partial_i\partial_j\dot\beta\partial^i\partial^j\beta\big)\bigg\}~.
\end{aligned}
\end{align}
Plugging these expressions into the Lagrangian \eqref{L_sGB} leads to the second-order Lagrangian \eqref{L_quad_SF}. Higher spatial derivatives of \eqref{R_SF} and \eqref{R_GB_SF} vanish upon integrating by parts the action, leaving at most two derivatives of the perturbations in \eqref{L_quad_SF}.

Next, we specify the intermediate steps in obtaining the Lagrangian \eqref{L_dphi_final} for $\delta\phi$ in the spatially flat gauge. We first eliminate $\alpha$ and $\beta$ from the Lagrangian \eqref{L_quad_SF} according to their EoM,
\begin{align}
\begin{aligned}\label{C_alpha_CPT}
    2(3H^2-\tfrac{1}{2}\dot{\bar\phi}^2-6\dot\xi H^3)\alpha+H(2-3\dot\xi H)\frac{\partial_i^2}{a^2}\beta+(\dot{\bar\phi}+3\xi_{,\phi}H^3)\delta\dot\phi\\
    +(3\xi_{,\phi\phi}H^3\dot{\bar\phi}+V_{,\phi})\delta\phi-\xi_{,\phi}H^2\frac{\partial_i^2}{a^2}\delta\phi=0~,
\end{aligned}\\
\begin{aligned}\label{C_beta_CPT}
    H(2-3\dot\xi H)\alpha-(\dot{\bar\phi}+\xi_{,\phi}H^3-\xi_{,\phi\phi}H^2\dot{\bar\phi})\delta\phi+\xi_{,\phi}H^2\delta\dot\phi=0~.
\end{aligned}
\end{align}
Inserting them back into the Lagrangian yields
\begin{equation}\label{L_dphi_prefinal}
    \mathcal{L}=\frac{a^3}{2}\Big[X_1\delta\dot\phi^2+X_2\delta\phi\frac{\partial^2_i}{a^2}\delta\phi-X_3\delta\phi^2\Big]~,
\end{equation}
where
\begin{align}
\begin{aligned}
    X_1 &\equiv\Big[1+\frac{3\omega^2}{4\ce(1-\omega)}\Big]\frac{(1-\omega)^2}{(1-\frac{3}{2}\omega)^2}~,\qquad\qquad\ce\equiv\frac{\dot{\bar\phi}^2}{2H^2}~,\\
    X_2 &\equiv\Big[1-\tfrac{1}{2}\omega+\frac{3\omega^2}{4\ce}(1-\epsilon)+\frac{3\omega^3\sigma}{8\ce(1-\frac{3}{2}\omega)}\Big]\frac{1}{1-\frac{3}{2}\omega}~,\\
    X_3 &\equiv V_{,\phi\phi}+3\xi_{,\phi\phi}H^4(1-\epsilon)-H^2\Big(\frac{\dot\ce}{H\ce}-2\epsilon-3\frac{\epsilon}{\ce}\omega\Big)\Big(\epsilon+\frac{\dot\ce\omega}{4H\ce(1-\frac{3}{2}\omega)}\Big)\\
    &-\frac{H^2}{\ce}(3+\ce-3\omega)\Big(\epsilon+\frac{\dot\ce\omega}{4H\ce(1-\frac{3}{2}\omega)}\Big)^2\\
    &-\frac{1}{2a^3}\partial_t\Big\{a^3H\Big[\epsilon\big(2+3\frac{\omega}{\ce}\big)+\frac{\dot\ce\omega}{2H\ce(1-\frac{3}{2}\omega)}+\big(2+3\frac{\omega}{\ce}\big)\frac{\dot\ce\omega(1-\omega)}{4H\ce(1-\frac{3}{2}\omega)^2}\Big]\Big\}~.
\end{aligned}
\end{align}
Using the background equations, one can show that
\begin{equation}\label{X3_eq}
    X_3=-\tfrac{1}{2}(a^3\sqrt{\ce})^{-1}\partial_t\Big(a^3\frac{\dot\ce}{\sqrt{\ce}}X_1\Big)~.
\end{equation}
Using \eqref{X3_eq} in Eq. \eqref{L_dphi_prefinal} leads to the final Lagrangian \eqref{L_dphi_final} for $\delta\phi$ in the main text.

\section{First-passage time calculation}

\subsection{``Saddle-point" expansion}\label{Saddle-point}

During slow-roll, taking $\dot{\bar\phi}\simeq 0$ implies ${V_{,\phi }} + 3{\xi _{,\phi }}{H^4}\simeq 0$. Thus, it is reasonable to expect that ${V_{,\phi }}$ and ${\xi _{,\phi }}$ have opposite signs. In most models of inflation, $v$ is monotonous during slow-roll. In this work, we assume that $v'>0$ and $\tilde{\xi}'<0$ during slow-roll. With these assumption, the term
\begin{equation}\label{Kernal_int}
	\int_{z_1}^{\bar \phi \left( {{\phi _1},{\phi _2}} \right)} {dz_2} \frac{1}{{v(z_2)}}\exp \left[ {\frac{1}{{{v_{{\text{eff}}}}(z_2)}}} \right]~,
\end{equation}
in Eq. \eqref{Gen_mN} varies exponentially. Thus, the largest contribution to the integral comes from the minimum of potential as well as the maximum of GB coupling $\xi$. Consequently, the integration of \eqref{Kernal_int} can be performed by expanding $1/v\left(z_2\right)\approx 1/v\left(z_1\right)-v^{\prime}(z_1) / v^2(z_1)(z_2-z_1)$ and $\tilde{\xi}\left(z_2\right)\approx \tilde{\xi}\left(z_1\right)+ \tilde{\xi}'\left(z_1\right)\left(z_2-z_1\right)$. Hence, Eq. \eqref{Gen_mN} leads to
\begin{equation}
	 \langle \mathcal{N}\rangle (\phi ) \simeq \int_{{\phi _{{\text{end}}}}}^\phi  {dz} \frac{{3v\left( z \right)}}{{3v'\left( z \right) + \tilde \xi '\left( z \right){v^2}\left( z \right)}}~,
\end{equation}
after neglecting the exponentially vanishing term $\exp\big[ -v^{\prime}(z_1) / v^2(z_1)(z_2-z_1)+ \tilde{\xi}'\left(z_1\right)\left(z_2-z_1\right) \big] $ because $v\left(\bar\phi\right)\gg v\left(z_1\right)$ and $\tilde{\xi}(\bar\phi)\ll \tilde{\xi}(z_1)$ due to our assumption.

Saddle-point expansion makes sense only when both $1/v(z_2)$ and $\tilde{\xi}(z_2)$ differ slightly from the values at point $x$, say, $|v(z_2)^{-1} -v(z_1)^{-1}| < R$ and $|{\tilde \xi \left( z_2 \right) - \tilde \xi \left( z_1 \right)}| < R$, where $R$ is a number much smaller than one. Using Taylor expansion up to first order yields $|z_2-z_1|<v^2R/v'$ and $|z_2-z_1|<-R/\tilde \xi '$. The requirement that the second order terms of the expansion are small translates to $|2v - v''v^2/v'^2| \ll 1$ and $|\tilde \xi ''/\tilde \xi'^2| \ll 1$. Hence, it is reasonable to define the ``classicality" criteria of $v$ and $\xi$ as
\begin{align}
	\eta _{{\text{cl}}}^v &= \left| {2v - \frac{{v''{v^2}}}{{v{'^2}}}} \right|~,\\
	\eta _{{\text{cl}}}^\xi  &= \left| {\frac{{\tilde \xi ''}}{{\tilde \xi {'^2}}}} \right|~.
\end{align}
The classical trajectory is a good approximation to the mean stochastic one iff both $v$ and $\xi$ satisfy the classicality criteria $\eta _{{\text{cl}}}^v \ll 1$ and $\eta _{{\text{cl}}}^\xi \ll 1$.

\subsection{First-passage time power spectrum calculation in USR limit}\label{Pow_spec_USR_FPT}

The second moment of the number of e-folds can be obtained from \eqref{chi_USR_cl_NLO} as
\begin{equation}
	\begin{aligned}
	\langle {\mathcal{N}^2}\rangle_{\rm NLO}&	= \frac{1}{9}{\left( {\ln \left( {1 - \frac{{{x_{{\text{in}}}}}}{{{y_{{\text{in}}}}}}} \right)} \right)^2} \\ + & \frac{{2\left( {1 - \ln \left( {1 - \frac{{{x_{{\text{in}}}}}}{{{y_{{\text{in}}}}}}} \right)} \right)}}{{27{{\left( {{y_{{\text{in}}}} - {x_{{\text{in}}}}} \right)}^2}}}\left( { - \nu \left( {{x_{{\text{in}}}}} \right)\ln \left( {1 - \frac{{{x_{{\text{in}}}}}}{{{y_{{\text{in}}}}}}} \right) + \int_0^{{x_{{\text{in}}}}} { {{\nu ^\prime }\left( x \right)\ln \left( {1 - \frac{x}{{{y_{{\text{in}}}}}}} \right)} d} x} \right)~.
	\end{aligned}
\end{equation}
Hence, 
\begin{equation}\label{USR_Delta_N2_cl_NLO}
	\begin{aligned}
		\langle \delta {\mathcal{N}^2}\rangle_{\rm NLO}  &= \langle {\mathcal{N}^2}\rangle_{\rm NLO}  - {{\langle \mathcal{N}\rangle_{\rm NLO}} ^2}  \\
		&= \frac{2}{{27{{\left( {{y_{{\text{in}}}} - {x_{{\text{in}}}}} \right)}^2}}}\left( { - \nu \left( {{x_{{\text{in}}}}} \right)\ln \left( {1 - \frac{{{x_{{\text{in}}}}}}{{{y_{{\text{in}}}}}}} \right) + \int_0^{{x_{{\text{in}}}}} {{{\nu ^\prime }\left( x \right)\ln \left( {1 - \frac{x}{y_{\rm in}}} \right)} d} x} \right)\\
		 &\ \ \ \  -\frac{{{{\left( { - \nu \left( {{x_{{\text{in}}}}} \right)\ln \left( {1 - \frac{{{x_{{\text{in}}}}}}{{{y_{{\text{in}}}}}}} \right) + \int_0^{{x_{{\text{in}}}}} {{{\nu ^\prime }\left( x \right)\ln \left( {1 - \frac{x}{y_{\rm in}}} \right)} d} x} \right)}^2}}}{{81{{\left( {{y_{{\text{in}}}} - {x_{{\text{in}}}}} \right)}^4}}}~.  
	\end{aligned} 
\end{equation}
Since the calculation is given in the small noise limit, it is reasonable to calculate the power spectrum \eqref{Gen_pow_spec} along the classical trajectory with $x=x_{\mathrm{in}}-y_{\mathrm{in}}\left[1-e^{-3\left(N-N_{\mathrm{in}}\right)}\right]$ and $y=y_{\text {in }} e^{-3\left(N-N_{\text {in }}\right)}$. Therefore, the power spectrum obtained via \eqref{USR_Delta_N2_cl_NLO} and \eqref{USR_N_cl_NLO} would take the form
\begin{equation}\label{Pow_spec_USR_cl_NLO_Gen}
	\begin{aligned}
		\cp_{\zeta{\rm (NLO)}}{\text{ = }}&\frac{{2\left( {3{{\left( {{y_{{\text{in}}}} - {x_{{\text{in}}}}} \right)}^2} + \nu \left( {{x_{{\text{in}}}}} \right)\ln \left( {1 - \frac{{{x_{{\text{in}}}}}}{{{y_{{\text{in}}}}}}} \right) - \int_0^{{x_{{\text{in}}}}} {{{\nu ^\prime }\left( x \right)\ln \left( {1 - \frac{x}{{{y_{{\text{in}}}}}}} \right)} d} x} \right)}}{{9{{\left( {{y_{{\text{in}}}} - {x_{{\text{in}}}}} \right)}^2}\left( {3{{\left( {{y_{{\text{in}}}} - {x_{{\text{in}}}}} \right)}^2} + \nu \left( {{x_{{\text{in}}}}} \right) + \int_0^{{x_{{\text{in}}}}} {{{\nu ^\prime }\left( x \right)\left( {\frac{x}{{{y_{{\text{in}}}} - x}}} \right)} d} x} \right)}} \\
		&\times \left( {\nu \left( {{x_{{\text{in}}}}} \right) + \int_0^{{x_{{\text{in}}}}} {{{\nu ^\prime }\left( x \right)\left( {\frac{x}{{{y_{{\text{in}}}} - x}}} \right)} d} x} \right).
	\end{aligned}
\end{equation}
In the absence of the GB correction $\nu^{\prime}(x)=0$, power spectrum \eqref{Pow_spec_USR_cl_NLO_Gen} indeed matches the GR result given in \cite{Vennin:2020kng}. 
 
Since our calculations based on the small-noise limit use \eqref{USR_cl_c}, we may expand \eqref{Pow_spec_USR_cl_NLO_Gen} to the linear order as
 \begin{equation}\label{Pow_spec_USR_cl_NLO_LO_App}
 		\cp_{\zeta{\rm (NLO)}}\approx\frac{{2\nu \left( {{x_{{\text{in}}}}} \right)}}{{9{y_{{\text{in}}}}^2}}\left( 1 + \frac{{{x_{{\text{in}}}}}}{{{3y_{{\text{in}}}}}} + \int_0^{{x_{{\text{in}}}}} {{\frac{{{\nu ^\prime }\left( x \right)}}{{\nu \left( {{x_{{\text{in}}}}} \right)}}\frac{x}{{{y_{{\text{in}}}}}}}d} x - \frac{{\nu \left( {{x_{{\text{in}}}}} \right)}}{{3{y_{{\text{in}}}}^2}}+O\left(\left(\frac{x_{\rm in}}{y_{\rm in}}\right)^2\right) \right)
 \end{equation}
where we assume $x_{\rm in}/y_{\rm in}$ and $\nu/{y_{\rm in}}^2$ are infinitesimal of the same order.

\clearpage

\providecommand{\href}[2]{#2}\begingroup\raggedright\endgroup


\begin{thebibliography}{10}

\bibitem{starobinsky1986stochastic}
A.~A. Starobinsky, ``{STOCHASTIC DE SITTER (INFLATIONARY) STAGE IN THE EARLY
  UNIVERSE},'' \href{http://dx.doi.org/10.1007/3-540-16452-9_6}{{\em Lect.
  Notes Phys.} {\bfseries 246} (1986) 107--126}.

\bibitem{Nambu:1987ef}
Y.~Nambu and M.~Sasaki, ``{Stochastic Stage of an Inflationary Universe
  Model},'' \href{http://dx.doi.org/10.1016/0370-2693(88)90974-4}{{\em Phys.
  Lett. B} {\bfseries 205} (1988) 441--446}.

\bibitem{Nambu:1988je}
Y.~Nambu and M.~Sasaki, ``{Stochastic approach to chaotic inflation and the
  distribution of universes},''
  \href{http://dx.doi.org/10.1016/0370-2693(89)90385-7}{{\em Phys. Lett. B}
  {\bfseries 219} (1989) 240--246}.

\bibitem{Kandrup:1988sc}
H.~E. Kandrup, ``{STOCHASTIC INFLATION AS A TIME DEPENDENT RANDOM WALK},''
  \href{http://dx.doi.org/10.1103/PhysRevD.39.2245}{{\em Phys. Rev. D}
  {\bfseries 39} (1989) 2245}.

\bibitem{Nakao:1988yi}
K.-i. Nakao, Y.~Nambu, and M.~Sasaki, ``{Stochastic Dynamics of New
  Inflation},'' \href{http://dx.doi.org/10.1143/PTP.80.1041}{{\em Prog. Theor.
  Phys.} {\bfseries 80} (1988) 1041}.

\bibitem{Nambu:1989uf}
Y.~Nambu, ``{Stochastic Dynamics of an Inflationary Model and Initial
  Distribution of Universes},''
  \href{http://dx.doi.org/10.1143/PTP.81.1037}{{\em Prog. Theor. Phys.}
  {\bfseries 81} (1989) 1037}.

\bibitem{Mollerach:1990zf}
S.~Mollerach, S.~Matarrese, A.~Ortolan, and F.~Lucchin, ``{Stochastic inflation
  in a simple two field model},''
  \href{http://dx.doi.org/10.1103/PhysRevD.44.1670}{{\em Phys. Rev. D}
  {\bfseries 44} (1991) 1670--1679}.

\bibitem{Linde:1993xx}
A.~D. Linde, D.~A. Linde, and A.~Mezhlumian, ``{From the Big Bang theory to the
  theory of a stationary universe},''
  \href{http://dx.doi.org/10.1103/PhysRevD.49.1783}{{\em Phys. Rev. D}
  {\bfseries 49} (1994) 1783--1826},
  \href{http://arxiv.org/abs/gr-qc/9306035}{{\ttfamily arXiv:gr-qc/9306035}}.

\bibitem{Starobinsky:1994bd}
A.~A. Starobinsky and J.~Yokoyama, ``{Equilibrium state of a selfinteracting
  scalar field in the De Sitter background},''
  \href{http://dx.doi.org/10.1103/PhysRevD.50.6357}{{\em Phys. Rev. D}
  {\bfseries 50} (1994) 6357--6368},
  \href{http://arxiv.org/abs/astro-ph/9407016}{{\ttfamily
  arXiv:astro-ph/9407016}}.

\bibitem{Campo:2005quantum}
D.~Campo and R.~Parentani, ``{Quantum correlations in inflationary spectra and
  violation of Bell inequalities},''
  \href{http://dx.doi.org/10.1590/S0103-97332005000700016}{{\em Braz. J. Phys.}
  {\bfseries 35} (2005) 1074--1079},
  \href{http://arxiv.org/abs/astro-ph/0510445}{{\ttfamily
  arXiv:astro-ph/0510445}}.

\bibitem{Campo:2005inflationary}
D.~Campo and R.~Parentani, ``{Inflationary spectra and violations of Bell
  inequalities},'' \href{http://dx.doi.org/10.1103/PhysRevD.74.025001}{{\em
  Phys. Rev. D} {\bfseries 74} (2006) 025001},
  \href{http://arxiv.org/abs/astro-ph/0505376}{{\ttfamily
  arXiv:astro-ph/0505376}}.

\bibitem{Gallicchio:2013testing}
J.~Gallicchio, A.~S. Friedman, and D.~I. Kaiser, ``{Testing
  Bell\textquoteright{}s Inequality with Cosmic Photons: Closing the
  Setting-Independence Loophole},''
  \href{http://dx.doi.org/10.1103/PhysRevLett.112.110405}{{\em Phys. Rev.
  Lett.} {\bfseries 112} no.~11, (2014) 110405},
  \href{http://arxiv.org/abs/1310.3288}{{\ttfamily arXiv:1310.3288
  [quant-ph]}}.

\bibitem{Maldacena:2015amodel}
J.~Maldacena, ``{A model with cosmological Bell inequalities},''
  \href{http://dx.doi.org/10.1002/prop.201500097}{{\em Fortsch. Phys.}
  {\bfseries 64} (2016) 10--23},
  \href{http://arxiv.org/abs/1508.01082}{{\ttfamily arXiv:1508.01082
  [hep-th]}}.

\bibitem{Green:2020signals}
D.~Green and R.~A. Porto, ``{Signals of a Quantum Universe},''
  \href{http://dx.doi.org/10.1103/PhysRevLett.124.251302}{{\em Phys. Rev.
  Lett.} {\bfseries 124} no.~25, (2020) 251302},
  \href{http://arxiv.org/abs/2001.09149}{{\ttfamily arXiv:2001.09149
  [hep-th]}}.

\bibitem{Nambu:2011classical}
Y.~Nambu and Y.~Ohsumi, ``{Classical and Quantum Correlations of Scalar Field
  in the Inflationary Universe},''
  \href{http://dx.doi.org/10.1103/PhysRevD.84.044028}{{\em Phys. Rev. D}
  {\bfseries 84} (2011) 044028},
  \href{http://arxiv.org/abs/1105.5212}{{\ttfamily arXiv:1105.5212 [gr-qc]}}.

\bibitem{Lim:2014quantum}
E.~A. Lim, ``{Quantum information of cosmological correlations},''
  \href{http://dx.doi.org/10.1103/PhysRevD.91.083522}{{\em Phys. Rev. D}
  {\bfseries 91} no.~8, (2015) 083522},
  \href{http://arxiv.org/abs/1410.5508}{{\ttfamily arXiv:1410.5508 [hep-th]}}.

\bibitem{martin2016quantum}
J.~Martin and V.~Vennin, ``Quantum discord of cosmic inflation: Can we show
  that cmb anisotropies are of quantum-mechanical origin?,'' {\em Physical
  Review D} {\bfseries 93} no.~2, (2016) 023505.

\bibitem{Mukhanov:1990theory}
V.~F. Mukhanov, H.~A. Feldman, and R.~H. Brandenberger, ``{Theory of
  cosmological perturbations. Part 1. Classical perturbations. Part 2. Quantum
  theory of perturbations. Part 3. Extensions},''
  \href{http://dx.doi.org/10.1016/0370-1573(92)90044-Z}{{\em Phys. Rept.}
  {\bfseries 215} (1992) 203--333}.

\bibitem{Ivanov:1997nonlinear}
P.~Ivanov, ``{Nonlinear metric perturbations and production of primordial black
  holes},'' \href{http://dx.doi.org/10.1103/PhysRevD.57.7145}{{\em Phys. Rev.
  D} {\bfseries 57} (1998) 7145--7154},
  \href{http://arxiv.org/abs/astro-ph/9708224}{{\ttfamily
  arXiv:astro-ph/9708224}}.

\bibitem{Garcia-Bellido:2017primordial}
J.~Garcia-Bellido and E.~Ruiz~Morales, ``{Primordial black holes from single
  field models of inflation},''
  \href{http://dx.doi.org/10.1016/j.dark.2017.09.007}{{\em Phys. Dark Univ.}
  {\bfseries 18} (2017) 47--54},
  \href{http://arxiv.org/abs/1702.03901}{{\ttfamily arXiv:1702.03901
  [astro-ph.CO]}}.

\bibitem{Pattison:2019stochastic}
C.~Pattison, V.~Vennin, H.~Assadullahi, and D.~Wands, ``{Stochastic inflation
  beyond slow roll},''
  \href{http://dx.doi.org/10.1088/1475-7516/2019/07/031}{{\em JCAP} {\bfseries
  07} (2019) 031}, \href{http://arxiv.org/abs/1905.06300}{{\ttfamily
  arXiv:1905.06300 [astro-ph.CO]}}.

\bibitem{Artigas:2021Hamiltonian}
D.~Artigas, J.~Grain, and V.~Vennin, ``{Hamiltonian formalism for cosmological
  perturbations: the~separate-universe approach},''
  \href{http://dx.doi.org/10.1088/1475-7516/2022/02/001}{{\em JCAP} {\bfseries
  02} no.~02, (2022) 001}, \href{http://arxiv.org/abs/2110.11720}{{\ttfamily
  arXiv:2110.11720 [astro-ph.CO]}}.

\bibitem{Artigas:2023kyo}
D.~Artigas, J.~Grain, and V.~Vennin, ``{Hamiltonian formalism for cosmological
  perturbations: fixing the gauge},''
  \href{http://dx.doi.org/10.1088/1475-7516/2025/01/083}{{\em JCAP} {\bfseries
  01} (2025) 083}, \href{http://arxiv.org/abs/2309.17184}{{\ttfamily
  arXiv:2309.17184 [gr-qc]}}.

\bibitem{Pinol:2020cdp}
L.~Pinol, S.~Renaux-Petel, and Y.~Tada, ``{A manifestly covariant theory of
  multifield stochastic inflation in phase space: solving the discretisation
  ambiguity in stochastic inflation},''
  \href{http://dx.doi.org/10.1088/1475-7516/2021/04/048}{{\em JCAP} {\bfseries
  04} (2021) 048}, \href{http://arxiv.org/abs/2008.07497}{{\ttfamily
  arXiv:2008.07497 [astro-ph.CO]}}.

\bibitem{Launay:2024qsm}
Y.~L. Launay, G.~I. Rigopoulos, and E.~P.~S. Shellard, ``{Stochastic inflation
  in general relativity},''
  \href{http://dx.doi.org/10.1103/PhysRevD.109.123523}{{\em Phys. Rev. D}
  {\bfseries 109} no.~12, (2024) 123523},
  \href{http://arxiv.org/abs/2401.08530}{{\ttfamily arXiv:2401.08530 [gr-qc]}}.

\bibitem{Salopek:1990nonlinear}
D.~S. Salopek and J.~R. Bond, ``{Nonlinear evolution of long wavelength metric
  fluctuations in inflationary models},''
  \href{http://dx.doi.org/10.1103/PhysRevD.42.3936}{{\em Phys. Rev. D}
  {\bfseries 42} (1990) 3936--3962}.

\bibitem{Wands:2000anew}
D.~Wands, K.~A. Malik, D.~H. Lyth, and A.~R. Liddle, ``{A New approach to the
  evolution of cosmological perturbations on large scales},''
  \href{http://dx.doi.org/10.1103/PhysRevD.62.043527}{{\em Phys. Rev. D}
  {\bfseries 62} (2000) 043527},
  \href{http://arxiv.org/abs/astro-ph/0003278}{{\ttfamily
  arXiv:astro-ph/0003278}}.

\bibitem{Vennin:2015correlation}
V.~Vennin and A.~A. Starobinsky, ``{Correlation Functions in Stochastic
  Inflation},'' \href{http://dx.doi.org/10.1140/epjc/s10052-015-3643-y}{{\em
  Eur. Phys. J. C} {\bfseries 75} (2015) 413},
  \href{http://arxiv.org/abs/1506.04732}{{\ttfamily arXiv:1506.04732
  [hep-th]}}.

\bibitem{Lovelock:1971yv}
D.~Lovelock, ``{The Einstein tensor and its generalizations},''
  \href{http://dx.doi.org/10.1063/1.1665613}{{\em J. Math. Phys.} {\bfseries
  12} (1971) 498--501}.

\bibitem{Padmanabhan:2013xyr}
T.~Padmanabhan and D.~Kothawala, ``{Lanczos-Lovelock models of gravity},''
  \href{http://dx.doi.org/10.1016/j.physrep.2013.05.007}{{\em Phys. Rept.}
  {\bfseries 531} (2013) 115--171},
  \href{http://arxiv.org/abs/1302.2151}{{\ttfamily arXiv:1302.2151 [gr-qc]}}.

\bibitem{Weinberg:2008hq}
S.~Weinberg, ``{Effective Field Theory for Inflation},''
  \href{http://dx.doi.org/10.1103/PhysRevD.77.123541}{{\em Phys. Rev. D}
  {\bfseries 77} (2008) 123541},
  \href{http://arxiv.org/abs/0804.4291}{{\ttfamily arXiv:0804.4291 [hep-th]}}.

\bibitem{Antoniadis:1993jc}
I.~Antoniadis, J.~Rizos, and K.~Tamvakis, ``{Singularity - free cosmological
  solutions of the superstring effective action},''
  \href{http://dx.doi.org/10.1016/0550-3213(94)90120-1}{{\em Nucl. Phys. B}
  {\bfseries 415} (1994) 497--514},
  \href{http://arxiv.org/abs/hep-th/9305025}{{\ttfamily arXiv:hep-th/9305025}}.

\bibitem{Kawai:1998ab}
S.~Kawai, M.-a. Sakagami, and J.~Soda, ``{Instability of one loop superstring
  cosmology},'' \href{http://dx.doi.org/10.1016/S0370-2693(98)00925-3}{{\em
  Phys. Lett. B} {\bfseries 437} (1998) 284--290},
  \href{http://arxiv.org/abs/gr-qc/9802033}{{\ttfamily arXiv:gr-qc/9802033}}.

\bibitem{Kawai:1999pw}
S.~Kawai and J.~Soda, ``{Evolution of fluctuations during graceful exit in
  string cosmology},''
  \href{http://dx.doi.org/10.1016/S0370-2693(99)00736-4}{{\em Phys. Lett. B}
  {\bfseries 460} (1999) 41--46},
  \href{http://arxiv.org/abs/gr-qc/9903017}{{\ttfamily arXiv:gr-qc/9903017}}.

\bibitem{Hwang:1999gf}
J.-c. Hwang and H.~Noh, ``{Conserved cosmological structures in the one loop
  superstring effective action},''
  \href{http://dx.doi.org/10.1103/PhysRevD.61.043511}{{\em Phys. Rev. D}
  {\bfseries 61} (2000) 043511},
  \href{http://arxiv.org/abs/astro-ph/9909480}{{\ttfamily
  arXiv:astro-ph/9909480}}.

\bibitem{Cartier:2001is}
C.~Cartier, J.-c. Hwang, and E.~J. Copeland, ``{Evolution of cosmological
  perturbations in nonsingular string cosmologies},''
  \href{http://dx.doi.org/10.1103/PhysRevD.64.103504}{{\em Phys. Rev. D}
  {\bfseries 64} (2001) 103504},
  \href{http://arxiv.org/abs/astro-ph/0106197}{{\ttfamily
  arXiv:astro-ph/0106197}}.

\bibitem{Guo:2006ct}
Z.-K. Guo, N.~Ohta, and S.~Tsujikawa, ``{Realizing scale-invariant density
  perturbations in low-energy effective string theory},''
  \href{http://dx.doi.org/10.1103/PhysRevD.75.023520}{{\em Phys. Rev. D}
  {\bfseries 75} (2007) 023520},
  \href{http://arxiv.org/abs/hep-th/0610336}{{\ttfamily arXiv:hep-th/0610336}}.

\bibitem{Satoh:2008ck}
M.~Satoh and J.~Soda, ``{Higher Curvature Corrections to Primordial
  Fluctuations in Slow-roll Inflation},''
  \href{http://dx.doi.org/10.1088/1475-7516/2008/09/019}{{\em JCAP} {\bfseries
  09} (2008) 019}, \href{http://arxiv.org/abs/0806.4594}{{\ttfamily
  arXiv:0806.4594 [astro-ph]}}.

\bibitem{Guo:2009uk}
Z.-K. Guo and D.~J. Schwarz, ``{Power spectra from an inflaton coupled to the
  Gauss-Bonnet term},''
  \href{http://dx.doi.org/10.1103/PhysRevD.80.063523}{{\em Phys. Rev. D}
  {\bfseries 80} (2009) 063523},
  \href{http://arxiv.org/abs/0907.0427}{{\ttfamily arXiv:0907.0427 [hep-th]}}.

\bibitem{Guo:2010jr}
Z.-K. Guo and D.~J. Schwarz, ``{Slow-roll inflation with a Gauss-Bonnet
  correction},'' \href{http://dx.doi.org/10.1103/PhysRevD.81.123520}{{\em Phys.
  Rev. D} {\bfseries 81} (2010) 123520},
  \href{http://arxiv.org/abs/1001.1897}{{\ttfamily arXiv:1001.1897 [hep-th]}}.

\bibitem{Jiang:2013gza}
P.-X. Jiang, J.-W. Hu, and Z.-K. Guo, ``{Inflation coupled to a Gauss-Bonnet
  term},'' \href{http://dx.doi.org/10.1103/PhysRevD.88.123508}{{\em Phys. Rev.
  D} {\bfseries 88} (2013) 123508},
  \href{http://arxiv.org/abs/1310.5579}{{\ttfamily arXiv:1310.5579 [hep-th]}}.

\bibitem{Koh:2014bka}
S.~Koh, B.-H. Lee, W.~Lee, and G.~Tumurtushaa, ``{Observational constraints on
  slow-roll inflation coupled to a Gauss-Bonnet term},''
  \href{http://dx.doi.org/10.1103/PhysRevD.90.063527}{{\em Phys. Rev. D}
  {\bfseries 90} no.~6, (2014) 063527},
  \href{http://arxiv.org/abs/1404.6096}{{\ttfamily arXiv:1404.6096 [gr-qc]}}.

\bibitem{Yi:2018gse}
Z.~Yi, Y.~Gong, and M.~Sabir, ``{Inflation with Gauss-Bonnet coupling},''
  \href{http://dx.doi.org/10.1103/PhysRevD.98.083521}{{\em Phys. Rev. D}
  {\bfseries 98} no.~8, (2018) 083521},
  \href{http://arxiv.org/abs/1804.09116}{{\ttfamily arXiv:1804.09116 [gr-qc]}}.

\bibitem{Odintsov:2018zhw}
S.~D. Odintsov and V.~K. Oikonomou, ``{Viable Inflation in Scalar-Gauss-Bonnet
  Gravity and Reconstruction from Observational Indices},''
  \href{http://dx.doi.org/10.1103/PhysRevD.98.044039}{{\em Phys. Rev. D}
  {\bfseries 98} no.~4, (2018) 044039},
  \href{http://arxiv.org/abs/1808.05045}{{\ttfamily arXiv:1808.05045 [gr-qc]}}.

\bibitem{Odintsov:2019clh}
S.~D. Odintsov and V.~K. Oikonomou, ``{Inflationary Phenomenology of Einstein
  Gauss-Bonnet Gravity Compatible with GW170817},''
  \href{http://dx.doi.org/10.1016/j.physletb.2019.134874}{{\em Phys. Lett. B}
  {\bfseries 797} (2019) 134874},
  \href{http://arxiv.org/abs/1908.07555}{{\ttfamily arXiv:1908.07555 [gr-qc]}}.

\bibitem{Gao:2020cvb}
T.-J. Gao, ``{Gauss\textendash{}Bonnet inflation with a constant rate of
  roll},'' \href{http://dx.doi.org/10.1140/epjc/s10052-020-08582-8}{{\em Eur.
  Phys. J. C} {\bfseries 80} no.~11, (2020) 1013},
  \href{http://arxiv.org/abs/2008.03976}{{\ttfamily arXiv:2008.03976 [gr-qc]}}.

\bibitem{Kawai:2021bye}
S.~Kawai and J.~Kim, ``{CMB from a Gauss-Bonnet-induced de Sitter fixed
  point},'' \href{http://dx.doi.org/10.1103/PhysRevD.104.043525}{{\em Phys.
  Rev. D} {\bfseries 104} no.~4, (2021) 043525},
  \href{http://arxiv.org/abs/2105.04386}{{\ttfamily arXiv:2105.04386
  [hep-ph]}}.

\bibitem{Kawai:2023nqs}
S.~Kawai and J.~Kim, ``{Probing the inflationary moduli space with
  gravitational waves},''
  \href{http://dx.doi.org/10.1103/PhysRevD.108.103537}{{\em Phys. Rev. D}
  {\bfseries 108} no.~10, (2023) 103537},
  \href{http://arxiv.org/abs/2308.13272}{{\ttfamily arXiv:2308.13272
  [astro-ph.CO]}}.

\bibitem{Addazi:2024gew}
A.~Addazi, Y.~Aldabergenov, and Y.~Cai, ``{Sound speed resonance of
  gravitational waves in Gauss-Bonnet-coupled inflation},''
  \href{http://dx.doi.org/10.1103/PhysRevD.110.123530}{{\em Phys. Rev. D}
  {\bfseries 110} no.~12, (2024) 123530},
  \href{http://arxiv.org/abs/2408.05091}{{\ttfamily arXiv:2408.05091 [gr-qc]}}.

\bibitem{Aldabergenov:2025oys}
Y.~Aldabergenov and D.~Berkimbayev, ``{Gauss\textendash{}Bonnet-Induced
  Symmetry Breaking/Restoration During Inflation},''
  \href{http://dx.doi.org/10.3390/universe11030098}{{\em Universe} {\bfseries
  11} no.~3, (2025) 98}, \href{http://arxiv.org/abs/2502.08986}{{\ttfamily
  arXiv:2502.08986 [hep-th]}}.

\bibitem{Nojiri:2005vv}
S.~Nojiri, S.~D. Odintsov, and M.~Sasaki, ``{Gauss-Bonnet dark energy},''
  \href{http://dx.doi.org/10.1103/PhysRevD.71.123509}{{\em Phys. Rev. D}
  {\bfseries 71} (2005) 123509},
  \href{http://arxiv.org/abs/hep-th/0504052}{{\ttfamily arXiv:hep-th/0504052}}.

\bibitem{Calcagni:2005im}
G.~Calcagni, S.~Tsujikawa, and M.~Sami, ``{Dark energy and cosmological
  solutions in second-order string gravity},''
  \href{http://dx.doi.org/10.1088/0264-9381/22/19/011}{{\em Class. Quant.
  Grav.} {\bfseries 22} (2005) 3977--4006},
  \href{http://arxiv.org/abs/hep-th/0505193}{{\ttfamily arXiv:hep-th/0505193}}.

\bibitem{Nojiri:2006je}
S.~Nojiri, S.~D. Odintsov, and M.~Sami, ``{Dark energy cosmology from
  higher-order, string-inspired gravity and its reconstruction},''
  \href{http://dx.doi.org/10.1103/PhysRevD.74.046004}{{\em Phys. Rev. D}
  {\bfseries 74} (2006) 046004},
  \href{http://arxiv.org/abs/hep-th/0605039}{{\ttfamily arXiv:hep-th/0605039}}.

\bibitem{Tsujikawa:2006ph}
S.~Tsujikawa and M.~Sami, ``{String-inspired cosmology: Late time transition
  from scaling matter era to dark energy universe caused by a Gauss-Bonnet
  coupling},'' \href{http://dx.doi.org/10.1088/1475-7516/2007/01/006}{{\em
  JCAP} {\bfseries 01} (2007) 006},
  \href{http://arxiv.org/abs/hep-th/0608178}{{\ttfamily arXiv:hep-th/0608178}}.

\bibitem{Koivisto:2006ai}
T.~Koivisto and D.~F. Mota, ``{Gauss-Bonnet Quintessence: Background Evolution,
  Large Scale Structure and Cosmological Constraints},''
  \href{http://dx.doi.org/10.1103/PhysRevD.75.023518}{{\em Phys. Rev. D}
  {\bfseries 75} (2007) 023518},
  \href{http://arxiv.org/abs/hep-th/0609155}{{\ttfamily arXiv:hep-th/0609155}}.

\bibitem{Neupane:2006dp}
I.~P. Neupane, ``{On compatibility of string effective action with an
  accelerating universe},''
  \href{http://dx.doi.org/10.1088/0264-9381/23/24/020}{{\em Class. Quant.
  Grav.} {\bfseries 23} (2006) 7493--7520},
  \href{http://arxiv.org/abs/hep-th/0602097}{{\ttfamily arXiv:hep-th/0602097}}.

\bibitem{Granda:2017oku}
L.~N. Granda and D.~F. Jimenez, ``{Dynamical analysis for a
  scalar\textendash{}tensor model with Gauss\textendash{}Bonnet and non-minimal
  couplings},'' \href{http://dx.doi.org/10.1140/epjc/s10052-017-5262-2}{{\em
  Eur. Phys. J. C} {\bfseries 77} no.~10, (2017) 679},
  \href{http://arxiv.org/abs/1710.04760}{{\ttfamily arXiv:1710.04760 [gr-qc]}}.

\bibitem{Chatzarakis:2019fbn}
N.~Chatzarakis and V.~K. Oikonomou, ``{Autonomous dynamical system of
  Einstein\textendash{}Gauss\textendash{}Bonnet cosmologies},''
  \href{http://dx.doi.org/10.1016/j.aop.2020.168216}{{\em Annals Phys.}
  {\bfseries 419} (2020) 168216},
  \href{http://arxiv.org/abs/1908.08141}{{\ttfamily arXiv:1908.08141 [gr-qc]}}.

\bibitem{Pozdeeva:2019agu}
E.~O. Pozdeeva, M.~Sami, A.~V. Toporensky, and S.~Y. Vernov, ``{Stability
  analysis of de Sitter solutions in models with the Gauss-Bonnet term},''
  \href{http://dx.doi.org/10.1103/PhysRevD.100.083527}{{\em Phys. Rev. D}
  {\bfseries 100} no.~8, (2019) 083527},
  \href{http://arxiv.org/abs/1905.05085}{{\ttfamily arXiv:1905.05085 [gr-qc]}}.

\bibitem{Vernov:2021hxo}
S.~Vernov and E.~Pozdeeva, ``{De Sitter Solutions in
  Einstein\textendash{}Gauss\textendash{}Bonnet Gravity},''
  \href{http://dx.doi.org/10.3390/universe7050149}{{\em Universe} {\bfseries 7}
  no.~5, (2021) 149}, \href{http://arxiv.org/abs/2104.11111}{{\ttfamily
  arXiv:2104.11111 [gr-qc]}}.

\bibitem{MohseniSadjadi:2023amn}
H.~Mohseni~Sadjadi, ``{Scalar-Gauss-Bonnet model, the coincidence problem and
  the gravitational wave speed},''
  \href{http://dx.doi.org/10.1016/j.physletb.2024.138508}{{\em Phys. Lett. B}
  {\bfseries 850} (2024) 138508},
  \href{http://arxiv.org/abs/2309.07816}{{\ttfamily arXiv:2309.07816 [gr-qc]}}.

\bibitem{Pinto:2024dnm}
M.~A.~S. Pinto and J.~a.~L. Rosa, ``{$\Lambda$CDM-like evolution in
  Einstein-scalar-Gauss-Bonnet gravity},''
  \href{http://arxiv.org/abs/2411.04066}{{\ttfamily arXiv:2411.04066 [gr-qc]}}.

\bibitem{TerenteDiaz:2023iqk}
J.~J. Terente~D{\'\i}az, K.~Dimopoulos, M.~Kar{\v{c}}iauskas, and A.~Racioppi,
  ``{Gauss-Bonnet Dark Energy and the speed of gravitational waves},''
  \href{http://dx.doi.org/10.1088/1475-7516/2023/10/031}{{\em JCAP} {\bfseries
  10} (2023) 031}, \href{http://arxiv.org/abs/2307.06163}{{\ttfamily
  arXiv:2307.06163 [astro-ph.CO]}}.

\bibitem{TerenteDiaz:2023kgc}
J.~J. Terente~D{\'\i}az, K.~Dimopoulos, M.~Kar{\v{c}}iauskas, and A.~Racioppi,
  ``{Quintessence in the Weyl-Gauss-Bonnet model},''
  \href{http://dx.doi.org/10.1088/1475-7516/2024/02/040}{{\em JCAP} {\bfseries
  02} (2024) 040}, \href{http://arxiv.org/abs/2310.08128}{{\ttfamily
  arXiv:2310.08128 [gr-qc]}}.

\bibitem{Kawai:2021edk}
S.~Kawai and J.~Kim, ``{Primordial black holes from Gauss-Bonnet-corrected
  single field inflation},''
  \href{http://dx.doi.org/10.1103/PhysRevD.104.083545}{{\em Phys. Rev. D}
  {\bfseries 104} no.~8, (2021) 083545},
  \href{http://arxiv.org/abs/2108.01340}{{\ttfamily arXiv:2108.01340
  [astro-ph.CO]}}.

\bibitem{Kawaguchi:2022nku}
R.~Kawaguchi and S.~Tsujikawa, ``{Primordial black holes from Higgs inflation
  with a Gauss-Bonnet coupling},''
  \href{http://dx.doi.org/10.1103/PhysRevD.107.063508}{{\em Phys. Rev. D}
  {\bfseries 107} no.~6, (2023) 063508},
  \href{http://arxiv.org/abs/2211.13364}{{\ttfamily arXiv:2211.13364
  [astro-ph.CO]}}.

\bibitem{Ashrafzadeh:2023ndt}
A.~Ashrafzadeh and K.~Karami, ``{Primordial Black Holes in Scalar Field
  Inflation Coupled to the Gauss\textendash{}Bonnet Term with Fractional
  Power-law Potentials},''
  \href{http://dx.doi.org/10.3847/1538-4357/ad293f}{{\em Astrophys. J.}
  {\bfseries 965} no.~1, (2024) 11},
  \href{http://arxiv.org/abs/2309.16356}{{\ttfamily arXiv:2309.16356
  [astro-ph.CO]}}.

\bibitem{Solbi:2024zhl}
M.~Solbi and K.~Karami, ``{Primordial black holes in non-minimal
  Gauss\textendash{}Bonnet inflation in light of the PTA data},''
  \href{http://dx.doi.org/10.1140/epjc/s10052-024-13271-x}{{\em Eur. Phys. J.
  C} {\bfseries 84} no.~9, (2024) 918},
  \href{http://arxiv.org/abs/2403.00021}{{\ttfamily arXiv:2403.00021 [gr-qc]}}.

\bibitem{Vennin:2020kng}
V.~Vennin, {\em {Stochastic inflation and primordial black holes}}.
\newblock PhD thesis, U. Paris-Saclay, 6, 2020.
\newblock \href{http://arxiv.org/abs/2009.08715}{{\ttfamily arXiv:2009.08715
  [astro-ph.CO]}}.

\bibitem{Press:1973iz}
W.~H. Press and P.~Schechter, ``{Formation of galaxies and clusters of galaxies
  by selfsimilar gravitational condensation},''
  \href{http://dx.doi.org/10.1086/152650}{{\em Astrophys. J.} {\bfseries 187}
  (1974) 425--438}.

\bibitem{Musco:2018rwt}
I.~Musco, ``{Threshold for primordial black holes: Dependence on the shape of
  the cosmological perturbations},''
  \href{http://dx.doi.org/10.1103/PhysRevD.100.123524}{{\em Phys. Rev. D}
  {\bfseries 100} no.~12, (2019) 123524},
  \href{http://arxiv.org/abs/1809.02127}{{\ttfamily arXiv:1809.02127 [gr-qc]}}.

\bibitem{Firouzjahi:2018vet}
H.~Firouzjahi, A.~Nassiri-Rad, and M.~Noorbala, ``{Stochastic Ultra Slow Roll
  Inflation},'' \href{http://dx.doi.org/10.1088/1475-7516/2019/01/040}{{\em
  JCAP} {\bfseries 01} (2019) 040},
  \href{http://arxiv.org/abs/1811.02175}{{\ttfamily arXiv:1811.02175
  [hep-th]}}.

\bibitem{bullock1997non}
J.~S. Bullock and J.~R. Primack, ``Non-gaussian fluctuations and primordial
  black holes from inflation,'' {\em Physical Review D} {\bfseries 55} no.~12,
  (1997) 7423.

\bibitem{Vennin_2017}
V.~Vennin, H.~Assadullahi, H.~Firouzjahi, M.~Noorbala, and D.~Wands, ``Critical
  number of fields in stochastic inflation,''
  \href{http://dx.doi.org/10.1103/physrevlett.118.031301}{{\em Physical Review
  Letters} {\bfseries 118} no.~3, (Jan., 2017) }.
  \url{http://dx.doi.org/10.1103/PhysRevLett.118.031301}.

\bibitem{Assadullahi_2016}
H.~Assadullahi, H.~Firouzjahi, M.~Noorbala, V.~Vennin, and D.~Wands, ``Multiple
  fields in stochastic inflation,''
  \href{http://dx.doi.org/10.1088/1475-7516/2016/06/043}{{\em Journal of
  Cosmology and Astroparticle Physics} {\bfseries 2016} no.~06, (June, 2016)
  043–043}. \url{http://dx.doi.org/10.1088/1475-7516/2016/06/043}.

\bibitem{Vennin_2025}
V.~Vennin and D.~Wands, {\em Quantum Diffusion and Large Primordial
  Perturbations from Inflation},
  \href{http://dx.doi.org/10.1007/978-981-97-8887-3_8}{p.~201–227}.
\newblock Springer Nature Singapore, 2025.
\newblock \url{http://dx.doi.org/10.1007/978-981-97-8887-3_8}.

\bibitem{Ezquiaga:2019ftu}
J.~M. Ezquiaga, J.~Garc{\'\i}a-Bellido, and V.~Vennin, ``{The exponential tail
  of inflationary fluctuations: consequences for primordial black holes},''
  \href{http://dx.doi.org/10.1088/1475-7516/2020/03/029}{{\em JCAP} {\bfseries
  03} (2020) 029}, \href{http://arxiv.org/abs/1912.05399}{{\ttfamily
  arXiv:1912.05399 [astro-ph.CO]}}.

\bibitem{Ashrafzadeh:2024oll}
A.~Ashrafzadeh, M.~Solbi, S.~Heydari, and K.~Karami, ``{Primordial black holes
  in SB SUSY Gauss-Bonnet inflation},''
  \href{http://dx.doi.org/10.1088/1475-7516/2025/01/025}{{\em JCAP} {\bfseries
  01} (2025) 025}, \href{http://arxiv.org/abs/2407.15445}{{\ttfamily
  arXiv:2407.15445 [hep-th]}}.

\bibitem{Carr:2020gox}
B.~Carr, K.~Kohri, Y.~Sendouda, and J.~Yokoyama, ``{Constraints on primordial
  black holes},'' \href{http://dx.doi.org/10.1088/1361-6633/ac1e31}{{\em Rept.
  Prog. Phys.} {\bfseries 84} no.~11, (2021) 116902},
  \href{http://arxiv.org/abs/2002.12778}{{\ttfamily arXiv:2002.12778
  [astro-ph.CO]}}.

\bibitem{Auffinger:2022khh}
J.~Auffinger, ``{Primordial black hole constraints with Hawking
  radiation{\textemdash}A review},''
  \href{http://dx.doi.org/10.1016/j.ppnp.2023.104040}{{\em Prog. Part. Nucl.
  Phys.} {\bfseries 131} (2023) 104040},
  \href{http://arxiv.org/abs/2206.02672}{{\ttfamily arXiv:2206.02672
  [astro-ph.CO]}}.

\bibitem{Honda:2024evc}
M.~Honda, R.~Jinno, and K.~Tokeshi, ``{Exactly solvable stochastic
  spectator},'' \href{http://dx.doi.org/10.1088/1475-7516/2024/12/044}{{\em
  JCAP} {\bfseries 12} (2024) 044},
  \href{http://arxiv.org/abs/2409.16272}{{\ttfamily arXiv:2409.16272
  [astro-ph.CO]}}.

\bibitem{Nasuda:2025gyd}
Y.~Nasuda, K.~Tokeshi, and Y.~Watanabe, ``{Scalar field stochastic dynamics in
  de Sitter spacetime from exact solutions of quantum deficient oscillators},''
  \href{http://arxiv.org/abs/2505.21429}{{\ttfamily arXiv:2505.21429
  [hep-th]}}.

\bibitem{Teimoori:2021thk}
Z.~Teimoori, K.~Rezazadeh, and K.~Karami, ``{Primordial Black Holes Formation
  and Secondary Gravitational Waves in Nonminimal Derivative Coupling
  Inflation},'' \href{http://dx.doi.org/10.3847/1538-4357/ac01cf}{{\em
  Astrophys. J.} {\bfseries 915} no.~2, (2021) 118},
  \href{http://arxiv.org/abs/2107.08048}{{\ttfamily arXiv:2107.08048 [gr-qc]}}.

\bibitem{Pattison_2017}
C.~Pattison, V.~Vennin, H.~Assadullahi, and D.~Wands, ``Quantum diffusion
  during inflation and primordial black holes,''
  \href{http://dx.doi.org/10.1088/1475-7516/2017/10/046}{{\em Journal of
  Cosmology and Astroparticle Physics} {\bfseries 2017} no.~10, (Oct., 2017)
  046–046}. \url{http://dx.doi.org/10.1088/1475-7516/2017/10/046}.

\bibitem{ozsoy2023inflation}
O.~{\"O}zsoy and G.~Tasinato, ``Inflation and primordial black holes,'' {\em
  Universe} {\bfseries 9} no.~5, (2023) 203.

\bibitem{Heydari:2023rmq}
S.~Heydari and K.~Karami, ``{Primordial black holes and secondary gravitational
  waves from generalized power-law non-canonical inflation with quartic
  potential},'' \href{http://dx.doi.org/10.1140/epjc/s10052-024-12489-z}{{\em
  Eur. Phys. J. C} {\bfseries 84} no.~2, (2024) 127},
  \href{http://arxiv.org/abs/2310.11030}{{\ttfamily arXiv:2310.11030 [gr-qc]}}.

\bibitem{Heydari:2024bxj}
S.~Heydari and K.~Karami, ``{Primordial Black Holes Generated by Fast-roll
  Mechanism in Noncanonical Natural Inflation},''
  \href{http://dx.doi.org/10.3847/1538-4357/ad7605}{{\em Astrophys. J.}
  {\bfseries 975} no.~1, (2024) 148},
  \href{http://arxiv.org/abs/2405.08563}{{\ttfamily arXiv:2405.08563 [gr-qc]}}.

\end{thebibliography}
\end{document}